\documentclass[10pt]{article}

\usepackage{lmodern}
\usepackage[latin1]{inputenc} 
\usepackage[T1]{fontenc}
\usepackage[french,english,francais]{babel}
\usepackage{amsmath}
\usepackage{amsthm}
\usepackage{amssymb}
\usepackage{amsfonts}
\usepackage{amscd}
\usepackage{latexsym}
\usepackage{graphicx}
\usepackage{hyperref}
\usepackage{epsfig}
\usepackage{bm}
\usepackage{dsfont}
\usepackage{color}
\usepackage{stmaryrd}
\usepackage{etoolbox}
\patchcmd{\thebibliography}{\section*}{\section}{}{}
\usepackage{tikz-cd}

\makeatletter
\@addtoreset{equation}{section}
\makeatother
\newtheorem{theorem}{Theorem}[section]

\newtheorem{remark}{Remark}[section]
\newtheorem{proposition}{Proposition}[section]
\newtheorem{lemma}{Lemma}[section]
\newtheorem{corollary}{Corollary}[section]
\newtheorem{definition}{Definition}[section]
\def\br{\begin{remark}\rm\small}
\def\er{\end{remark}}
\def\bt{\begin{theorem}}
\def\et{\end{theorem}}
\def\bd{\begin{definition}}
\def\ed{\end{definition}}
\def\bp{\begin{proposition}}
\def\ep{\end{proposition}}
\def\bl{\begin{lemma}}
\def\el{\end{lemma}}
\def\bc{\begin{corollary}}
\def\ec{\end{corollary}}
\def\beaq{\begin{eqnarray}}
\def\eeaq{\end{eqnarray}}

\newcommand\encadremath[1]{\vbox{\hrule\hbox{\vrule\kern8pt
\vbox{\kern8pt \hbox{$\displaystyle #1$}\kern8pt}
\kern8pt\vrule}\hrule}}
\def\enca#1{\vbox{\hrule\hbox{
\vrule\kern8pt\vbox{\kern8pt \hbox{$\displaystyle #1$}
\kern8pt} \kern8pt\vrule}\hrule}}

\newcommand{\eq}[1]{eq.~(\ref{#1})}

\newcommand{\Res}{\mathop{\,\rm Res\,}}

\newcommand{\beq}{\begin{equation}}
\newcommand{\eeq}{\end{equation}}
\newcommand{\beqq}{\begin{equation*}}
\newcommand{\eeqq}{\end{equation*}}
\newcommand{\bea}{\begin{eqnarray}}
\newcommand{\eea}{\end{eqnarray}}
\newcommand{\beaa}{\begin{eqnarray*}}
\newcommand{\eeaa}{\end{eqnarray*}}

\newcommand\eop{\vspace*{\fill}\pagebreak}

\newcommand{\Tr}{\operatorname{Tr}}

\newcommand{\om}{\omega}

\begin{document}
\selectlanguage{english}
\begin{center}
\vspace{1cm}

{\Large \bf {Quantization of hyper-elliptic curves from isomonodromic systems and topological recursion}}

\vspace{0.5cm}

{Olivier Marchal}$^\dagger$\footnote{olivier.marchal@univ-st-etienne.fr},
{Nicolas Orantin}$^\ddagger$\footnote{nicolas.orantin@unige.ch},

\vspace{5mm}
$^\dagger$\, Universit\'e de  Lyon, Universit\'{e} Jean Monnet Saint-\'{E}tienne, CNRS UMR 5208, Institut Camille Jordan, F-42023 Saint-Etienne, France
\vspace{5mm}

$^\ddagger$\, University of Geneva , 2-4 rue du Li\`evre, 1211 Gen\`eve 4, Switzerland.

\vspace{5mm}

\end{center}
\vspace{1cm}

{\bf Abstract :}
We prove that the topological recursion formalism can be used to compute the WKB expansion of solutions of second order differential operators obtained by quantization of any hyper-elliptic curve. We express this quantum curve in terms of spectral Darboux coordinates on the moduli space of meromorphic $\mathfrak{sl}_2$-connections on $\mathbb{P}^1$ and argue that the topological recursion produces a $2g$-parameter family of associated tau functions, where $2g$ is the dimension of the moduli space considered. We apply this procedure to the 6 Painlev\'e equations which correspond to $g=1$ and consider a $g=2$ example.

\tableofcontents


\section{Introduction}

Since the pioneering works of Kontsevich \cite{Kont} proving Witten conjecture \cite{Witten} on  intersection numbers on the moduli space of Riemann surfaces, it is known that there is a big interplay between the theory of integrable systems and enumerative geometry, going through mirror symmetry. The original presentation states that there exists a generating function for intersections of $\psi$ classes on $\overline{\mathcal{M}}_{g,n}$ which is a KdV tau function. However, one can present it in a slightly different way as follows. Let us define a different generating series by
\beq
\Psi^K(x,\hbar):= \exp \left[ \sum_{g=0}^\infty\sum_{n=1}^\infty \frac{\hbar^{2g-2+n}}{n!} \sum_{\mathbf{k} \in \mathbb{N}^n} \int_{\overline{\mathcal{M}}_{g,n}} \prod_{j=1}^n \left[\psi_j^{k_j} \, \frac{(2k_i-1)!!}{x^{k_i+\frac{1}{2}}} \right]\right].
\eeq
The Virasoro constraints satisfied by the corresponding KdV tau function are equivalent to the Airy equation
\beq
\left(\hbar^2 \frac{\partial^2}{\partial x^2} - \frac{x}{4} \right) \Psi^K(x,\hbar) = 0.
\eeq
This differential operator is related to our enumerative problem by mirror symmetry. Indeed, $\Psi^K(x,\hbar)$ is a generating series for Gromov-Witten invariants of the point. This Gromov-Witten theory turns out to have a Landau-Ginzburg model defined on the Riemann surface $\{y^2 = \frac{x}{4}\} \subset \mathbb{C}^2$. The Airy equation can thus be interpreted as a quantization of the curve mirror symmetric to the Gromov-Witten theory of the point. 

One may wonder if this procedure mixing mirror symmetry and quantization of algebraic curves can be applied to other problems of enumerative geometry. More precisely, given a problem of enumerative geometry, can one build a generating series $\Psi(x,\hbar)$ for the numbers of interest such that it is annihilated by a differential operator
\beq
\hat{P}\left(x,\hbar \frac{\partial}{\partial x} \right) \Psi(x,\hbar) = 0
\eeq
whose classical limit $P(x,y) := \underset{\hbar\to 0}{\lim} \hat{P}(x,y) $ defines a Riemann surface $\{P(x,y) = 0\} \subset \mathbb{C}^2$ which is mirror to our enumerative problem? The topological recursion formalism \cite{EO} developed in the last 10 years and giving a universal solution to semi-simple Gromov-Witten theories \cite{DBOSS} is conjectured to give a positive answer to this question in a large setup. 

This problem consists in proving that the application of the topological recursion to a classical curve $\Sigma:=\{P(x,y) = 0\}$ allows to build a generating series $\Psi(x,\hbar)$ solution to a differential equation $\widehat{P}\left(x,\hbar \frac{\partial}{\partial x}\right)\Psi(x,\hbar) = 0$ \cite{BEInt}. This claim has been proved in the case when $\Sigma$ has genus zero \cite{Reconstruction,MO} and in a variety of examples (see \cite{ReviewNorbury} for a nice review of this topic) when considering variables in $\mathbb{C}^*$ instead of $\mathbb{C}$ as well. Unfortunately, until recently no example with higher genus Riemann surface $\Sigma$ was worked out. Some attempts in this direction have been made in the context of Painlev\'e equation where the expected genus is equal to 1. In \cite{Eynbook,P2,MarchalIwaki,IS}, solutions to Painlev\'e equations where built using topological recursion  starting from a singular genus 0 curve, namely by considering a singular point in a corresponding moduli space of quadratic differentials. These works use the general result of \cite{LoopLie} which is valid only for genus 0 Riemann surfaces.

A breakthrough has been made in \cite{Iwaki} where the author proved that the topological recursion can be used to quantize the Weierstrass curve $y^2 = x^3 + \alpha x + \beta$ for generic values of $\alpha$ and $\beta$. On the way, this allows to provide with a 2-parameter solution of Painlev\'e 1 equation.

This quantization procedure may not only be used to find solutions to Painlev\'e type equations but also to compute enumerative invariants as Gromov-Witten invariants or Hurwitz numbers \cite{DBGW} as explained above. The research activity on this topic is nowadays very active for its possible applications to the computation of knot invariants in the context of the volume conjecture \cite{BEKnot,DKnot,LiuHurwitz}.

In the present article, we generalize Iwaki's wonderful result to the quantization of any hyper-elliptic curve, paving the way to a possible generalization to any algebraic curve. Let us now summarize how this is done.

In Section \ref{sec-TR}, we recall the topological recursion formalism in the case of hyper-elliptic curves. Given a meromorphic quadratic differential $\phi_0$ on $\mathbb{P}^1$, let $\Sigma_{\phi_0}:=\overline{\{(x,y) \in \mathbb{C} \,\slash\, y^2 (dx)^2  = \phi_0\}} $ be an associated compact Riemann surface. We explain how the recursion associates a set of multilinear forms $\om_{h,n}$ on $\Sigma_{\phi_0}^n$  to such a quadratic differential together with a Torelli marking of $\Sigma_{\phi_0}$.

In Section \ref{sec-quad-diff}, we recall some well-known facts about the moduli space of quadratic differentials and explain how the output $(\om_{h,n})_{h,n}$ varies when $\phi_0$ moves in this space.

After these background sections, we present the first important result of this paper. Given the same data as above, one can collect the result of the topological recursion into a single generating series (see definition \ref{DefPerturbativeWaveFunctions})
\beq
\psi(x,\hbar) = \exp\left[ \sum_{h\geq 0} \sum_{n\geq 1} \frac{\hbar^{2h-2+n}}{n!} \int_{\gamma(x)} \cdots \int_{\gamma(x)} \om_{h,n} \right]
\eeq
where $\gamma(x)$ is a well chosen integration path in $\Sigma_{\phi_0}$ with end points in the fiber above a point $x$ in the base curve $\mathbb{P}^1$. If this function does not satisfy any differential equation by quantization of $\Sigma_{\phi_0}$, we prove in Theorem \ref{th-PDE} that it is a solution to a PDE with respect to $x$ and a subset of coordinates on the moduli space of quadratic differentials. The proof of this first important result follows the line of \cite{Quantum} taking into account the additional contributions involved by the non-vanishing genus of $\Sigma_{\phi_0}$.

In order to obtain a function annihilated by a quantum curve, one needs to correct $\psi(x,\hbar)$ by exponentially small corrections in $\hbar$ to build a ``non-perturbative'' analog. This non-perturbative wave function $\Psi(x,\hbar)$ is built in Definition \ref{def-non-pert} as a Fourier transform of $\psi(x,\hbar)$. We finally prove the main result of this article in Theorem \ref{QuantumCurveTheorem}. The latter proves that $\Psi(x,\hbar)$ is annihilated by a ``quantum curve''
\beq
\left[\hbar^2 \frac{\partial^2}{\partial x^2} - \hbar^2 R(x) \frac{\partial}{\partial x}  - \hbar Q(x)  -\mathcal{H}(x)\right] \Psi(x,\hbar) = 0.
\eeq
In this expression, all functions are rational functions of $x$ with poles at the poles of $\phi_0$ together with simple poles at a set of ``apparent singularities'' $(q_i)_{i=1}^g$ where $g$ is the genus of $\Sigma_{\phi_0}$. Theorem \ref{QuantumCurveTheorem} provides explicit expressions of these different functions. The proof of this theorem mainly relies on the fact that $\Psi(x,\hbar)$ is a solution to a PDE similar to the one satisfied by $\psi(x,\hbar)$ but that additionally it also has ``good'' monodromies along non-trivial cycles as explained in Lemma \ref{lemma-non-pert-monodromy}. This simple property allows to prove that some corresponding Wronskian functions are rational in $x$ eventually leading to the result.

In order to have a better geometrical understanding of the quantum curve thus produced, we linearize the system in Section \ref{SectionLaxRepresentation} replacing the quantum curve by a $\mathfrak{sl}_2$-connection on $\mathbb{P}_1$. We then study the corresponding characteristic variety, the $\hbar$-deformed spectral curve in Theorem \ref{th-sp-curve}. From this point of view, the topological recursion produces some flows in the $\hbar$-direction in a moduli space of quadratic differentials starting from the initial value $\phi_0$. Finally, in Section  \ref{sec-isomonodromic}, we embed the result in a corresponding isomonodromic system explaining how our procedure allows to build isomonodromic tau functions.

In Section \ref{sec-examples}, we apply the procedure to the simplest cases. We thus get 2-parameter solutions to the 6 Painlev\'e equations as well as corresponding tau functions. We also consider a genus 2 example leading to the second element in Painlev\'e 2 hierarchy.

\medskip

The present work gives one possible quantization of a spectral curve and is partly motivated by applications in mathematical physics. Indeed, recent progresses have been made in the computation of isomonodromic tau functions motivated by the possibility of interpreting the latter in terms of conformal blocks in associated Conformal Field Theories \cite{CGL,Teschner-and-co,Teschner}. From this perspective, our work only considers a base curve equal to $\mathbb{P}^1$ and the Lie algebra $\mathfrak{sl}_2$. We shall naturally consider a generalization of our work to the higher genus base curves and arbitrary semi-simple Lie algebras. We hope that the present article could generalize nicely to these cases. Indeed, the PDE of Theorem \ref{th-PDE} can be thought of as a loop equation in the general topological recursion formalism and should possibly be obtained for this general setup. The second step starting from this PDE to the quantum curve only uses the monodromy properties of the non-perturbative wave functions along cycles in the spectral cover. We believe that this step can be adapted to the general setup as well.

On another hand, we build only formal trans-series solutions to differential equations in the present article. It is fundamental to understand if, and when, these formal objects admit some Borel summability properties in order to study their Stokes properties in the spirit of the exact WKB analysis of \cite{IwakiExactWKB,IwakiExactWKB2}. We hope to address this issue as well as the related questions of the dependance of the wave functions to a choice of Torelli marking in a future work. This second question can be interpreted as asking if our quantization procedure depends on a choice of polarization. A choice of Torelli marking can indeed probably be interpreted as a choice of real polarization. Changing such a polarization will probably lead to modular properties related to cluster transformations as in the recent works on exact WKB computations mentioned above.

\bigskip

{\bf Acknowledgments:} 
Both authors are grateful to Kohei Iwaki for sharing with them a preliminary version of \cite{Iwaki} and explaining it in details.

N.O. would like to thank Philip Boalch, Ga\"etan Borot, Tom Bridgeland, Andrea Brini, Mattia Cafasso, John Harnad and Volodya Roubstov for helpful discussions on the subject as well as Universit\'e de Lyon for its hospitality. The work of N.O. is supported by the European grant ERC SYNERGY ``ReNewQuantum'', ERC-2018-SyG  810573, as a member of M. Mari\~no's team in Geneva University.

O.M. would like to thank Universit\'e Lyon $1$, Universit\'e Jean Monnet and Institut Camille Jordan for material support. This work was supported by the LABEX MILYON (ANR-10-LABX-0070) of Universit\'e de Lyon, within the program ``Investissements d'Avenir'' (ANR-11-IDEX-0007) operated by the French National Research Agency (ANR). 

As this article was being written, Bertrand Eynard and Elba Garcia-Failde informed us that they were working on a similar problem and propose an independent solution using different methods. We would like to thank them for letting us know about their work.

\bigskip

\section{Quadratic differentials and topological recursion}
\label{sec-TR}

In this section, we recall the formalism of topological recursion \cite{EO,Borot-review} starting from the data of a quadratic differential on the Riemann sphere.

\subsection{Quadratic differentials and initial data}

Let $\phi$ be a meromorphic quadratic differential on $\mathbb{P}^1$, namely it reads
\beq
\phi(x) = f_\phi(x) \left(dx\right)^2
\eeq
where $f_\phi(x)$ is a rational function of $x$. Let us denote by $\Sigma_\phi$ the compact Riemann surface:
\beq
\Sigma_\phi = \overline{ \left\{(x,y) \in \mathbb{C}^2| y^2 =  f_\phi(x) \right\}}
\eeq
where $\overline{\cdot}$ denotes the compactification at $x=\infty$. For reasons explained below, we sometimes call $\Sigma_\phi$ a \emph{classical spectral curve}.

\bd[Admissible curves]
A quadratic differential $\phi$ is called admissible if $\Sigma_\phi$ is a smooth algebraic curve such that, away from $x=\infty$, the poles of $\phi$ are distinct from the critical values of the map $x : \Sigma_\phi \to \mathbb{P}^1$.

Let us denote by $P_\phi$ the set of poles of $\phi$ on $\mathbb{P}^1$ and by $\mathcal{P}_\phi$ the set composed of the pre-images by $x$ of those poles  on $\Sigma_\phi$.

Let us denote by $\mathcal{R}_\phi:=\{a_i\}$ the set of finite ramification points of the map $x$ defined by 
\beq
\left\{
\begin{array}{l}
x(a_i) = u_i \neq \infty \cr
dx(a_i) = 0 \cr
\end{array}
\right. .
\eeq
In addition, we will the set  $\{u_i=x(a_i)\}$ will be referred to as critical points.
\ed

In its original version \cite{EO}, the topological recursion is a procedure taking as input an algebraic curve together with a Torelli marking. For this purpose, in the present paper, our input is defined in the following way.

\bd[Admissible initial data]
\sloppy{An admissible initial data is the data of a pair $\left(\phi_0,\left(\mathcal{A}_i,\mathcal{B}_i\right)_{i=1}^{g(\phi_0)} \right)$ where $\phi_0$ is an admissible quadratic differential, $g(\phi_0)$ denotes the genus of $\Sigma_{\phi_0}$
and $\left(\mathcal{A}_i,\mathcal{B}_i\right)_{i=1}^{g(\phi_0)}$ is a symplectic basis of $H_{1}\left(\Sigma_{\phi_0},\mathbb{Z}\right)$.

To any such initial data, one associates the initial values $\om_{0,1} \in H^0\left(\Sigma_{\phi_0}\setminus \mathcal{P}_\phi, K_{\Sigma_{\phi_0}\setminus \mathcal{P}_\phi}\right)$ and $\om_{0,2} \in H^0\left[\Sigma_{\phi_0} \times \Sigma_{\phi_0}, \left(p_1^* K_{\Sigma_{\phi_0}} \otimes p_2^*K_{\Sigma_{\phi_0}}\right) (-2 \Delta) \right]$ where $p_1$ and $p_2$ the projections $\Sigma_{\phi_0} \times \Sigma_{\phi_0} \to \Sigma_{\phi_0} $ on the first and second factor respectively and $\Delta$ is the diagonal divisor in $ \Sigma_{\phi_0} ^2$ in the following way.

By choosing a branch of the square root once and for all, one defines
\beq
\om_{0,1}[\phi_0]:= \left[ \phi_0\right]^{\frac{1}{2}}:= y dx
\eeq
where
$y^2 = f_\phi$.

$\om_{0,2}\left[\phi_0,\left(\mathcal{A}_i,\mathcal{B}_i\right)_{i=1}^g\right] $ is defined as the unique differential on $\Sigma_{\phi_0}\times \Sigma_{\phi_0}$ whose only singularities are double poles without residue on the diagonal and normalized by
\beq
\om_{0,2}(z_1,z_2) =\frac{dz_1 \otimes dz_2}{(z_1-z_2)^2} + O(dz_1 \otimes dz_2)
\eeq
in any local coordinates as $z_1 \to z_2$ and 
\beq
\forall\, i\in\llbracket1, g\rrbracket \, : \; \oint_{z_1 \in \mathcal{A}_i} \om_{0,2}(z_1,z_2)  = 0.
\eeq}
\ed

It is worth noticing that $\om_{0,1}$ does not depend on the Torelli marking while the latter fixes $\om_{0,2}$ uniquely.

\subsection{Topological recursion}\label{SectionTopRec}

Let us now present the topological recursion formalism in this simple setup. It is an inductive procedure associating to any admissible initial data $\left(\phi_0,\left(\mathcal{A}_i,\mathcal{B}_i\right)_{i=1}^{g(\phi_0)} \right)$, a set of differential forms $\om_{h,n}\left[\phi_0,\left(\mathcal{A}_i,\mathcal{B}_i\right)_{i=1}^{g(\phi_0)} \right] \in H^0\left(\Sigma_{\phi_0}^n , K_{\Sigma_{\phi_0}}^{\otimes n}(- (6h-6+4n) \mathcal{R}^{[n]}) \right)$ where $\mathcal{R}^{[n]} = \underset{i=1}{\overset{n}{\bigsqcup}} p_i^{-1}  (\mathcal{R})$, $p_i: \Sigma_{\phi_0}^n \to \Sigma_{\phi_0}$ being the projection along the $i^{\text{th}}$ component. It is defined as follows:

\bd
For any admissible initial data $\left(\phi_0,\left(\mathcal{A}_i,\mathcal{B}_i\right)_{i=1}^{g(\phi_0)} \right)$, let us define $\om_{0,1}[\phi_0]$ and $\om_{0,2}\left[\phi_0,\left(\mathcal{A}_i,\mathcal{B}_i\right)\right]$ as above and, for $2h-2+n \geq 0$, $\om_{h,n}\left[\phi_0,\left(\mathcal{A}_i,\mathcal{B}_i\right)_{i=1}^{g(\phi_0)} \right] \in H^0\left(\Sigma_{\phi_0}^n , K_{\Sigma_{\phi_0}}^{\otimes n}(- (6g-6+4n) \mathcal{R}^{[n]}) \right)$ is defined inductively by
\beaa
\om_{h,n}(z_1,\dots,z_n) &:=& \sum_{p \in \mathcal{R}} \Res_{z \to p} \frac{\int_{\sigma(z)}^z \om_{0,2}(z_1, \cdot)}{2 \left(\om_{0,1}(z) - \om_{0,1}(\sigma(z))\right)}
\Bigg[ \om_{h-1,n+1}(z, \sigma(z), z_2,\dots , z_n) \cr
&& + \sum_{\begin{array}{c} h_1+h_2 = h \cr A \sqcup B = \{z_2,\dots,z_n\} \cr (h_1,|A|) \notin \{(0,0),(h,n-1)\}  \cr \end{array}} \om_{h_1,|A|+1}(z,A) \,\om_{h_2,|B|+1}(z,B) \Bigg] 
\eeaa
where $\sigma: \Sigma_{\phi_0} \to \Sigma_{\phi_0}$ is the hyper-elliptic involution, namely, it is defined by 
\beq
\forall\, z \in \Sigma_{\phi_0} \setminus \mathcal{R}  \, , \; x(z)= x(\sigma(z)) \quad  \hbox{and} \quad \sigma(z) \neq z.
\eeq

For $h \geq 2$, we define the free energies $\om_{h,0}\left[\phi_0,\left(\mathcal{A}_i,\mathcal{B}_i\right)_{i=1}^{g(\phi_0)} \right] \in \mathbb{C}$ by
\beqq
\om_{h,0} := \frac{1}{2-2h} \sum_{p \in \mathcal{R}} \Res_{z \to p} \om_{h,1}(z) \,  \int_{o}^z \om_{0,1}
\eeqq
where $o \in \Sigma_{\phi_0}$ is an arbitrary base point of which $\om_{h,0}$ is independent.\\

Finally, for $h\in\{0,1\}$, we can define $\om_{0,0}$ and $\om_{1,0}$. Their explicit expressions being technical and useless for our purpose, we refer the reader to \cite{EO} for them. The only important point is that they are defined in such a way that they satisfy the upcoming variational formulas of Lemma \ref{LemmaVariations}. 

\ed

In the expression above and in the following, we do not write the dependence of $\om_{h,n}$ in the initial data except when we want to emphasize it.

\section{Space of quadratic differentials and deformations}
\label{sec-quad-diff}

An important property of the objects built by topological recursion is that they have nice properties under variations of the initial data. One can actually think of $\om_{h,n}$ as a generating series of $n^{\text{th}}$ order derivatives of $\om_{h,0}$ with respect to an infinite number of parameters in the space of initial data. In order to make this statement more precise, we shall now consider a finite dimensional sub-space of the space of initial data.

\subsection{Space of quadratic differentials and coordinates}

Let $n\geq 0$ be a positive integer and let $(X_\nu)_{\nu=1}^n$ be a set of distinct points in $\mathbb{C}$.
Let $(r_\nu)_{\nu=1}^n\in \left(\mathbb{N}\setminus \{0\}\right)^{n}$  be a set of degrees associated to the points $(X_\nu)_{\nu=1}^n$. Let us denote by $r_\infty \in \mathbb{N}\setminus\{0\}$ a degree at infinity. Without loss of generality, we shall assume the existence of a pole at infinity in this article to avoid cumbersome notations.
One defines by $D = \underset{\nu=1}{\overset{n}{\sum}} r_{\nu} (X_\nu) + r_\infty (\infty)$ the resulting divisor. 

\bd[Space of quadratic differentials $\mathcal{Q}(\mathbb{P}^1,D,n_\infty)$]
Given a divisor $D = \underset{\nu=1}{\overset{n}{\sum}} r_{\nu} (X_\nu)+ r_\infty (\infty)$ and $n_\infty \in \{0,1\}$, let $\mathcal{Q}(\mathbb{P}^1,D,n_\infty)$ be the moduli space of quadratic differentials on $\mathbb{P}^1$ such that any $\phi \in \mathcal{Q}(\mathbb{P}^1,D,n_\infty)$ has a pole of order at most $2 r_{\nu}$ at the finite pole $X_\nu \in \mathcal{P}^{\text{finite}}$ and a pole of order at most $2 r_\infty - n_\infty$ at infinity.
\ed

Note in particular that \textbf{the degree of the pole at infinity may be even or odd while for finite poles we assume that the degrees are always even}. This requirement will be more transparent from the integrable systems perspective where we may allow at most only one odd pole to avoid degeneracy. Using a trivial reparameterization, this pole may always be chosen to be $\infty$, a widespread convention used in many examples like Painlev\'{e} equations.

$\mathcal{Q}(\mathbb{P}^1,D,n_\infty)$ is a finite dimensional space equipped with a Poisson structure (see for example \cite{Bridgeland,Bertola-Korotkin} for a recent account close to our presentation). In the present paper, we shall be interested only in the formal neighborhood of a point in this space hence we shall not discuss any global property of such spaces. On the contrary, we will now describe local coordinates around a quadratic differential $\phi_0$. 

A first set of coordinates is given by the coefficients of the partial fraction decomposition of $f_\phi$
\beq
f_\phi = \sum_{k=0}^{2(r_\infty - 2) - n_\infty} H_{\infty,k} x^k + \sum_{\nu=1}^n \sum_{k=1}^{2r_\nu} \frac{H_{\nu,k}}{(x-X_\nu)^k}.
\eeq

As a moduli space, this space can be equipped with a Poisson structure in such a way that part of these coefficients are Casimirs (see for example \cite{AHH2,Darboux}). Fixing them allows restricting to a symplectic leaf of this Poisson manifold. We shall now present and use different coordinates that are natural from the topological recursion perspective, allowing fixing the values of these Casimirs.

\bd

Given $\phi \in \mathcal{Q}(\mathbb{P}^1,D,n_\infty)$ the associated one form $\om_{0,1}:= \phi^{\frac{1}{2}}$ is meromorphic on $\Sigma_\phi$ with poles along the pre-image of the points in the divisor $D$. Let us define the pre-images of the poles by $\left\{b_{\nu}^+,b_\nu^-\right\} := x^{-1}\left(X_\nu\right)$ for $\nu \in\llbracket 1,n\rrbracket$ and $\left\{b_{\infty}^+,b_\infty^-\right\} := x^{-1}\left(\infty \right)$ if $n_\infty  = 0$ and $\left\{b_\infty \right\}:= x^{-1}\left(\infty \right)$ if $n_\infty  = 1$. Let us denote by $\mathcal{P} := \mathcal{P}^{\text{finite}} \sqcup  \mathcal{P}^{\infty}$ the set of poles on $\Sigma_\phi$ where $ \mathcal{P}^{\text{finite}}:=\{b_\nu^+,b_\nu^-\}_{\nu=1}^n$ and $\mathcal{P}^{\infty} := \{b_\infty^+,b_\infty^-\}$ (resp. $\mathcal{P}^{\infty} := \{b_\infty\}$) if $n_\infty = 0$ (resp. $n_\infty  = 1$).

\ed

To obtain a symplectic leaf of our system, one shall fix the singular behavior of $\om_{0,1}$ around its poles, i.e. its residue and singular type defined below.

\bd[Times]
For any $\mathbf{T} \in \mathbb{C}^{r_\infty+\underset{\nu=1}{\overset{n}{\sum}} r_\nu-n_\infty}$, where $\mathbf{T}$ has components labeled $T_{\infty,k}$ with $k \in\llbracket 1,,  r_\infty\rrbracket$ (resp. $k \in\llbracket 2,  r_\infty\rrbracket$) if $n_\infty = 0$ (resp. if $n_\infty =1$) and $T_{\nu,k}$ for $\nu\in\llbracket1, n\rrbracket$ and $k\in \llbracket 1,r_\nu\rrbracket$, let $\mathcal{Q}(\mathbb{P}^1,D ,n_\infty, \mathbf{T}) \subset \mathcal{Q}(\mathbb{P}^1,D,n_\infty)$ be the space of quadratic differentials (known as the Whitham-Krichever differentials) such that $\om_{0,1}[\phi]$ has the following Laurent expansions 
\begin{itemize}
\item  around $b_\nu^\pm$,
\beqq
\om_{0,1}[\phi] = \pm \sum_{k=1}^{r_{\nu}} T_{\nu,k} \frac{dx}{(x-X_\nu)^{k}} + O\left(dx\right);
\eeqq

\item around $b_\infty^\pm$,
\beqq
\om_{0,1}[\phi] = \pm \sum_{k=1}^{r_\infty} T_{\infty,k} (x^{-1})^{-k} d(x^{-1}) + O(d(x^{-1}))= \mp \sum_{k=1}^{r_\infty} T_{\infty,k} x^{k-2} dx + O(x^{-2} dx)
\eeqq
 if $n_\infty =0$;
 
 \item around $b_\infty$,
\beqq
\om_{0,1}[\phi] = \sum_{k=2}^{r_\infty} T_{\infty,k} x^{k-1} d(x^{-\frac{1}{2}}) + O( d(x^{-\frac{1}{2}}) )= - \sum_{k=2}^{r_\infty} \frac{T_{\infty,k}}{2} x^{k-\frac{5}{2}} dx + O( d(x^{-\frac{1}{2}}) )
\eeqq
 if $n_\infty =1$.

\end{itemize}

\ed

For generic values of the times $\mathbf{T}$, $\mathcal{Q}(\mathbb{P}^1,D,n_\infty, \mathbb{T})$ is a symplectic manifold of real dimension equal to twice the genus of $\Sigma_\phi$
\beq
g(\Sigma_\phi) = r_\infty+ \sum_{\nu=1}^n r_\nu -3.
\eeq
One can understand this by remarking that fixing the value of $\mathbf{T}$ fixes the values of the coefficients $\left\{H_{\infty,k}\right\}_{k=r_\infty-3}^{2 r_\infty-4-n_\infty}$ and
$\left\{H_{\nu,k}\right\}_{k=r_\nu+1}^{2r_\nu}$ for $\nu\in \llbracket1,n\rrbracket$ unambiguously. The latter are Casimirs of our system. The remaining $g(\Sigma_\phi)$ coefficients $\left\{H_{\infty,k}\right\}_{k=0}^{r_\infty-4}$ and $\left\{H_{\nu,k}\right\}_{k=1}^{r_\nu}$ for $\nu\in\llbracket 1,n\rrbracket$ are Hamiltonians of the residual integrable system. 

We will not use these coefficients to parametrize the space $\mathcal{Q}(\mathbb{P}^1,D,n_\infty, \mathbf{T})$ but rather homological coordinates given by the periods of $\om_{0,1}[\phi]$ in the spirit of the original topological recursion \cite{EO} and more recently \cite{Bertola-Korotkin,Bridgeland,Eyn-periods}.

\bd[Periods]
For admissible initial data $\left(\phi, (\mathcal{A}_i,\mathcal{B}_i)_{i=1}^{g(\Sigma_\phi)}\right)$, let the period vector $\boldsymbol{\epsilon} \in \mathbb{C}^{g(\Sigma_\phi)}$ be defined by
\beq
\forall\, i\in\llbracket 1, g(\Sigma_\phi)\rrbracket \, : \; \epsilon_i := \oint_{\mathcal{A}_i} \om_{0,1}.
\eeq
\ed

Remark that, if the times $\mathbf{T}$ depend only on the quadratic differential, the periods depend on a choice of Torelli marking. This choice of Torelli marking can be interpreted as a choice of real polarization \cite{Bertola-Korotkin}.

Hence, any $\phi \in \mathcal{Q}(\mathbb{P}^1,D,n_\infty)$ takes the form
\beq
\phi(\mathbf{T},\boldsymbol{\epsilon}) =  \left[
\sum_{k=r_\infty-3}^{2(r_\infty - 2) - n_\infty} H_{\infty,k}(\mathbf{T}) \, x^k + \sum_{k=0}^{r_\infty - 4} H_{\infty,k}(\mathbf{T},\boldsymbol{\epsilon}) \, x^k  + \sum_{\nu=1}^n \left(\sum_{k=r_\nu+1}^{2r_\nu} \frac{H_{\nu,k}(\mathbf{T})}{(x-X_\nu)^k} + \sum_{k=1}^{r_\nu} \frac{H_{\nu,k}(\mathbf{T},\boldsymbol{\epsilon})}{(x-X_\nu)^k} \right)
\right]
\eeq
where $\mathbf{T}$ and $\boldsymbol{\epsilon}$ are local coordinates.

\subsection{Variational formulas}

Fixing a divisor $D$, the topological recursion taking as initial data a quadratic differential $\phi_0 \in \mathcal{Q}(\mathbb{P}^1,D,n_\infty)$ is a procedure generating functions on a neighborhood of $\phi_0 \in \mathcal{Q}(\mathbb{P}^1,D,n_\infty)$ with value in different spaces of differentials. Using the local coordinates given by the times and periods described above, one can study them as functions of the latter. The general theory developed for the topological recursion provides a nice way to compute the derivatives of such functions \cite{EO}.

Remark that, because the coefficients of the expansion around $b_\infty^+$ and $b_\infty^-$ are not independent, the variational formulas for $T_{\infty,k}$ include residues both at $b_\infty^+$ and $b_{\infty}^-$. The same subtlety arises for the coefficients of the expansion around $b_{\nu}^\pm$.

\bl\label{LemmaVariations}[Variational formulas]

Given admissible initial data $(\phi_0,(\mathcal{A}_i,\mathcal{B}_i)_{i=1}^{g(\Sigma_{\phi_0})})$, the output of the topological recursion $\left(\om_{h,n}\right)_{n\geq 0,h\geq 0}$ satisfies

\begin{itemize}

\item For the times associated to $\infty$:
\beq
\forall\, k \geq 2 \, , \; \frac{\partial \om_{h,n}(\mathbf{z})}{\partial T_{\infty,k}} = \Res_{ p \to b_\infty^+} \om_{h,n+1}(p,\mathbf{z}) \frac{x(p)^{k-1}}{k-1} - \Res_{ p \to b_\infty^-} \om_{h,n+1}(p,\mathbf{z}) \frac{x(p)^{k-1}}{k-1}
\eeq
if $n_\infty =0$  and
\beq
\forall \,k \geq 2 \, , \; \frac{\partial \om_{h,n}(\mathbf{z})}{\partial T_{\infty,k}} = \Res_{p \to b_\infty} \om_{h,n+1}(p,\mathbf{z}) \frac{x(p)^{k-1-\frac{1}{2}}}{2k-3}
\eeq
otherwise.

\item For the times associated to $b_\nu^\pm$:
\beq
\forall \,k \geq 2 \, , \; \frac{\partial \om_{h,n}(\mathbf{z})}{\partial T_{\nu,k}} = \Res_{p \to b_\nu^+} \om_{h,n+1}(p,\mathbf{z}) \frac{(x(p)-X_\nu)^{-k+1}}{k-1} - \Res_{p \to b_\nu^-} \om_{h,n+1}(p,\mathbf{z}) \frac{(x(p)-X_\nu)^{-k+1}}{k-1}
\eeq
and
\beq
\frac{\partial \om_{h,n}(\mathbf{z})}{\partial T_{\nu,1}}  = \int_{b_\nu^+}^p \om_{h,n+1}(\cdot,\mathbf{z}) - \int_{b_\nu^-}^p \om_{h,n+1}(\cdot,\mathbf{z}) .
\eeq

\item For the periods:
\beq
\forall \,j \in\llbracket1,g\rrbracket \, : \, \frac{\partial \om_{h,n}(\mathbf{z})}{\partial \epsilon_j} = \frac{1}{2 \pi i } \oint_{\mathcal{B}_j} \om_{h,n+1}(\cdot,\mathbf{z}).
\eeq
\end{itemize}
\el

Remark that one can perform variations with respect to any time by considering $\phi_0$ as a point in $\mathcal{Q}(\mathbb{P}^1,D,n_\infty)$ for $r_\nu$ and $r_\infty$ large enough.

From this point of view, $\om_{h,n+m}$ can be thought of as a generating function for the $m^{\text{th}}$ derivative of $\om_{h,n}$ with respect to the parameters $\mathbf{T}$ and $\boldsymbol{\epsilon}$. In particular, one can express the coefficients of the expansion of $\om_{0,1}$ around any of these poles in these terms.

\bc
The expansion of $\om_{0,1}$ in local coordinates around its poles reads
\begin{itemize}
\item  around $b_\nu^\pm$,
\beqq
\forall \, l \geq 2 \, : \; \om_{0,1}[\phi] = \pm \sum_{k=1}^{r_{\nu}} T_{\nu,k} \frac{dx}{(x-X_\nu)^{k}} \pm \sum_{k=2}^l \frac{k-1}{2}  \frac{\partial \om_{0,0}}{\partial T_{\nu,k}} (x-X_\nu)^{k-2} dx + O\left((x-X_\nu)^{l-1} dx\right)
\eeqq

\item around $b_\infty^\pm$,
\beqq
\forall \, l \geq 2 \, : \;  \om_{0,1}[\phi] = \mp \sum_{k=1}^{r_\infty} T_{\infty,k} x^{k-2} dx \mp \sum_{k=2}^l \frac{k-1}{2} \frac{\partial \om_{0,0}}{\partial T_{\infty,k}} x^{-k} dx + O(x^{-l-1} dx)
\eeqq
 if $n_\infty =0$;
 
 \item around $b_\infty$,
\beqq
\forall\, l \geq 2 \, : \; \om_{0,1}[\phi])= - \sum_{k=2}^{r_\infty} \frac{T_{\infty,k}}{2} x^{k-\frac{5}{2}} dx - \sum_{k=2}^l \frac{2k-3}{2} \frac{\partial \om_{0,0}}{\partial T_{\infty,k}} x^{-k+ \frac{1}{2}} dx
+O\left( x^{-l-\frac{1}{2}} dx \right)
\eeqq
 if $n_\infty =1$.
\end{itemize}
\ec

Thanks to this simple corollary, one can get some expressions of the quadratic differential emphasizing the dependence on the times and the periods.

\bl\label{Decompositionfphi} 
A quadratic differential $\phi \in \mathcal{Q}(\mathbb{P}^1,D,n_\infty,\mathbf{T})$ reads
\bea
f_\phi &=& \left[\left( \sum_{k=1}^{r_\infty} T_{\infty,k} x^{k-2} \right)^2 \right]_{\infty,+} + \sum_{\nu=1}^n \left[\left( \sum_{k=1}^{r_{\nu}} T_{\nu,k} \frac{dx}{(x-X_\nu)^{k}} \right)^2 \right]_{X_\nu,-} \cr
&&+  \sum_{k\in K_\infty} U_{\infty,k}(x) \frac{\partial \om_{0,0} }{\partial T_{\infty,k}} 
+ \sum_{\nu =1}^n\sum_{k\in K_{\nu}} U_{\nu,k}(x) \frac{\partial \om_{0,0}}{\partial T_{\nu,k}}
\eea
if $n_\infty = 0$ and
\bea
f_\phi &=& \left[\left( \sum_{k=2}^{r_\infty} \frac{T_{\infty,k}}{2} x^{k-\frac{5}{2}} \right)^2 \right]_{\infty,+} + \sum_{\nu=1}^n \left[\left( \sum_{k=1}^{r_{\nu}} T_{\nu,k} \frac{dx}{(x-X_\nu)^{k}} \right)^2 \right]_{X_\nu,-} \cr
&&+  \sum_{k\in K_\infty} U_{\infty,k}(x) \frac{\partial \om_{0,0} }{\partial T_{\infty,k}} 
+ \sum_{\nu =1}^n\sum_{k\in K_{\nu}} U_{\nu,k}(x) \frac{\partial \om_{0,0}}{\partial T_{\nu,k}} 
\eea
if $n_\infty =1$.\\
Here, $[f(x)]_{\infty,+}$ (resp.  $[f(x)]_{X_\nu,-}$) refers to the positive part of the expansion in $x$ of a function $f(x)$ around $\infty$, including the constant term, (resp. the strictly negative part of the expansion in $(x-X_\nu)$ around $X_\nu$) and we have defined 
\begin{itemize}
\item $K_\infty=\llbracket 2,r_\infty-2\rrbracket$  and $\forall \, k\in K_\infty$:
\beq
U_{\infty,k}(x) := (k-1) \sum_{l=k+2}^{r_\infty} T_{\infty,l} \, x^{l-k-2} \,,\,\text{ if } n_\infty=0
\eeq
and
\beq
U_{\infty,k}(x) := \left( k- \frac{3}{2} \right) \sum_{l=k+2}^{r_\infty} T_{\infty,l} \, x^{l-k-2}\,,\, \text{ if } n_\infty=1
\eeq

\item  $K_{\nu}=\llbracket 2, r_{\nu}+1\rrbracket$ and $\forall \,k\in K_{\nu}$:
\beq
U_{\nu,k}(x) := (k-1) \sum_{l=k-1}^{r_{\nu}} T_{\nu,l} \, (x-X_\nu)^{-l+k-2}
\eeq
\end{itemize}
\el

\begin{proof}
The proof immediately follows from the partial fraction decomposition of $f_\phi$ and the expression of the first holomorphic terms of the expansion of $\om_{0,1}$ around one of its poles using the variational formulas of Lemma \ref{LemmaVariations}.
\end{proof}

\begin{remark}
Note that in the expression of Lemma \ref{Decompositionfphi}, $\phi$ depends on the periods only through $\om_{0,0}$.
\end{remark}

\subsection{Symmetries}

In addition to the variational formulas, the output of the topological recursion is skew-symmetric under the hyper-elliptic involution $\sigma$, i.e.
\beq\label{eq-skew-sym}
\forall \,h \geq 0 \, , \, \forall \,n \geq 1 \, : \; \om_{h,n}(z_1,\dots,z_n) + \om_{h,n}(\sigma(z_1),z_2,\dots,z_n) = \delta_{h,0} \delta_{n,2} \frac{dx(z_1) \, dx(z_2)}{(x(z_1)-x(z_2))^2}.
\eeq

One also has 
\beq\label{eq-sym-berg}
\forall \,(z_1,z_2) \in (\Sigma_\phi)^2  \setminus \Delta \, , \; \om_{0,2}(z_1,z_2) =  \om_{0,2}(\sigma(z_1),\sigma(z_2)).
\eeq

One can use these symmetry properties to obtain easily a few equalities that we shall use repetitively in the following:
\beq\label{eq-skew-sym-int}
\forall\, (z_1,z_2) \in \Sigma_\phi^2 \setminus \Delta \, , \; \int_{\sigma(z_1)}^{z_1} \om_{0,2}(z_2,\cdot) = - \int_{\sigma(z_1)}^{z_1} \om_{0,2}(\sigma(z_2),\cdot) 
\eeq
which implies, for any ramification point $a$ (thus satisfying $\sigma(a)=a$),
\beq\label{eq-skew-sym-int-bp}
\forall\, (z_1,z_2) \in \Sigma_\phi^2 \setminus \Delta \, , \; \int_{a}^{z_1} \int_{\sigma(z_2)}^{z_2} \om_{0,2}= - \int_{a}^{\sigma(z_1)} \int_{\sigma(z_2)}^{z_2} \om_{0,2}.
\eeq

\section{The perturbative world}

\subsection{Perturbative partition function}

Given an admissible initial data $(\phi_0, (\mathcal{A}_i,\mathcal{B}_i))$, one can build generating functions collecting the quantities  $(\omega_{h,n})_{h\geq 0, n\geq 0}$ defined in Section \ref{SectionTopRec}. Since the variational formulas allow to think of $\om_{h,n}$ as generating functions for variations of $\om_{h,0}$ it makes sense to collect the information obtained from the topological recursion only in the latter. For this reason, one defines a partition function as

\bd[Perturbative partition function]
Given an admissible initial data, one defines the perturbative partition function as a function of a formal parameter $\hbar$ and the initial data by
\beq
Z^{\text{pert}}(\hbar,\mathbf{T},\boldsymbol{\epsilon}):= \exp \left( \sum_{h=0}^\infty \hbar^{2h-2} \om_{h,0}(\mathbf{T},\boldsymbol{\epsilon}) \right).
\eeq
\ed

It follows from this definition that $Z^{\text{pert}}(\hbar,\mathbf{T},\boldsymbol{\epsilon}) \exp(-\hbar^{-2} \om_{0,0})$ is a formal power series in $\hbar^2$.

\subsection{Perturbative wave functions}

In order to quantize the classical spectral curve, we would like to define some wave functions as some particular generating series of the correlators $\om_{h,n}$ for $n \geq 1$. We first define:

\bd[Definition of $(F_{h,n})_{h\geq 0,n\geq 1}$ by integration of the correlators] 
For $n \geq 1$ and $h\geq 0$ such that $2h-2+n \geq 1$, let us define 
\beqq F_{h,n}(z_1,\dots,z_n) = \frac{1}{2^n} \int_{\sigma(z_1)}^{z_1} \dots \int_{\sigma(z_n)}^{z_n} \om_{h,n}\eeqq
where one integrates each of the $n$ variables along paths linking two Galois conjugate points inside a fundamental domain cut out by the chosen symplectic basis $\left(\mathcal{A}_j,\mathcal{B}_j\right)_{1\leq j\leq g(\Sigma_\phi)}$.\\
For $(h,n)=(0,1)$ we define similarly
\beqq F_{0,1}(z):= \frac{1}{2} \int_{\sigma(z)}^z \om_{0,1} .\eeqq
Finally, for $(h,n) = (0,2)$ one cannot define $F_{0,2}$ in the exact same way since $\om_{0,2}$ has poles on the diagonal $\Delta$. One thus needs to regularize it by removing the polar part. Hence, we define 
\beqq
F_{0,2}(z_1,z_2):= \frac{1}{4} \int_{\sigma(z_1)}^{z_1} \int_{\sigma(z_2)}^{z_2} \om_{0,2}  - \frac{1}{2} \ln\left(x(z_1)-x(z_2)\right)
\eeqq
which also reads, in terms of theta functions,
\beqq
F_{0,2}(z_1,z_2) = \frac{1}{4}  \ln \left(\frac{\Theta( \mathbf{v}(z_1) - \mathbf{v}(z_2) + \mathbf{c}) \, \Theta(\mathbf{v}(\sigma(z_1)) - \mathbf{v}(\sigma(z_2)) + \mathbf{c})  }{\Theta( \mathbf{v}(z_1) - \mathbf{v}(\sigma(z_2)) + \mathbf{c}) \, \Theta( \mathbf{v}(\sigma(z_1)) - \mathbf{v}(z_2) + \mathbf{c})} \right) - \frac{1}{2} \ln\left(x(z_1)-x(z_2)\right) 
\eeqq
where $\mathbf{v}$ denotes the Abel-Jacobi map and $\mathbf{c}$ is a non-singular half-integer odd characteristic. 
\ed

\br Note that since $\mathbf{v}(\sigma(z)) = - \mathbf{v}(z)$,  and by skew-symmetry of the theta function, $F_{0,2}(z_1,z_2)$ may alternatively be written as
\beq
F_{0,2}(z_1,z_2) = \frac{1}{2}  \ln \left(\frac{\Theta( \mathbf{v}(z_1) - \mathbf{v}(z_2) + \mathbf{c})}{\Theta( \mathbf{v}(z_1) - \mathbf{v}(\sigma(z_2)) + \mathbf{c}) }\right) - \frac{1}{2} \ln\left(x(z_1)-x(z_2)\right)
\eeq
\er

Another useful way to rewrite $F_{0,2}(z_1,z_2)$ is the following lemma.

\bl\label{RewritingF02}
For any pair of distinct ramification points $(a_i,a_j)$, one has
\beq
F_{0,2}(z_1,z_2) = - \int_{a_i}^{\sigma(z_1)} \int_{a_j}^{z_2} \om_{0,2} - \frac{1}{2} \log \left( \frac{(u_i-x(z_2)) (x(z_1)-u_j)}{u_i-u_j} \right) 
\eeq
where we recall that $u_i := x(a_i)$. In particular, this reformulation shows that $F_{0,2}(z,z)$ is well-defined.
\el

\begin{proof}
Since $(a_i,a_j)$ are ramification points, they satisfy $\sigma(a_i)=a_i$ and $\sigma(a_j)=a_j$. Then, we have
\bea
\frac{1}{4} \int_{\sigma(z_1)}^{z_1} \int_{\sigma(z_2)}^{z_2} \om_{0,2}  
&=& \frac{1}{4} \left[ \int_{a_i}^{z_1} \int_{\sigma(z_2)}^{z_2} \om_{0,2}  -  \int_{a_i}^{\sigma(z_1)} \int_{\sigma(z_2)}^{z_2} \om_{0,2}  \right] 
\stackrel{(\eqref{eq-skew-sym-int-bp})}{=} -\frac{1}{2} \int_{a_i}^{\sigma(z_1)} \int_{\sigma(z_2)}^{z_2} \om_{0,2}  \cr
& = &  -\frac{1}{2} \int_{a_i}^{\sigma(z_1)}  \left[ \int_{a_j}^{z_2} \om_{0,2} - \int_{a_j}^{\sigma(z_2)} \om_{0,2} \right]\cr
&\stackrel{z\to \sigma(z)}{=}& -\frac{1}{2} \int_{a_i}^{\sigma(z_1)} \int_{a_j}^{z_2} \left[ \om_{0,2}(\cdot,\cdot) - \om_{0,2}(\cdot,\sigma(\cdot)) \right] \cr
& \stackrel{(\eqref{eq-skew-sym})}{=} & - \int_{a_i}^{\sigma(z_1)} \int_{a_j}^{z_2} \om_{0,2} + \frac{1}{2} \int_{z = a_i}^{\sigma(z_1)} \int_{z' = a_j}^{z_2} \frac{dx(z) dx(z')}{(x(z)-x(z'))^2} \cr
&&\eea
leading to the result.
\end{proof}

One can now collect functions $(F_{h,n})_{h\geq 0, n\geq 1}$ into generating series.

\bd[Definition of the perturbative wave functions]\label{DefPerturbativeWaveFunctions}
We define:
\beaa
S_{-1}^{\pm\text{ pert}}(x) &:=& \pm F_{0,1}(z(x))\cr
S_0^{\pm\text{ pert}}(x) &:=& \frac{1}{2} F_{0,2}(z(x),z(x))\cr
\forall \, k \geq 1 \, , \;  S_k^{\pm\text{ pert}}(x)&:=& \sum_{\substack{h\geq 0,n\geq 1\\ 2h-2+n = k}} \frac{(\pm 1)^n}{n!} F_{h,n}(z(x),\dots,z(x)) 
\eeaa
where, for any $\lambda \in \mathbb{P}^1$, we define $z(\lambda) \in \Sigma_\phi$ as the unique point such that $x(z(\lambda)) = \lambda$ and $\om_{0,1}(z(\lambda))= \sqrt{\phi(\lambda)}$\footnote{Let us stress that, as an argument of the function, $x$ refers to a point in $\mathbb{P}^1$ and not the map $x:\Sigma_{\phi_0} \to \mathbb{P}^1$. We hope that the reader will not be confused by this notation.}. Remark that the $\pm$ sign refers to the choice of sheet when choosing a point in the pre-image of $\lambda$. Note that $S_0^{\pm\text{ pert}}$ makes sense since Lemma \ref{RewritingF02} allows the evaluation of $F_{0,2}$ on the diagonal.\\
Eventually, we define the perturbative wave functions $\psi_{\pm}$ by:
\beqq {\psi}_\pm(x,\hbar,\mathbf{T},\boldsymbol{\epsilon}):= \exp \left(\sum_{k\geq -1}  \hbar^k S_k^{\pm\text{ pert}}(x)\right)\eeqq
\ed

\br
The perturbative partition function $\psi$ should actually be a function of the integration path on $\Sigma_{\phi_0}$ between two points. Since, for a given $x \in \mathbb{P}^1$ there are only two such paths up to homotopy in a given fundamental domain linking the two points in the fiber above $x$, we abusively write $\psi(x)$ as a function of $x$. The $\pm$ index referring to a choice of orientation of the path, i.e. to the choice of one of the two possible integration contours.

By abuse of notations, we shall sometimes denote the perturbative wave functions as ${\psi}_\pm(x,\hbar)$ forgetting about the dependance on the other parameters. Remark that the perturbative wave functions satisfy 
\beq
{\psi}_-(x,\hbar) = {\psi}_+(x,-\hbar).
\eeq

\er

\br Definition \ref{DefPerturbativeWaveFunctions} is identical to the one proposed in \cite{Quantum} and used in many papers like \cite{Reconstruction, Eyn-periods, IwakiExactWKB, IwakiExactWKB2, MO}.
\er

\subsection{Properties and PDE satisfied by the perturbative wave functions}

The perturbative wave functions have non-trivial monodromies along elements of $H_1(\Sigma_\phi,\mathbb{Z})$ that can be computed explicitly.

\bl\label{MonodromyPerturbative}
The perturbative wave functions $\psi_\pm$ satisfy the following properties.
\begin{itemize}
\item For $i\in\llbracket 1,g\rrbracket$, the function $\psi_{\pm}(x,\hbar,\mathbf{T},\boldsymbol{\epsilon})$ has a formal monodromy along  $\mathcal{A}_i$ given by
\beq
\psi_{\pm}(x,\hbar,\mathbf{T},\boldsymbol{\epsilon}) \mapsto e^{\pm 2 \pi i \frac{\epsilon_i}{\hbar}} \psi_{\pm}(x,\hbar,\mathbf{T},\boldsymbol{\epsilon}) .
\eeq

\item For $i\in\llbracket 1,g\rrbracket$, the function $\psi_{\pm}(x,\hbar,\mathbf{T},\boldsymbol{\epsilon})$ has a formal monodromy along  $\mathcal{B}_i$ given by
\beq
\psi_{\pm}(x,\hbar,\mathbf{T},\boldsymbol{\epsilon}) \mapsto \frac{Z^{\text{pert}}(\hbar,\mathbf{T},\boldsymbol{\epsilon}\pm \hbar \,\mathbf{e}_i)}{Z^{\text{pert}}(\hbar,\mathbf{T},\boldsymbol{\epsilon})} \psi_{\pm}(x,\hbar,\mathbf{T},\boldsymbol{\epsilon}\pm \hbar \,\mathbf{e}_i) 
\eeq
where $\mathbf{e}_i \in \mathbb{C}^g$ is the vector with the $i^{\text{th}}$ component equal to $1$ and all others vanishing.
\end{itemize}
\el

\begin{proof}

Reminding that the $\mathcal{A}$-periods of the $\om_{h,n}$ are vanishing unless for $(h,n) = (0,1)$ where
\beq
\forall \,j\in\llbracket 1,g\rrbracket \, : \; \epsilon_j = \oint_{\mathcal{A}_j} \om_{0,1},
\eeq
one immediately gets the first claim.

The second claim follows a simple computation similar to the one for Painlev\'e 1 written in \cite{Iwaki}. 

The analytic continuation of the perturbative wave function along the cycle ${\mathcal B}_j$ reads 
\bea
&&\exp \left[ \sum_{h\geq 0} \sum_{n \geq 1} \frac{\hbar^{2h-2} (\pm \hbar)^n}{n! \, 2^n} \sum_{n_1+n_2= n}\left(\begin{array}{c} n \cr n_1\cr \end{array} \right) 2^{n_1} \overbrace{\oint_{\mathcal{B}_j} \dots \oint_{\mathcal{B}_j}}^{n_1} \overbrace{\int_{\sigma(z)}^z \dots \int_{\sigma(z)}^z}^{n_2} \om_{h,n}
\right] \cr
&& =\exp \left[ \sum_{h\geq 0} \sum_{n \geq 1} \hbar^{2h-2} (\pm \hbar)^n \sum_{n_1+n_2= n}  \frac{1}{2^{n_2} n_1! \, n_2!} \frac{\partial^{n_1}}{\partial \epsilon_j^{n_1}} \overbrace{\int_{\sigma(z)}^z \dots \int_{\sigma(z)}^z}^{n_2} \om_{h,n_2}
\right] .
\eea
Factoring out the terms with $n_2=0$ gives
\beq
\exp \left[  \sum_{h\geq 0} \sum_{n \geq 1} \frac{ \hbar^{2h-2} (\pm \hbar)^n}{n! } \frac{\partial^{n}}{\partial \epsilon_j^{n}} \om_{h,0} \right]
\exp \left[\sum_{n_1\geq 0}  \frac{(\pm \hbar)^{n_1}}{n_1!}  \frac{\partial^{n_1}}{\partial \epsilon_j^{n_1}} \sum_{h\geq 0} \sum_{n_2 \geq 1} \hbar^{2h-2} (\pm \hbar)^{n_2}   \frac{1}{2^{n_2} \, n_2!} \overbrace{\int_{\sigma(z)}^z \dots \int_{\sigma(z)}^z}^{n_2} \om_{h,n_2}
\right] 
\eeq
leading to the result.

\end{proof}

The perturbative wave functions are built to be solutions of a PDE. Indeed, the main theorem of this section states that

\bt\label{th-PDE}
The perturbative wave functions are solutions of the PDE 
\beq\label{eq-PDE}
\left[\hbar^2 \frac{\partial^2}{\partial x^2}  - \hbar^2 \sum_{k\in K_\infty} U_{\infty,k}(x) \frac{\partial}{\partial T_{\infty,k}} 
- \hbar^2 \sum_{\nu =1}^n\sum_{k\in K_{b_\nu}} U_{\nu,k}(x) \frac{\partial}{\partial T_{\nu,k}} - {H}(x) \right] \psi_\pm(x,\hbar) = 0
\eeq
where
\beq \label{HHH}
{H}(x) = \left[\hbar^2 \sum_{k\in K_\infty} U_{\infty,k}(x) \frac{\partial}{\partial T_{\infty,k}} 
+ \hbar^2 \sum_{\nu =1}^n\sum_{k\in K_{b_\nu}} U_{\nu,k}(x) \frac{\partial}{\partial T_{\nu,k}} \right] \left[ \log  Z^{\text{pert}}(\hbar) - \hbar^{-2} \om_{0,0} \right] + \frac{\phi_0(x)}{(dx)^2} .
\eeq
\et

\begin{remark}The dependence in $x$ is completely explicit in \eqref{eq-PDE} and \eqref{HHH}. Coefficients in \eqref{HHH} may be computed from the knowledge of  $Z^{\text{pert}}$ given by topological recursion.
\end{remark}

\begin{proof}

The proof follows from the combination of a few lemmas. We first prove in Section \ref{app-proof-PDE-1} that
\bl\label{lemma-PDE-1}
For any $h \geq 0$ and $n \geq 1$ satisfying $2h-2+n \geq 2$,  the combination
\beq\label{eq-holo-1}
\begin{array}{l}
d_{z_1} F_{h,n}(z_1, \dots,z_n) +
{\displaystyle \sum_{j=2}^n} \int_{\sigma(z_j)}^{z_j} \om_{0,2}(z_1,\cdot)  \left[\frac{ d_{z_1} F_{h,n-1}(z_1,\mathbf{z}_{\{2,\dots,n\}\setminus \{j\}}) }{2 \om_{0,1}(z_1)} 
- \frac{ d_{z_j} F_{h,n-1}(z_j,\mathbf{z}_{\{2,\dots,n\}\setminus \{j\}}) }{2 \om_{0,1}(z_j)} \right]  \cr
  + \frac{1}{2 \om_{0,1}(z_1)} d_{u_1} d_{u_2} \Bigg[ F_{h-1,n+1}(u_1,u_2,z_2,\dots,z_n) \cr
 \qquad + {\displaystyle \overset{\mathrm{stable}}{\sum_{\begin{array}{c}
h_1+h_2 = h \cr
A\sqcup B = \{z_2,\dots,z_n\} \cr
\end{array}}}} 
F_{h_1,|A|+1}(u_1,\mathbf{z}_A) \, F_{h_2,|B|+1}(u_2,\mathbf{z}_B)
\Bigg]_{u_1 = u_2 = z_1} 
\cr
\end{array}
\eeq
is a holomorphic form in the first variable $z_1$. In this expression, ${\displaystyle \overset{\mathrm{stable}}{\sum_{\begin{array}{c}
h_1+h_2 = h \cr
A\sqcup B = \{z_2,\dots,z_n\} \cr
\end{array}}}} $ refers to the sum where $2h_1-2+|A| >0$ and $2h_2-2+|B| >0$.

In the same way, 
\footnotesize{\beq \label{eq-holo-2}
d_{z_1} F_{0,3}(z_1,z_2,z_3) + \frac{\int_{\sigma(z_2)}^{z_2} \om_{0,2}(z_1,\cdot) \; \int_{\sigma(z_3)}^{z_3} \om_{0,2}(z_1,\cdot)}{4 \om_{0,1}(z_1)}
 - \frac{\int_{\sigma(z_2)}^{z_2} \om_{0,2}(z_1,\cdot) \; \int_{\sigma(z_3)}^{z_3} \om_{0,2}(z_2,\cdot)}{4 \om_{0,1}(z_2)}
- \frac{\int_{\sigma(z_3)}^{z_3} \om_{0,2}(z_1,\cdot) \; \int_{\sigma(z_2)}^{z_2} \om_{0,2}(z_3,\cdot)}{4 \om_{0,1}(z_3)} 
\eeq}
\normalsize{and}
\beq
d_{z_1} F_{1,1}(z_1) - \frac{\om_{0,2}(z_1,\sigma(z_1))}{2 \om_{0,1}(z_1)}
\eeq
are holomorphic forms in the first variable $z_1$.
\el

Then, we have the following lemma.
\bl\label{eq-res-holo} For any  holomorphic differential $\om$ on $\Sigma_\phi$, one has
\beqq
- 2 \frac{y(z_1)}{dx(z_1)}  \om(z_1) = \sum_{p \in \mathcal{P}} \Res_{z_2 \to p} \frac{\om(z_2) \, y(z_2)}{x(z_2)-x(z_1)}.
\eeqq
\el
\begin{proof}
We first remark that for any holomorphic differential $\om$ on $\Sigma_\phi$, one has
\beqq
\sum_{p \in \mathcal{P}} \Res_{z_2 \to p} \frac{\om(z_2) \, y(z_2)}{x(z_2)-x(z_1)} = - \Res_{z_2 \to z_1,\sigma(z_1)}  \frac{\om(z_2) \, y(z_2)}{x(z_2)-x(z_1)}.
\eeqq
Indeed, writing $\om(z) = \frac{P(x(z))dx(z)}{y(z)}$ where $P(x)$ is a rational function, one sees that there is no contribution from the boundary of a fundamental domain when moving the integration contour and that $\om(z)  = - \om(\sigma(z))$. Computing the residue gives
\beq
\sum_{p \in \mathcal{P}} \Res_{z_2 \to p} \frac{\om(z_2) \, y(z_2)}{x(z_2)-x(z_1)} = \frac{y(z_1)}{dx(z_1)} \left( \om(\sigma(z_1)) - \om(z_1) \right),
\eeq
so that we obtain the proof of Lemma \ref{eq-res-holo}.
\end{proof}

\bigskip

Eventually, application of Lemma \ref{eq-res-holo} to the holomorphic differentials of Lemma \ref{lemma-PDE-1} before considering the diagonal specialization $z_1 = \dots = z_n = z$ gives, after some elementary but lengthy computations presented in Appendix \ref{app-proof-PDE-2}, the results of Theorem \ref{th-PDE}.
\end{proof}

\section{The non-perturbative world and the quantum curve}

In the previous section, we have seen that the perturbative wave functions are not annihilated by the naive quantization of the classical spectral curve. This differential operator in the variable $x$ needs to be corrected by a combination of linear operators in the times. We shall now see that these perturbative wave functions can be corrected by exponentially small corrections as $\hbar \to 0$ to produce some non-perturbative analogs that are annihilated by a quantization of the classical spectral curve.

\subsection{Definitions}

Out of the perturbative wave function, one can build a non-perturbative analog that will be the main character of the present article. The latter is defined as a discrete Fourier transform of the perturbative wave function with respect to the $\mathcal{A}$-periods as originally expected from \cite{Eyn-np}. In this section, we pick an admissible initial data $(\phi_0,(\mathcal{A}_j,\mathcal{B}_j)_{j=1}^{g(\Sigma_\phi)})$ where $\phi_0 \in \mathcal{Q}(\mathbb{P}^1,D,n_\infty,\mathbf{T})$. For simplicity, we denote by $g:=g(\Sigma_{\phi_0})$ the genus of the classical spectral curve and by $\boldsymbol{\epsilon}=(\epsilon_i)_{i=1}^g$ the associated periods.

\bd\label{def-non-pert}
Let the non-perturbative partition function be the Fourier transform
\beqq
Z(\mathbf{T},\boldsymbol{\epsilon} , \boldsymbol{\rho}) := {\sum_{\mathbf{k} \in \mathbb{Z}^g} e^{\frac{ 2  \pi i}{\hbar} {\displaystyle \sum_{j=1}^g} k_j \rho_j} Z^{pert}(\hbar,\mathbf{T},\boldsymbol{\epsilon}+ \hbar \mathbf{k}).}
\eeqq
In the same way, one can define the non-perturbative wave function by
\beqq
\Psi_{\pm}(x,\mathbf{T},\boldsymbol{\epsilon} , \boldsymbol{\rho}) := \frac{1}{Z(\mathbf{T},\boldsymbol{\epsilon} , \boldsymbol{\rho})}{\displaystyle \sum_{\mathbf{k} \in \mathbb{Z}^g}} e^{ \frac{ 2  \pi i}{\hbar}{\displaystyle \sum_{j=1}^g} k_j \rho_j}Z^{pert}(\hbar,\mathbf{T},\boldsymbol{\epsilon}+ \hbar \mathbf{k}) \;  \psi_{\pm}(x,\hbar, \mathbf{T}, \boldsymbol{\epsilon} + \hbar \mathbf{k})
\eeqq
\ed

As functions of $\hbar$, these non-perturbative objects are of very different nature compared to their perturbative counterparts whose logarithms admit a formal series expansion in $\hbar$.  Considering the $\hbar$ expansion of the perturbative partition function, one can observe that
\beq
\frac{Z^{pert}(\hbar,\mathbf{T},\boldsymbol{\epsilon}+ \hbar \mathbf{k})}{Z^{pert}(\hbar,\mathbf{T},\boldsymbol{\epsilon})} = \exp \left[ \frac{ 2  \pi i}{\hbar} \sum_{j=1}^g k_j \phi_j + {\displaystyle \sum_{j,l=1}^g} \pi i k_l k_j \tau_{l,j} \right] 
\exp \left[ \sum_{h\geq 0} \sum_{n\geq \max(1,3-2h)} \sum_{\mathbf{i} \in \llbracket 1,g \rrbracket^n} \frac{\hbar^{2h-2+n}}{\text{Aut}(\mathbf{i})} \prod_{j=1}^n k_{i_j} \frac{\partial^n \om_{h,0}}{\partial \epsilon_{i_1} \dots \partial \epsilon_{i_n}}
\right]
\eeq
where, denoting by $l(\mathbf{i})$ the number of times $l$ appears in the vector $\mathbf{i}$, 
$\text{Aut}(\mathbf{i}) := \underset{l}{\prod} l(\mathbf{i})!$,
$\phi_i:=\frac{\partial \om_{0,0}}{\partial \epsilon_i}$ and
$\tau_{i,j}:= \frac{\partial^2 \om_{0,0}}{\partial \epsilon_i \, \partial \epsilon_j}$.

This allows to recombine the Fourier transform under the form
\beqq
\frac{Z(\mathbf{T},\boldsymbol{\epsilon} , \boldsymbol{\rho})}{Z^{pert}(\hbar,\mathbf{T},\boldsymbol{\epsilon})}  = \sum_{\mathbf{k} \in \mathbb{Z}^g} e^{\frac{ 2  \pi i}{\hbar}{\displaystyle \sum_{j=1}^g} k_j \left(\rho_j + \phi_j\right) + {\displaystyle \sum_{l,j=1}^g} \pi i k_l k_j \tau_{l,j} } \exp \left[ \sum_{h\geq 0} \sum_{n\geq \max(1,3-2h)} \sum_{\mathbf{i} \in \llbracket 1,g \rrbracket^n} \frac{\hbar^{2h-2+n}}{\text{Aut}(\mathbf{i})} \prod_{j=1}^n k_{i_j} \frac{\partial^n \om_{h,0}}{\partial \epsilon_{i_1} \dots \partial \epsilon_{i_n}} \right] .
\eeqq
Hence, it takes the form
\beq
Z(\mathbf{T},\boldsymbol{\epsilon} , \boldsymbol{\rho}) = Z^{pert}(\hbar,\mathbf{T},\boldsymbol{\epsilon}) \sum_{m=0}^\infty \hbar^m \Theta_m(\hbar, \mathbf{T}, \boldsymbol{\epsilon}, \boldsymbol{\rho})
\eeq
where the coefficients $\Theta_m(\hbar, \mathbf{T}, \boldsymbol{\epsilon}, \boldsymbol{\rho})$ are obtained as finite linear combinations of derivatives of theta functions of the form
$
\left. \frac{\partial^n \theta(\mathbf{v}, \boldsymbol{\tau})}{\partial v_{i_1} \dots \partial v_{i_n}} \right|_{\mathbf{v} = \frac{\mathbf{\phi} + \boldsymbol{\rho}}{\hbar}}
$
with
\beqq
\theta(\mathbf{v}, \boldsymbol{\tau}):= \sum_{\mathbf{k} \in \mathbb{Z}^g} e^{2 i \pi {\displaystyle \sum_{j=1}^g} k_j  v_j  + {\displaystyle \sum_{l,j=1}^g} \pi i k_l k_j \tau_{l,j} } .
\eeqq

In the same way, the non-perturbative wave function takes the form
\beq
\Psi_{\pm}(x,\mathbf{T},\boldsymbol{\epsilon} , \boldsymbol{\rho}) = \psi_{\pm}(x,\mathbf{T},\boldsymbol{\epsilon}) \frac{{\displaystyle \sum_{m=0}^\infty} \hbar^m \, \Xi_m(x,\hbar, \mathbf{T}, \boldsymbol{\epsilon}, \boldsymbol{\rho})}{{\displaystyle \sum_{m=0}^\infty} \hbar^m \, \Theta_m(\hbar, \mathbf{T}, \boldsymbol{\epsilon}, \boldsymbol{\rho})}
\eeq
where $\Xi_m(x,\hbar, \mathbf{T}, \boldsymbol{\epsilon}, \boldsymbol{\rho})$ are combinations of derivatives of theta functions of the form
$
\left. \frac{\partial^n \theta(\mathbf{v}, \boldsymbol{\tau})}{\partial v_{i_1} \dots \partial v_{i_n}} \right|_{\mathbf{v} = \frac{\boldsymbol{\phi} + \boldsymbol{\rho}}{\hbar} + \boldsymbol{\mu}(x)}
$
with 
\beq
\mu_j(x) := \frac{\partial S_{-1}(x)}{\partial \epsilon_j}.
\eeq

It is important to remark that these non-perturbative objects are not formal power series in $\hbar$ but \textbf{formal trans-series in $\hbar$}.

\subsection{Properties and quantum curve}

One of the main motivations for the definition of the non-perturbative wave functions is the simplicity of its monodromies compared to its perturbative counterpart.

\bl\label{lemma-non-pert-monodromy}

The non-perturbative wave functions satisfy the following properties.

\begin{itemize}
\item For $j\in\llbracket 1,g\rrbracket$, the function $\Psi_{\pm}(x,\mathbf{T},\boldsymbol{\epsilon} , \boldsymbol{\rho})$ has a formal monodromy along  $\mathcal{A}_j$ given by
\beq
\Psi_{\pm}(x,\mathbf{T},\boldsymbol{\epsilon} , \boldsymbol{\rho}) \mapsto e^{\pm 2 \pi i \frac{\epsilon_j}{\hbar}} \Psi_{\pm}(x,\mathbf{T},\boldsymbol{\epsilon} , \boldsymbol{\rho}) .
\eeq
\item For $j \in\llbracket 1,g\rrbracket$, the function $\Psi_{\pm}(x,\mathbf{T},\boldsymbol{\epsilon} , \boldsymbol{\rho})$ has a formal monodromy along  $\mathcal{B}_j$ given by
\beq
\Psi_{\pm}(x,\mathbf{T},\boldsymbol{\epsilon} , \boldsymbol{\rho}) \mapsto e^{\mp 2 \pi i \frac{\rho_j}{\hbar}} \Psi_{\pm}(x,\mathbf{T},\boldsymbol{\epsilon} , \boldsymbol{\rho}) .
\eeq
\end{itemize}
\el

\begin{proof} The proof easily follows from the monodromies of the perturbative wave functions.
\end{proof}

The main result of this article is that, unlike its perturbative partners, the non-perturbative wave functions are solutions of a second order ODE that quantizes the classical spectral curve. The second order differential operator annihilating both non-perturbative wave functions is thus often referred to as the corresponding \emph{quantum curve}.

\bt[Quantum curve]\label{QuantumCurveTheorem}
The non-perturbative wave functions satisfy
\beq
\left[\hbar^2 \frac{\partial^2}{\partial x^2} - \hbar^2 R(x) \frac{\partial}{\partial x}  - \hbar Q(x)  -\mathcal{H}(x)\right] \Psi_\pm = 0
\eeq
with
\beq \label{DefHHH}
\mathcal{H}(x) =\left[\hbar^2 \sum_{k\in K_\infty} U_{\infty,k}(x) \frac{\partial}{\partial T_{\infty,k}} 
+ \hbar^2 \sum_{\nu =1}^n\sum_{k\in K_{b_\nu}} U_{\nu,k}(x) \frac{\partial}{\partial T_{\nu,k}} \right] \left[ \log  Z(\mathbf{T},\boldsymbol{\epsilon},\boldsymbol{\rho}) - \hbar^{-2} \om_{0,0} \right] + \frac{\phi_0(x)}{(dx)^2}
\eeq
and
\beq
R(x) = \frac{\partial \log W(x)}{\partial x}
\eeq
where the Wronskian
\beq
W(x) := \hbar \left[ \frac{\partial \Psi_+(x)}{\partial x}  \Psi_-(x) -  \frac{\partial \Psi_-(x)}{\partial x}  \Psi_+(x) \right]
\eeq
is a rational function of the form
\beq\label{eq-shape-wronskian}
W(x)  = w \frac{\underset{j=1}{\overset{g}{\prod}} (x-q_j)}{\underset{\nu=1}{\overset{n}{\prod}} (x-X_\nu)^{r_\nu}}
\eeq
with
\beq\label{eq-coef-wronskian}
w :=  \left\{
\begin{array}{l}
-2 T_{\infty,r_\infty} \exp \left(A_{\infty,0}^+ + A_{\infty,0}^-\right) \quad \hbox{if} \quad n_\infty = 0 \cr
- T_{\infty,r_\infty} \exp \left(A_{\infty,0}^+ + A_{\infty,0}^-\right) \quad \hbox{if} \quad n_\infty = 1 \cr
\end{array}
\right. 
\eeq
and 
\beq \label{DefQQQ}
Q(x) = \sum_{j=1}^g \frac{p_j}{x-q_j} +  \frac{\hbar}{2} \left[ \sum_{k \in K_\infty} U_{\infty,k}(x) \frac{\partial (S_+(x)+S_-(x))}{\partial T_{\infty,k}} \right]_{\infty,+}
+  \frac{\hbar}{2}\sum_{\nu=1}^n \left[ \sum_{k \in K_\nu} U_{\nu,k}(x) \frac{\partial (S_+(x)+S_-(x))}{\partial T_{\nu,k}} \right]_{X_\nu,-}
\eeq
with
\beq
\forall\, j\in \llbracket1,g\rrbracket \, , \; p_j:= - \hbar \left. \frac{\partial \log \Psi_+(x)}{\partial x} \right|_{x=q_j} = - \hbar \left. \frac{\partial \log \Psi_-(x)}{\partial x} \right|_{x=q_j}.
\eeq

In addition,  the pairs $\left(q_i,p_i\right)_{i=1}^g$ satisfy
\beq
\forall\, i\in \llbracket1,g\rrbracket \, , \; p_i^2 = \mathcal{H}(q_i)- \hbar p_i \left[ {\displaystyle \sum_{j \neq i}} \frac{1}{q_i-q_j} - {\displaystyle\sum_{\nu=1}^n} \frac{r_\nu}{q_i-X_\nu} \right]  + \hbar 
\left[  \left(Q(x) - \frac{p_i}{x-q_i}\right) \right]_{x=q_i} .
\eeq
\et

Before proving this theorem, let us mention that the pairs $\left(q_i,p_i\right)_{i=1}^g$, which depend on $\hbar$, form a set of Darboux coordinates on some associated symplectic space. We shall discuss this point in Section \ref{SectionLaxRepresentation}.

\begin{remark} The sums involved in \eqref{DefHHH} and \eqref{DefQQQ} are finite. Moreover, the dependence in $x$ is explicit in the definition \eqref{DefQQQ} of $Q$. The dependence in $x$ is also explicit in $\mathcal{H}(x)$ though the coefficients involve time derivatives of the partition function. These quantities may be seen in two different ways: either they are obtained by topological recursion (that computes the partition function) or they can be seen as undetermined coefficients (independent of $x$) that are in one-to-one correspondence with the Hamiltonians and their expressions in terms of the Darboux coordinates $\left(q_i,p_i\right)_{i=1}^g$. The second aspect shall be developed below in section \ref{SectionLaxRepresentation} and in the examples presented in section \ref{sec-examples}.
\end{remark}

\begin{proof}

The details of the proof are given in Appendix \ref{app-proof-ODE}.
Let us give the main steps in the remaining of this section, referring to the appendix for the technical details. 

Let us first remark that, as a linear combination of perturbative wave functions, $\Psi_{\pm}(x,\mathbf{T},\boldsymbol{\epsilon} , \boldsymbol{\rho})$ are solutions of the PDE \eq{eq-PDE}.

We shall prove that the non-perturbative wave functions are solutions of a linear second order differential equation following \cite{Iwaki}.

Let us first introduce a few useful functions built out of $\Psi_\pm$. We denote the Wronskian with respect to $x$ by
\beq
W(x) := \hbar \left(\frac{\partial \Psi_+}{\partial x} \Psi_- - \Psi_+ \frac{\partial \Psi_-}{\partial x}\right) 
\eeq
and the Wronskian with respect to the times by
\beq
\forall \, p \in \{\infty\}\cup\llbracket 1,n\rrbracket \, , \; \forall k \in K_p \, : \; W_{T_{p,k}}(x) := \hbar \left(\frac{\partial \Psi_+}{\partial T_{p,k}} \Psi_- - \Psi_+ \frac{\partial \Psi_-}{\partial T_{p,k}}\right) .
\eeq

One can use them to define
\beq
\forall \, p \in \{\infty\}\cup\llbracket 1,n\rrbracket \, , \; \forall k \in K_p \, : \;  R_{p,k} := \frac{W_{T_{p,k}}(x)}{W(x)}
\qquad
\hbox{and }
\qquad
Q_{p,k} :=  \hbar^2 \frac{ \frac{\partial \Psi_+}{\partial x}\frac{\partial \Psi_-}{\partial {T}_{p,k}}- \frac{\partial \Psi_-}{\partial x}\frac{\partial \Psi_+}{\partial {T}_{p,k}}}{W(x)} .
\eeq

They are defined in such a way that 
\beq
\forall \, p \in \{\infty\}\cup\llbracket 1,n\rrbracket \, , \; \forall k \in K_p \, : \;   \left(
\begin{matrix}
\Psi_+ & \Psi_- \cr
\hbar \frac{\partial \Psi_+}{\partial T_{p.k}}  & \hbar \frac{\partial \Psi_-}{\partial T_{p.k}}  \cr
\end{matrix}
\right) = 
\left(\begin{matrix}
1&0 \cr
Q_{p,k} & R_{p,k}
\end{matrix}
\right)
\left(
\begin{matrix}
\Psi_+ & \Psi_- \cr
\hbar \frac{\partial \Psi_+}{\partial x}  & \hbar \frac{\partial \Psi_-}{\partial x}  \cr
\end{matrix}
\right) .
\eeq 

To simplify notations, we shall identify the set of poles $\mathcal{P}$ with the corresponding index set and write $\mathcal{P}:= \{\infty\}\cup\llbracket 1,n\rrbracket$ when no confusion may arise.

By linearity, one can consider arbitrary linear combinations of the functions defined above and observe that they satisfy similar equations. In particular, defining
\beq
R(x) :=  \sum_{p \in \mathcal{P}} \sum_{k \in K_p} U_{p,k}(x) R_{p,k}(x)
\qquad \hbox{and} \qquad
Q(x) :=  \sum_{p \in \mathcal{P}} \sum_{k \in K_p}  U_{p,k}(x) Q_{p,k}(x),
\eeq
one has
\beq
\left(
\begin{matrix}
\Psi_+ & \Psi_- \cr
 \hbar {\displaystyle \sum_{p \in \mathcal{P}} \sum_{k \in K_p}}  U_{p,k}(x) \frac{\partial \Psi_+}{\partial T_{p.k}}  &  \hbar  {\displaystyle \sum_{p \in \mathcal{P}} \sum_{k \in K_p}}  U_{p,k}(x)  \frac{\partial \Psi_-}{\partial T_{p.k}}  \cr
\end{matrix}
\right) = 
\left(\begin{matrix}
1&0 \cr
Q & R 
\end{matrix}
\right)
\left(
\begin{matrix}
\Psi_+ & \Psi_- \cr
\hbar \frac{\partial \Psi_+}{\partial x}  & \hbar \frac{\partial \Psi_-}{\partial x}  \cr
\end{matrix}
\right) .
\eeq

First of all, the fact that $\Psi_\pm$ are linear combinations of solutions of \eq{eq-PDE}, and taking into account the contributions of the coefficients of these combinations, one observes that we have
\beq\label{eq-PDE-np}
\left[\hbar^2 \frac{\partial^2}{\partial x^2} -  \hbar^2 {\displaystyle \sum_{p \in \mathcal{P}} \sum_{k \in K_p}}  U_{p,k}(x) \frac{\partial }{\partial T_{p,k}} -\mathcal{H}(x)\right] \Psi_{\pm}(x) = 0
\eeq
where $\mathcal{H}(x)$ is obtained by replacing the perturbative partition function by its non-perturbative partner in $H(x)$.

By the definition of $Q$ and $R$ together, $\Psi_+$ and $\Psi_-$ are thus solutions to the compatible system
\beq\label{eq-syst-np}
\left\{
\begin{array}{l}
\left[\hbar^2 \frac{\partial^2}{\partial x^2} -  \hbar^2 {\displaystyle \sum_{p \in \mathcal{P}} \sum_{k \in K_p}}  U_{p,k}(x) \frac{\partial }{\partial T_{p,k}} -\mathcal{H}(x)\right] \Psi = 0 \cr
\left[ \hbar^2 {\displaystyle \sum_{p \in \mathcal{P}} \sum_{k \in K_p}}  U_{p,k}(x) \frac{\partial }{\partial T_{p,k}} - \hbar Q(x) - \hbar^2 R(x) \frac{\partial}{\partial x}
\right] \Psi = 0 \cr
\end{array}
\right. .
\eeq
Plugging the second equation into the first one implies that the non-perturbative wave functions are solutions to the ODE
\beq\label{eq-quant-curve-1}
\left[\hbar^2 \frac{\partial^2}{\partial x^2} - \hbar^2 R(x) \frac{\partial}{\partial x}  - \hbar Q(x)  -\mathcal{H}(x)\right] \Psi_\pm = 0.
\eeq
This is our quantum curve. We now need to study the functions $R(x)$ and $Q(x)$ that may a priori be multi-valued functions of $x$ with singularities at the zeros of $W(x)$, the poles $\infty$ and $\{X_\nu\}$ as well as the critical values $u_i = x(a_i)$. We shall now see that these functions are rational functions of $x$ without poles at the critical values.

We first prove in Appendix \ref{app-lemma-wronsk-log} that one can simplify the expression of $R(x)$.
\bl\label{lemma-wronsk-log}
The function $R(x)$ reads
\beq
R(x) =  \frac{1}{W(x)} \frac{\partial W(x)}{\partial x}.
\eeq
\el

One can also remark that, by definition, 
\beq\label{eq-Q-times}
Q(x) = \frac{\hbar}{2} \frac{ {\displaystyle \sum_{p \in \mathcal{P}} \sum_{k \in K_p} } U_{p,k}(x) \left( \frac{\partial W(x)}{\partial T_{p,k}} - \frac{\partial W_{T_{p,k}}(x)}{\partial x}\right)}{ W(x)}.
\eeq 

Let us now study the properties of the functions $R_{p,k}$ and $Q_{p,k}$.

First of all, let us remind that $\Psi_{\pm}$ are combinations of terms which are derivatives of theta functions evaluated at $\mathbf{v} = \frac{\phi+\rho}{\hbar}+ \mathbf{\mu}(x)$. On the other hand, because the combination $\theta(\mathbf{v}+\mathbf{u},\boldsymbol{\tau}) \theta(\mathbf{v}-\mathbf{u},\boldsymbol{\tau})$ is a theta function of order 2 in $\mathbf{v}$, it can be decomposed in a basis of squares of theta functions with different characteristics (see for example \cite{Fay}). Hence, the Wronskians involve only combinations of derivatives of theta functions evaluated at  $\mathbf{v} = \frac{\phi+\rho}{\hbar}$, their dependence in $x$ appearing only in the coefficients.

This means that, for all pair $(p,k)$, $R_{p,k}$ and $Q_{p,k}$ admit an expansion in $\hbar$ of the form
\beq
R_{p,k}(x,\mathbf{T},\hbar)  = \sum_{m=0}^\infty \hbar^m {R}_{p,k}^{(m)} (x,\mathbf{T},\hbar)
\qquad \hbox{and} \qquad Q_{p,k}(x,\mathbf{T},\hbar) = \sum_{m=0}^\infty \hbar^m Q_{p,k}^{(m)}(x,\mathbf{T},\hbar) 
\eeq
where the coefficients take the form
\beq
{R}_{p,k}^{(m)} (x,\mathbf{T},\hbar) = \left. \hat{R}_{p,k}^{(m)} (x,\mathbf{T},\mathbf{v})\right|_{\mathbf{v} = \frac{\phi+\rho}{\hbar}}
\qquad \hbox{and} \qquad   Q_{p,k}^{(m)}(x,\mathbf{T},\hbar)  = \left. \hat{Q}_{p,k}^{(m)}(x,\mathbf{T},\mathbf{v})\right|_{\mathbf{v} = \frac{\phi+\rho}{\hbar}} .
\eeq

Thanks to Lemma \ref{lemma-non-pert-monodromy}, one can also check that the monodromies of the non-perturbative wave functions around elements of $H_1(\Sigma_\phi,\mathbb{Z})$ ensure that the Wronskians do not have any monodromy and thus
\bc
The Wronskians and the functions $\hat{R}_{p,k}^{(h)}$ and $\hat{Q}_{p,k}^{(h)}$ are rational functions of $x$.
\ec

Indeed, the essential singularities cancel in the definition of the Wronskians.

We can even go further and prove in Appendix \ref{app-lemma-no-pole-at-branch-points-1} the following important result.

\bl \label{lemma-no-pole-at-branch-points-1}
The rational functions $R_{p,k}$ and $Q_{p,k}$ have no pole at the ramification points.
\el

This implies that
\bc
The Wronskians $W(x) $ and $W_{{T}_{p,K}}(x)$ are rational functions of $x$ with poles only at $p \in \mathcal{P}$.
\ec

\begin{proof}
From Lemma \ref{lemma-wronsk-log}, one knows that
\beqq
R(x) = \partial_x \left[\log W(x)\right].
\eeqq
Lemma \ref{lemma-no-pole-at-branch-points-1} ensures that it does not have any pole at the ramification points. On the other hand, from the properties of $\Psi_\pm$, one knows that the RHS is a rational function of $x$ with possible  poles only at  $p \in \mathcal{P}$ and the critical values. Hence, this combination has no poles at the ramification points and $W(x)$ is a rational function of $x$ with poles only at  $p \in \mathcal{P}$.

By definition, one has
\beq
W_{{T}_{p,K}}(x) = R_{p,K}(x) W(x).
\eeq
The left hand side might have poles at the ramification points and $p \in \mathcal{P}$ only. The right hand side does not have any pole at the ramification points leading to the result.

\end{proof}

The asymptotic expansions around poles recalled in Appendix \ref{app-asymptot} ensure that $W(x)$ takes the form
\beq
W(x)  = w \frac{{\displaystyle \prod_{j=1}^g} (x-q_j)}{\underset{\nu=1}{\overset{n}{\prod}} (x-X_\nu)^{r_\nu}}
\eeq
where, using the notations of Appendix \ref{app-asymptot}, one defines
\beq
w :=  \left\{
\begin{array}{l}
-2 T_{\infty,r_\infty} \exp \left(A_{\infty,0}^++A_{\infty,0}^-\right) \quad \hbox{if} \quad n_\infty = 0 \cr
- T_{\infty,r_\infty} \exp \left(A_{\infty,0}^+ + A_{\infty,0}^-\right) \quad \hbox{if} \quad n_\infty = 1 \cr
\end{array}
\right. .
\eeq

To conclude the proof, let us study the properties of $Q(x)$. By its definition \eq{eq-Q-times} and Lemma \ref{lemma-no-pole-at-branch-points-1}, it is a rational function of $x$ with simple poles at the $q_i$ and poles at $p \in \mathcal{P}$ whose degree follows from the asymptotics given in Appendix \ref{app-asymptot}. It thus reads
\beq
Q(x) = \sum_{i=1}^g \frac{Q_{q_i}}{x-q_i} + \sum_{k=0}^{r_\infty-4} Q_{\infty,k} x^k + \sum_{\nu=1}^n \sum_{k=1}^{r_\nu+1} \frac{Q_{\nu,k}}{(x-X_\nu)^k}.
\eeq
One can compute the coefficients of this partial fraction expansion by expanding the differential equations satisfied by the wave functions around the different poles.

\begin{itemize}
\item The leading and subleading orders in the expansion of the quantum curve \eq{eq-quant-curve-1} around $x = q_i$, for $i\in \llbracket 1,g\rrbracket$, read
\beq\label{eq-Q(q_i)}
\left\{
\begin{array}{l}
Q_{q_i} =  p_i \cr
p_i^2 = \mathcal{H}(q_i)- \hbar p_i \left[ {\displaystyle \sum_{j \neq i}} \frac{1}{q_i-q_j} - {\displaystyle\sum_{\nu=1}^n} \frac{r_\nu}{q_i-X_\nu} \right] + \hbar 
\left[  \left(Q(x) - \frac{p_i}{x-q_i}\right) \right]_{x=q_i}
\cr
\end{array}
\right.
\eeq
where
\beq
\forall \, j\in \llbracket 1,g\rrbracket \, :\; p_j:= -\hbar \left. \frac{\partial \log \Psi_+(x)}{\partial x} \right|_{x=q_j} = -\hbar \left. \frac{\partial \log \Psi_-(x)}{\partial x} \right|_{x=q_j}.
\eeq
As we shall see, the second equation means that each pair $\left(q_i,p_i\right)_{i=1}^g$ defines a point in a $\hbar$-deformation of the classical spectral curve.

\item Let us consider the sum of the second equation of the system \eq{eq-syst-np} for $\Psi_+$ and $\Psi_-$. The expansion of this symmetric version around $x = \infty$ reads
\beq
 \left[Q(x)\right]_{\infty,+} = \frac{\hbar}{2} \left[ \sum_{k \in K_\infty} U_{\infty,k}(x) \frac{\partial (S_+(x)+S_-(x))}{\partial T_{\infty,k}} \right]_{\infty,+}.
\eeq

\item The expansion of the same expression around $x = X_\nu$ reads
\beq
\left[Q(x)\right]_{X_\nu,-} = \frac{\hbar}{2} \left[ \sum_{k \in K_\nu} U_{\nu,k}(x) \frac{\partial (S_+(x)+S_-(x))}{\partial T_{\nu,k}} \right]_{X_\nu,-}.
\eeq
\end{itemize}
\end{proof}

\section{Lax representation\label{SectionLaxRepresentation}}

After building a quantization of the classical spectral curve, it is natural to study a linearization of the corresponding differential equation. In particular, in this section we associate a $\mathfrak{sl}_2(\mathbb{C})$ connection depending on $\hbar$ to such a quantum curve and describe it as a point in the corresponding moduli space of irregular connections. We compute the corresponding quadratic differential which can be understood as a $\hbar$-deformation of the initial data $\phi_0$ in the moduli space of meromorphic connections $\mathcal{Q}(\mathbb{P}^1,D,n_\infty,\mathbf{T})$. We finally study the corresponding isomonodromic system when it can be obtained as de-autonomization of an isospectral system.

\subsection{$\mathfrak{sl}_2$ connection from quantum curve}

Let us linearize the quantum curve. For this purpose, we define
\beq
\hat{\Psi}_{\pm}(x) := \frac{1}{W(x)} \left[ P(x) \Psi_\pm(x) + \hbar \frac{\partial \Psi_\pm(x)}{\partial x} \right] 
\eeq
where $P(x)$ is a rational function of $x$ with poles at $x(p)$ for $p \in \mathcal{P}$ to be fixed.

The ODE given in Theorem \ref{QuantumCurveTheorem} implies that
\beq
\hbar \frac{\partial }{\partial x} \left(\begin{array}{c} \hat{\Psi}_\pm \cr \Psi_\pm \cr \end{array} \right) =
\left( 
\begin{matrix}
P(x) & M(x) \cr
W(x) & -P(x) \cr
\end{matrix}
\right) \; \left(\begin{array}{c} \hat{\Psi}_\pm \cr \Psi_\pm \cr \end{array} \right) 
\eeq
where
\beq
M(x) =  \frac{\hbar \frac{\partial P(x)}{\partial x} - \hbar \frac{\partial \log W(x)}{\partial x} P(x) - P(x)^2 + \hbar Q(x) +  \mathcal{H}(x) }{W(x)} .
\eeq

This rank 2 differential system defines a connection on $\mathbb{P}^1$ which has poles only on $\mathcal{P}$ by imposing a simple condition on the function $P(x)$.

\bl
If 
\beq
\forall\, i\in\llbracket 1,g\rrbracket \, : \; P(q_i) = p_i
\eeq
then 
$M(x)$
is a rational function of x with poles only at $x \in \mathcal{P}$. 
\el

\begin{proof}
To prove this lemma, one only needs to prove that the $M(x)$ does not have any pole at $x = q_i$ for $i\in \llbracket 1,g\rrbracket$.

Let us compute the two leading terms of $\frac{\hbar \frac{\partial P}{\partial x} - \hbar \frac{\partial \log Wr^x(\Psi_+,\Psi_-)}{\partial x} P - P^2 + \hbar Q + \mathcal{H} }{Wr^x(\Psi_+,\Psi_-)}$ as $x \to q_i$. Using \eq{eq-Q(q_i)}, they read
\beq
- \hbar P(q_i) + \hbar p_i
\eeq
and
\beq
-\hbar P(q_i) \left[ \sum_{j\neq i}  \frac{1 }{q_i-q_j} - \sum_{\nu = 1}^n \frac{r_{\nu}}{q_i-X_\nu} \right] - P(q_i)^2 + \hbar \left[Q(x)- \frac{P(q_i)}{x-q_i} \right]_{x=q_i}+ \mathcal{H}(q_i).
\eeq
These two terms, which are the coefficients of the singular part of $\frac{\hbar \frac{\partial P}{\partial x} - \hbar \frac{\partial \log Wr^x(\Psi_+,\Psi_-)}{\partial x} P - P^2 + \hbar Q + \mathcal{H} }{Wr^x(\Psi_+,\Psi_-)}$ vanish if $P(q_i) = p_i$ thanks to Theorem \ref{QuantumCurveTheorem}.
\end{proof}

Hence $ \left(\begin{array}{c} \hat{\Psi}_\pm \cr \Psi_\pm \cr \end{array} \right)$ is a basis of solutions to the Lax system
\beq
 \hbar \frac{\partial }{\partial x} \left(\begin{array}{c} \hat{\Psi}_\pm \cr \Psi_\pm \cr \end{array} \right) = L(x) \left(\begin{array}{c} \hat{\Psi}_\pm \cr \Psi_\pm \cr \end{array} \right)
 \eeq
 where
 \beq
 L(x) = \left( 
\begin{matrix}
P(x) & M(x) \cr
W(x) & -P(x) \cr
\end{matrix}
\right)
\eeq
is a rational function of $x$ with value in $\mathfrak{sl}_2$ with poles only in $D$ if the degrees of the poles of $P(x)$ are kept small enough. It defines a connection $\hbar \text{d}- L(x) dx$ on $\mathbb{P}^1$ that can be considered as a point in the corresponding moduli space of $\mathfrak{sl}_2$ connections with irregular singularities along $D$. Fixing the value of $\mathbf{T}$ amounts to fixing its residues and irregular type thus leading to a point in a symplectic leaf of dimension $2g$. The definition of $\left(q_i,p_i\right)_{i=1}^g$ by
\beq
\forall \, i\in \llbracket1, g\rrbracket \, , \;\left\{
\begin{array}{l}
W(q_i) = 0 \cr
P(q_i) = p_i \cr
\end{array}
\right.
\eeq
actually identifies them with the spectral Darboux coordinates defined by \cite{Darboux} on this symplectic leaf.

It is important to notice that the Lax matrix $L(x)$ depends on $\hbar$ only through $\left(q_i,p_i\right)_{i=1}^g$. The latter being Darboux coordinates, they depend on $\hbar$ only through their time evolution as we will see in the next section.

\subsection{Deformed spectral curve}

Before explaining how to choose a particular gauge in order to make the explicit computation of examples easier, let us describe the spectral curve of the matrix $L(x)$ defined by
\beq
0 = \det(ydx-L(x) dx): = y^2 (dx)^2 - \phi_\hbar.
\eeq
 It defines a quadratic differential
\beq
\phi_\hbar = \left[\hbar \frac{\partial P(x)}{\partial x} - \hbar \frac{\partial \log W(x)}{\partial x} P(x)  + \hbar Q(x) +  \mathcal{H}(x)\right] (dx)^2 .
\eeq
We have thus defined a flow along the $\partial_\hbar$ direction in the corresponding moduli space of quadratic differentials. Let us present a complete expression for our $\hbar$-deformed quadratic differential using the expression of $Q(x)$ of Theorem \ref{QuantumCurveTheorem}.

\bt \label{th-sp-curve}
The $\hbar$-deformed spectral curve reads
\bea
\frac{\phi_\hbar}{(dx)^2} &=& \mathcal{H}(x) + \hbar \sum_{j=1}^g \frac{p_j}{x-q_j} + \frac{\hbar^2}{2} \left[ \sum_{k \in K_\infty} U_{\infty,k}(x) \frac{\partial (S_+(x)+S_-(x))}{\partial T_{\infty,k}} \right]_{\infty,+}
\cr&&+ \frac{\hbar^2}{2}\sum_{\nu=1}^n  \left[ \sum_{k \in K_\nu} U_{\nu,k}(x) \frac{\partial (S_+(x)+S_-(x))}{\partial T_{\nu,k}} \right]_{X_\nu,-} 
+ \hbar \frac{\partial P(x)}{\partial x} - \hbar \frac{\partial \log W(x)}{\partial x} P(x). \cr
&&
\eea
\et

Even if this expression seems to have simple poles at $x=q_i$, they cancel due to the condition $P(q_i) = p_i$.

\subsection{Gauge choice}
\label{sec-gauge}

The characterization of $P(x)$ by its values at $(q_i)_{i=1}^g$ does not fix it without ambiguity. Fixing a specific form of the rational function $P$ corresponds to a gauge choice for the system under consideration, i.e. a choice of representative of the reduced coadjoint orbit under consideration. Let us give an example of gauge choice in this section. We follow the choice considered in \cite{MO}. The gauge depends on whether the point above infinity is a ramification point or not (i.e. if $n_\infty=0$ or $n_\infty=1$).

\subsubsection{Non-degenerate case: $n_\infty =0$}

If $\infty$ is not a critical value for the map $x$, one fixes a gauge by writing the system under the form
\beq
\hbar \frac{\partial }{\partial x}  \left(\begin{array}{c} \hat{\Psi}_\pm \cr \Psi_\pm \cr \end{array} \right)  = L(x)   \left(\begin{array}{c} \hat{\Psi}_\pm \cr \Psi_\pm \cr \end{array} \right) 
\eeq
where
\beq
L(x) = \sum_{k \leq r_\infty} L_{\infty,k} x^{k-2}
\eeq
as $x \to \infty$ with 
\beq
L_{\infty,r_\infty} = \left(\begin{matrix} \alpha & 0 \cr 0 & -\alpha \cr \end{matrix} \right) .
\eeq
As we shall see from the examples, the value of $\alpha$ and $\beta$ can be computed from the Casimirs. More precisely, for $r_\infty \geq 2$, one has $\alpha = T_{\infty,r_\infty}$ and $\beta = T_{\infty, r_\infty} + \frac{\hbar}{2}$.

This requires that $P(x)$ takes the form
\beq
P(x) = \frac{\alpha x^{g+1}+ \beta x^g+ {\displaystyle \sum_{l=0}^{g-1}} \alpha_l x^l}{\underset{\nu=1}{\overset{n}{\prod}}(x-X^\nu)^{r_{b_\nu}}}.
\eeq

The last coefficients $\left(\alpha_l\right)_{l=0}^{g-1}$ are fixed by the $g$ conditions
\beq
\forall \, i\in\llbracket1,g\rrbracket \, : \; P(q_i) = p_i.
\eeq

We call this case non-degenerate since the leading order of the Lax matrix $L(x)$ at infinity as full rank in this coadjoint orbit.

\subsubsection{Degenerate case: $n_\infty =1$}

This degenerate case corresponds to the case when the leading order of the Lax matrix at infinity is of rank 1 instead of rank 2. As in \cite{MO}, we look for a representative of the coadjoint orbit satisfying
\beq
L_{\infty,r_\infty} = 
\left(\begin{matrix}
0 & 1 \cr
0 & 0 \cr
\end{matrix}
\right) 
\qquad \hbox{and} \qquad
L_{\infty,r_\infty-1} = 
\left(\begin{matrix}
0 & \beta \cr
A & 0 \cr
\end{matrix}
\right)
\eeq 
for arbitrary $\beta$. This can be obtained by imposing the form
\beq
P(x) =  \frac{{\displaystyle \sum_{l=0}^{g-1}} \alpha_l x^l}{\underset{\nu=1}{\overset{n}{\prod}} (x-X^\nu)^{r_{b_\nu}}}
\eeq
satisfying
\beq
\forall\, i\in\llbracket1,g\rrbracket \, : \; P(q_i) = p_i.
\eeq

\section{Isomonodromic deformations}
\label{sec-isomonodromic}

It is often very useful to consider isomonodromic deformations of our Lax matrix $L(x)$. In particular, it is a way to access a set of equations defining the function $(p_i,q_i)_{i=1}^g$. In examples of Section \ref{sec-examples}, we use Hamiltonian representations of the corresponding isomonodromic systems to get 2-parameters solutions of the six Painlev\'e equations. In this section, we recall one way to obtain isomonodromic deformations starting from an isospectral system following the work of Montr\'{e}al school \cite{RMatrixHarnad2}. This section does not present any new material but should be taken as a guidebook for the computations of the example section. 

\subsection{Isospectral deformations}

Using the classical $R$-matrix formalism on loop algebras, one can define a Poisson structure on the space of Lax matrices $L(x)$ as before. Indeed, one can define a set of commuting flows generated by spectral invariants. In our case, the set of spectral invariants is generated by the following Hamiltonians:
\beq
 \forall\, k \in \llbracket 1, r_\infty-3\rrbracket \, : \; H_{p,k} := \frac{1}{2} \Res_{x \to \infty} x^{-k} \Tr L(x)^2 dx
\eeq
and
\beq
\forall\, \nu \in\llbracket 1,n\rrbracket \, , \;  \forall \,k \in\llbracket 0, r_\nu-1\rrbracket \,: \; H_{\nu,k} := \frac{1}{2} \Res_{x \to X_\nu} (x-X_\nu)^{k} \Tr L(x)^2 dx.
\eeq
The associated Hamilton's equations read
\beq
\forall \,p \in \{\infty\}\cup\llbracket1,n\rrbracket \, , \; \forall\, k \, :\;  \frac{\partial L(x) }{\partial t_{p,k}} = \left[ L_{t_{p,k}}(x), L(x) \right]
\eeq
where
\beq
 \forall\, k \in\llbracket 1 , r_\infty-3\rrbracket \, : \; L_{t_{\infty,k}}:=  \left[ x^{-k} L(x) \right]_{\infty,+}
 \eeq
 and
 \beq
\forall\, \nu \in\llbracket1 ,n\rrbracket \, , \;  \forall \,k \in\llbracket 0 , r_\nu-1\rrbracket \, : \; L_{t_{\nu,k}}:=  \left[ (x-X_\nu)^{k} L(x) \right]_{X_{\nu},-}.
\eeq
These Hamiltonian flows are isospectral since they preserve the spectrum of $L(x)$.

\subsection{Non-autonomous system and isomonodromic deformations}

Let us now consider that, in addition, $L(x)$ depends explicitly on $t_{p,k}$ in such a way that
\beq \label{isomono-condition}
\frac{\delta L(x)}{\delta t_{p,k}} = \frac{\partial L_{t_{p,k}}}{\partial x}
\eeq
where $\frac{\delta L(x)}{\delta t_{p,k}}$ denotes the variation of $L(x)$ with respect to its explicit dependence on $t_{p,k}$ only. Then, Hamilton's equations are replaced by
\beq
\frac{\partial L(x) }{\partial t_{p,k}} = \left[ L_{t_{p,k}}(x), L(x) \right] + \frac{\partial L_{t_{p,k}}}{\partial x}.
\eeq
This new equation is equivalent to the commutation relation
\beq
\left[\frac{\partial}{\partial x}- L(x), \frac{\partial}{\partial t_{p,k}}- L_{t_{p,k}}(x)\right] = 0
\eeq
It ensures that the flow along $\frac{\partial}{\partial t_{p,k}}$ is isomonodromic. For this reason, we refer to \eq{isomono-condition} as the isomonodromic condition. From now on, we assume that it is satisfied in this section for all pairs $(p,k)$ described above\footnote{To our knowledge, a general procedure to prove the isomonodromic condition \eq{isomono-condition} from some general isospectral deformations formalism is not known}.

In such a case, an isomonodromic tau function $\tau$ is defined by the condition
\beq
d \ln \tau = \sum_{p,k} H_{p,k} \, dt_{p,k}.
\eeq

One can also use the isomonodromic times $t_{p,k}$ as new coordinates replacing the spectral times $T_{p,k}$.  One can decompose our vector field $\frac{\partial }{\partial T_{\nu,k}}$ in this basis by comparing the corresponding Wronskians, or more precisely, the action of the vector fields on the singular part of the logarithm of the wave functions. The change of basis is more naturally expressed by decomposing the usual isomonodromic flows in terms of our vector fields, leading to
\beq \label{change-times-nu}
\forall \,\nu \in\llbracket 1, n\rrbracket \, , \; \forall\, l\in \llbracket 0, r_\nu-1\rrbracket \, : \; \frac{\partial}{\partial t_{\nu,l}} =  \sum_{k=2}^{r_\nu-l+1} (k-1) T_{\nu,k+l-1} \frac{\partial}{\partial T_{\nu,k}},
\eeq
\beq
\forall\,  l\in\llbracket1, r_\infty-3\rrbracket \, : \; \frac{\partial}{\partial t_{\infty,l}} =  \sum_{k=2}^{r_\infty-l-1} (k-1) T_{\infty,k+l+1} \frac{\partial}{\partial T_{\infty,k}}\,\text{ if } n_\infty=0
\eeq
and 
\beq
\forall\, l\in \llbracket 1, r_\infty-3\rrbracket \, : \; \frac{\partial}{\partial t_{\infty,l}} =  \sum_{k=2}^{r_\infty-l-1} \frac{2k-3}{2} T_{\infty,k+l+1} \frac{\partial}{\partial T_{\infty,k}} \,\text{ if } n_\infty=1
\eeq

In addition, one can use this decomposition to express the function $\mathcal{H}(x)$ in terms of the variations with respect to the usual isomonodromic times. One has
\beq \label{eq-change-times-U-1}
\sum_{k\in K_\infty} U_{\infty,k}(x) \frac{\partial }{\partial T_{\infty,k}} = \sum_{l=1}^{r_\infty-3} x^{l-1} \frac{\partial}{\partial t_{\infty,l}}
\eeq
and
\beq \label{eq-change-times-U-2}
\forall\, \nu\in \llbracket 1,n\rrbracket \, : \; \sum_{k\in K_\nu} U_{\nu,k}(x) \frac{\partial }{\partial T_{\nu,k}} = \sum_{l=0}^{r_\nu-1} \frac{1}{(x-X_\nu)^{l+1}} \frac{\partial}{\partial t_{\nu,l}}.
\eeq

Using these relations, one can build a tau function in terms of the non-perturbative partition function. Rather than writing a lengthy general formula, we prefer referring the reader to Section \ref{sec-examples} for examples.

\subsection{Flows of Darboux coordinates}

The  vector $\left(\begin{array}{c} \hat{\Psi}_\pm \cr \Psi_\pm \cr \end{array} \right) $ is subject to a set of compatible equations with respect to the isomonodromic times (denoted by the generic letter $t$ for clarity in the following equations)
\beq
\frac{\partial }{\partial t} \left(\begin{array}{c} \hat{\Psi}_\pm \cr \Psi_\pm \cr \end{array} \right) 
 = \left(\begin{matrix} P_t(x) & M_t(x) \cr
 W_t(x) & - P_t(x) \cr
 \end{matrix}
 \right)
 \left(\begin{array}{c} \hat{\Psi}_\pm \cr \Psi_\pm \cr \end{array} \right) 
 \eeq
 where
 \beq
 L_t(x) =  \left(\begin{matrix} P_t(x) & M_t(x) \cr
 W_t(x) & - P_t(x) \cr
 \end{matrix}
 \right).
 \eeq

 The compatibility of the system evaluated at $x=q_i$ imposes that, for an arbitrary isomonodromic time $t$, one has 
 \beq \label{eq-compatibility}
\forall \,i\in \llbracket1,g\rrbracket \, : \;
\left\{
\begin{array}{l}
 \hbar \frac{\partial p_i}{\partial t} = \hbar\left.  \frac{\partial P_{t}(x)}{\partial x} \right|_{x=q_i} 
 - \frac{  \hbar\left.  \frac{\partial W_{t}(x)}{\partial x} \right|_{x=q_i}   \left.  \frac{\partial P(x)}{\partial x} \right|_{x=q_i} + W_t(q_i)  \left.  \frac{\partial \phi_{\hbar}(x)}{\partial x} \right|_{x=q_i}  }{\left.  \frac{\partial W(x)}{\partial x} \right|_{x=q_i} } 
 \cr
 \hbar \frac{\partial q_i}{\partial t}=-  \frac{ \hbar\left.  \frac{\partial W_{t}(x)}{\partial x} \right|_{x=q_i} + 2 p_i W_t(q_i)}{  \left. \frac{\partial W(x)}{\partial x} \right|_{x=q_i}} \cr 
\end{array}
\right. .
\eeq

\begin{proof}
Given any isomonodromic time $t$, the compatibility condition between the time derivative and the $x$ derivative implies
\beq
\left\{
\begin{array}{l}
\hbar \frac{\partial P(x)}{\partial t} = \hbar \frac{\partial P_t(x)}{\partial x} + M_t(x) W(x) - M(x) W_t(x) \cr
\hbar \frac{\partial W(x)}{\partial t} = \hbar \frac{\partial W_t(x)}{\partial x} +2  P(x) W_t(x) - 2 P_t(x) W(x) \cr
\end{array} .
\right.
\eeq

Let us remark that, for any rational function $f(x)$,
\beq
\frac{\partial f(q_i)}{\partial t} = \left. \frac{\partial f(x)}{\partial t}\right|_{x=q_i} + \left. \frac{\partial f(x)}{\partial x}\right|_{x=q_i}  \frac{\partial q_i}{\partial t}.
\eeq

The evaluation of the compatibility conditions at $x=q_i$ then reads
\beq
\left\{
\begin{array}{l}
\hbar \frac{\partial p_i}{\partial t} = \hbar \left. \frac{\partial P(x)}{\partial x} \right|_{x=q_i}   \frac{\partial q_i}{\partial t} + \hbar \left. \frac{\partial P_t(x)}{\partial x} \right|_{x=q_i}  - M(q_i) W_t(q_i) \cr
- \hbar \left. \frac{\partial W(x)}{\partial x} \right|_{x=q_i} \frac{\partial q_i}{\partial t}  = \hbar \left. \frac{\partial W_t(x)}{\partial x} \right|_{x=q_i}+2  P(q_i) W_t(q_i)  \cr
\end{array} .
\right.
\eeq
Combining these equations and the equation of the spectral curve, one obtains the result.
\end{proof}

Remark that this result is valid even if the coefficients of the elements of the Lax matrix depend explicitly on the isomonodromic time $t$.

\section{Examples \label{sec-examples}}

In this section, we present a few examples of application of our general formalism. In each case, we follow the same procedure. We first compute the Wronskian $W(x)$. It always takes the form of \eq{eq-shape-wronskian} where the leading order is given by \eq{eq-coef-wronskian}.  The value of the zeros of $W(x)$ gives half of the spectral Darboux coordinates and can be computed using the asymptotics of Appendix \ref{app-asymptot}.

One can then compute the quantum curve thanks to Theorem \ref{QuantumCurveTheorem}.

We then compute the element $P(x)$ of the Lax matrix in order to have the gauge considered in Section \ref{sec-gauge}. After that, we can compute the $\hbar$-dependent quadratic differential $\phi_\hbar \in \mathcal{Q}(\mathbb{P}^1,D,n_\infty,\mathbf{T})$.  First, we express its coefficients in terms of the times and the partition function thanks to Theorem \ref{th-sp-curve} and the asymptotics of Appendix \ref{app-asymptot}. On the other hand, we express these coefficients in terms of the spectral Darboux coordinates by writing that any pair $\left(q_i,p_i,\right)_{i=1}^g$ defines a point on the $\hbar$-deformed spectral curve.
This allows relating the partition function to an isomonodromic tau function. Indeed, in all the cases considered, it is known that the isomonodromic condition \eq{isomono-condition} can be fulfilled (see \cite{RMatrixHarnad,LoopAlgebraHarnad} or \cite{MO}).

In each case, we write an identification of times allowing to solve this condition. After this identification, we write the associated Hamilton's equations obtained through the compatibility condition \eq{eq-compatibility}. In the first cases, we show how it leads to the six Painlev\'e equations, showing that we have produced 2-parameters solutions for the latter. This also provides us an expression of the coordinates $p_i$'s in terms of the variations of the $q_i$'s.

\subsection{Painlev\'e 1}
\label{sec-ex-P1}

Let us consider the case $r_\infty = 4$, $n_\infty = 1$ and $n=0$. Let us first observe that the PDE \eq{eq-PDE-np} imposes that $ A_{\infty,1}^+ + A_{\infty,1}^- = 0$.

Hence, the Wronskian reads
\beq
W(x) = -T_{\infty,4} (x-q)
\eeq
where
\beq
q = - \frac{T_{\infty,3}}{T_{\infty,4}} - (A_{\infty,2}^+- A_{\infty,2}^-).
\eeq
From \eq{eq-PDE-np}, one obtains that
\beq
A_{\infty,2}^+ - A_{\infty,2}^- = - \frac{1}{2} \frac{\partial A_{\infty,1}^+ - A_{\infty,1}^-}{\partial T_{\infty,2}} = - \hbar^2 \frac{\partial^2 \log Z}{\partial T_{\infty,2}^2},
\quad 
\hbox{i.e.}
\quad
q = - \frac{T_{\infty,3}}{T_{\infty,4}} + \hbar^2 \frac{\partial^2 \log Z}{\partial T_{\infty,2}^2}.
\eeq
From Theorem \ref{QuantumCurveTheorem}, one can compute 
\beq
Q(x) = \frac{p}{x-q},
\eeq
and the quantum curve reads
\beq
\left[\hbar^2 \frac{\partial^2}{\partial x^2} - \frac{\hbar^2}{x-q} \frac{\partial }{\partial x} - \frac{\hbar p}{x-q} - T_{\infty,4}^2 x^3 - 2 T_{\infty,3} T_{\infty,4} x^2 - \left(T_{\infty,3}^2 + 2 T_{\infty,2} T_{\infty,4}\right) x - H_0 \right] \Psi_\pm(x) = 0
\eeq
where
\beq
H_0 = \frac{T_{\infty,3} T_{\infty,2}}{4} + \hbar^2 \frac{T_{\infty,4}}{2} \frac{\partial \ln Z}{\partial T_{\infty,2}}.
\eeq

The diagonal element $P(x)$ of the Lax matrix is a constant so that
\beq
P(x) = p.
\eeq

The $\hbar$-deformed spectral curve reads
\beq
\frac{\phi_\hbar}{(dx)^2} = \frac{T_{\infty,4}^2}{4} x^3 + \frac{ T_{\infty,3} T_{\infty,4}}{2} x^2 + \frac{1}{4} \left(T_{\infty,3}^2 + 2 T_{\infty,2} T_{\infty,4}\right) x + H_0
\eeq
where the condition on $(p,q)$ implies that
\beq
 H_0 =  p^2-  \frac{1}{4}T_{\infty,4}^2 q^3 - \frac{1}{2} T_{\infty,3} T_{\infty,4} q^2 - \frac{1}{4}\left(T_{\infty,3}^2 + 2 T_{\infty,2} T_{\infty,4}\right) q.
\eeq

Breaking the autonomy of the Hamiltonian system with respect to $t_{\infty,1}$ by setting $T_{\infty,2}:=\frac{t_{\infty,1}}{2 T_{\infty,4}} $ as our isomonodromic time, the compatibility condition \eq{eq-compatibility} gives the Hamiltonian system
\beq
\left\{
\begin{array}{l}
2 \hbar \frac{\partial q}{\partial t_{\infty,1}} = - \frac{\partial H_0}{\partial p} = - 2 p \cr
2 \hbar \frac{\partial p}{\partial t_{\infty,1}} =  \frac{\partial H_0}{\partial q} = - \frac{3}{4} T_{\infty,4}^2 q^2 - T_{\infty,3} T_{\infty,4} q - \frac{T_{\infty,3}^2}{4} - t_{\infty,1} \frac{T_{\infty,4}^2}{2} \cr
\end{array}
\right. 
\eeq
which means that $q$ is solution to 
\beq
\hbar^2 \frac{\partial^2 q}{\partial t_{\infty,1}^2} = \frac{3 T_{\infty,4}^2}{8} q^2  + \frac{ T_{\infty,3} T_{\infty,4}}{2} q + \frac{T_{\infty,3}^2 }{8} + t_{\infty,1} \frac{T_{\infty,4} }{4}.
\eeq
Setting $T_{\infty,4} = \frac{1}{2}$ and $T_{\infty,3} = 0$, one gets Painlev\'e 1 equation
\beq
\hbar^2 \frac{\partial^2 \tilde{q}}{\partial t^2} = 6 \tilde{q}^2 + t
\eeq
for $\tilde{q} = \frac{q}{4}$ and $t = \frac{t_{\infty,1}}{4}$.

This also gives
\beq
p =- \hbar \frac{\partial q}{\partial t_{\infty,1}} .
\eeq

\subsection{Painlev\'e 2}

Let us now consider the case $n=0$, $n_\infty = 0$ and $r_\infty=4$. The Wronskian reads
\beq
W(x) = - 2 T_{\infty,4} \exp \left(A_{\infty,0}^+ + A_{\infty,0}^-\right) (x-q)
\eeq
where
\beq
q=- \frac{T_{\infty,3}}{T_{\infty,4}} - (A_{\infty,1}^+ + A_{\infty,1}^-)
\eeq

From Theorem \ref{QuantumCurveTheorem}, one can compute 
\beq
Q(x) = \frac{p}{x-q} + \frac{\hbar}{2} T_{\infty,4} \frac{\partial A_{\infty,0}^+ + A_{\infty,0}^-}{\partial T_{\infty,2}}  ,
\eeq
and the quantum curve reads
\bea
\left[\hbar^2 \frac{\partial^2}{\partial x^2} - \frac{\hbar^2}{x-q} \frac{\partial }{\partial x} - \frac{\hbar p}{x-q} -  T_{\infty,4}^2 x^4 - 2 T_{\infty,3}  T_{\infty,4}  x^3 - \left(  T_{\infty,3}^2 + 2   T_{\infty,4} T_{\infty,2} \right) x^2 \right. \cr
\qquad \qquad \left.  - \left[ 2 T_{\infty,3} T_{\infty,2} + 2 T_{\infty,4} T_{\infty,1}  \right] x - H_0  - \hbar T_{\infty,4} q  \right] \Psi_\pm(x) = 0.
\eea
where
\beq
H_0 = \hbar^2 T_{\infty,4} \frac{\partial \ln \hat{Z}}{\partial T_{\infty,2}} + 2 T_{\infty,1} T_{\infty,3} + T_{\infty,2}^2 - \hbar T_{\infty,4} q
\eeq
with
\beq
\hat{Z}:= \exp \left( \frac{A_{\infty,0}^+ + A_{\infty,0}^-}{2} \right) Z .
\eeq

One has
\beq
P(x) = T_{\infty,4} x^2 + T_{\infty,3} x + p - T_{\infty,4} q^2 - T_{\infty,3} q.
\eeq
 So that the $\hbar$-deformed spectral curve reads, by Theorem \ref{th-sp-curve},
\beq
\frac{\phi_\hbar}{(dx)^2} = T_{\infty,4}^2 x^4 + 2 T_{\infty,3}  T_{\infty,4}  x^3 + \left(  T_{\infty,3}^2 + 2   T_{\infty,4} T_{\infty,2} \right) x^2 + \left[ 2 T_{\infty,3} T_{\infty,2} + 2 T_{\infty,4} \left(T_{\infty,1} + \frac{\hbar}{2}\right) \right] x + H_0 .
\eeq
The fact that $(p,q)$ belongs to the $\hbar$-deformed spectral curve means that
\beq
H_0 = p^2 - T_{\infty,4}^2 q^4 - 2 T_{\infty,3}  T_{\infty,4}  q^3 - \left(  T_{\infty,3}^2 + 2   T_{\infty,4} T_{\infty,2} \right) q^2 - \left[ 2 T_{\infty,3} T_{\infty,2} + 2 T_{\infty,4}\left( T_{\infty,1} + \frac{\hbar}{2} \right) \right] q.
\eeq 

Using the change of coordinates $2 T_{\infty,2} =  t_{\infty,1}$, the evolution equations of the spectral Darboux coordinates read
\beq
2\hbar \frac{\partial p}{\partial t_{\infty,1}} =  \frac{\partial H_0}{\partial q}
\qquad \hbox{and} \qquad 2 \hbar \frac{\partial q}{\partial t_{\infty,1}} = - \frac{\partial H_0 }{\partial p}.
\eeq
One recovers a classical representation of Painlev\'e 2 by setting $T_{\infty,4} = 1$, $T_{\infty,3} = 0$, $T_{\infty,1}  = - \theta$
\beq
\hbar^2 \frac{\partial^2 q}{\partial t_{\infty,1}^2}  = 2 q^3 + t_{\infty,1} q + \frac{\hbar}{2} - \theta.
\eeq

It also gives
\beq
 p = - \hbar \frac{\partial q}{\partial t_{\infty,1}} .
 \eeq

\subsection{Painlev\'e 3}

Let us now consider the case $n=1$, $n_\infty = 0$, $r_\infty=2$ and $r_1 = 2$. This case being more subtle because $r_\infty \leq2$, we shall describe it in greater details. The Wronskian reads
\beq
W(x) =w\frac{(x-q)}{(x-X_1)^2}=\frac{w}{x-X_1}+\frac{w(X_1-q)}{(x-X_1)^2}
\eeq
with 
\bea w&=& - 2 T_{\infty,2} \exp \left(A_{\infty,0}^+ + A_{\infty,0}^-\right), \cr
(X_1-q)w&=&2 T_{1,2} \exp \left(A_{1,0}^+ + A_{1,0}^-\right) .
\eea
Let us work with $X_1$ fixed and set its  value to $X_1 = 0$.

We have also
\beq
q = 2  - \frac{T_{\infty,1}}{T_{\infty,2}} - \left(A_{\infty,1}^+ + A_{\infty,1}^-\right) \qquad \hbox{and} \qquad 
w  = - wq \frac{T_{1,1}}{T_{1,2}} - wq \left( A_{1,1}^+ + A_{1,1}^-\right) . 
\eeq

This allows expressing
\beq
\left\{
\begin{array}{l}
A_{1,0}^+ + A_{1,0}^- = \log \left(-\frac{qw}{2 T_{1,2}} \right)\cr
 A_{1,1}^+ + A_{1,1}^- = - \frac{1}{q} - \frac{T_{1,1}}{T_{1,2}} \cr
\end{array}
\right. .
 \eeq
 
 Working with fixed value of $X_1$ means that the action of the vector field $\frac{\partial }{\partial X_1}$ is vanishing. This can be translated into the vanishing of the vector field $T_{1,1} \frac{\partial }{\partial T_{1,2}} + T_{1,2} \frac{\partial }{\partial T_{1,3}} $ thanks to \eq{change-times-nu}.

From Theorem \ref{QuantumCurveTheorem}, one has
\beq
Q(x) = \frac{p}{x-q} + \frac{\hbar T_{1,2}}{2 x^2} \frac{\partial A_{1,0}^+ + A_{1,0}^-}{\partial T_{1,2}} + \frac{\hbar}{2 x} T_{1,2}  \frac{\partial A_{1,1}^+ + A_{1,1}^-}{\partial T_{1,2}}. 
\eeq
On the other hand \eq{eq-Q-times} and the asymptotics from Appendix \ref{app-asymptot} give that $Q(x)  = O(x^{-2})$ as $x \to \infty$. Thus
\beq
p = - \frac{\hbar}{2 } T_{1,2}  \frac{\partial A_{1,1}^+ + A_{1,1}^-}{\partial T_{1,2}},
\eeq
i.e.
\beq
p = - \frac{\hbar T_{1,2}}{2 q^2} \frac{\partial q}{\partial T_{1,2}} - \frac{\hbar}{2} \frac{T_{1,1}}{T_{1,2}}
\eeq
and 
\beq
Q(x) = \frac{p}{x-q} + \frac{1}{x^2} \left[ \frac{\hbar T_{1,2}}{2 q} \frac{\partial q}{\partial T_{1,2}}+ \frac{\hbar T_{1,2}}{2 } \frac{\partial \log w}{\partial T_{1,2}} - \frac{\hbar}{2} \right]
-  \frac{p}{x}. 
\eeq

The quantum curve thus reads
\beq
\left\{\hbar^2 \frac{\partial^2}{\partial x^2} - \hbar^2 \left[\frac{1}{x-q} - \frac{2}{x} \right] \frac{\partial }{\partial x} - \frac{\hbar p}{x-q}  - T_{\infty,2}^2 - \frac{C_1}{x} - \frac{C_2}{x^2}  - \frac{2 T_{1,1} T_{1,2}}{x^3} - \frac{T_{1,2}^2}{x^4} \right\} \Psi_\pm(x) = 0 
\eeq
where
\beq
C_1:= 2 T_{\infty,1} T_{\infty,2} - \hbar p
\eeq
and
\beq
C_2:= \hbar^2 T_{1,2} \frac{\partial \log \left(w^{\frac{1}{2}} Z\right)}{\partial T_{1,2}} + T_{1,1}^2 -  \hbar pq - \frac{\hbar}{2} - \frac{\hbar^2q}{2} \frac{T_{1,1}}{T_{1,2}}.
\eeq

One can now compute $P(x)$. As in all cases where $r_\infty \leq 2$, its computation is slightly more complicated. Let us write it under the form
\beq
P(x) = \frac{ax^2 + bx + c}{x^2}.
\eeq
The constraint $P(q) = p$ implies that
\beq
c = p q^2 -a q^2 - b q.
\eeq
In terms of these coefficients, the $\hbar$-deformed spectral curve reads
\beq
\frac{\phi_\hbar}{(dx)^2} =   T_{\infty,2}^2 + \frac{C_1}{x} + \frac{C_2}{x^2}  + \frac{2 T_{1,1} T_{1,2}}{x^3} + \frac{T_{1,2}^2}{x^4} + \hbar \frac{a+p}{x} + \hbar \frac{(p-a) q}{x^2}.
\eeq
On the other hand, $\frac{\phi_\hbar}{(dx)^2} = P(x)^2 + M(x) W(x)$ implies that
\beq
\frac{\phi_\hbar}{(dx)^2} = a^2 + \frac{2ab}{x} + O(x^{-2})
\eeq
since both $M(x)$ and $W(x)$ behave as $x^{-1}$ as $x \to \infty$. Equating the two leading orders gives
\beq
\left\{
\begin{array}{l}
a = T_{\infty,2} \cr
b = T_{\infty,1} + \frac{\hbar}{2}\cr
\end{array}
\right. .
\eeq

Hence the $\hbar$-deformed spectral curve reads
\beq
\frac{\phi_\hbar}{(dx)^2} =   T_{\infty,2}^2 + \frac{H_{-1}}{x} + \frac{H_{-2}}{x^2}  + \frac{2 T_{1,1} T_{1,2}}{x^3} + \frac{T_{1,2}^2}{x^4} 
\eeq
where
\beq
H_{-1}:=2 T_{\infty,2} \left(T_{\infty,1}+\frac{\hbar}{2} \right)
\qquad \hbox{and}
\qquad
H_{-2} := \hbar^2 T_{1,2} \frac{\partial \log \left(w^{\frac{1}{2}} Z\right)}{\partial T_{1,2}} + T_{1,1}^2 - \frac{\hbar}{2} - \hbar T_{\infty,2} q - \hbar^2 \frac{q T_{1,1}}{2 T_{1,2}}.
\eeq

The fact that the spectral Darboux coordinates belong to the $\hbar$-deformed spectral curve implies that
\beq
H_{-2} = p^2 q^2 - T_{\infty,2}^2 q^2 - H_{-1} q - \frac{2 T_{1,1} T_{1,2}}{q} - \frac{T_{1,2}^2}{q^2} .
\eeq

Let us now explain how to get an isomonodromic system in this case. For this purpose, one has to impose an explicit dependence of the Casimirs on the isomonodromic time $t:=t_{1,1}$ by identifying our Lax matrix with the one of \cite{MarchalIwaki}. This is done by comparing the elements $P(x)$ and $W(x)$ of our Lax matrix with the ones of \cite{MarchalIwaki} as well as the coefficients of $x^{-3}$ and $x^{-4}$ of their determinant. One obtains the identification 
\beq
\left\{
\begin{array}{l}
T_{\infty,2} := \frac{t}{2}\cr
T_{\infty,1} + \frac{\hbar}{2}:= - \frac{\theta_{\infty}}{2} \cr
T_{1,1}  :=  - \frac{\theta_{0}}{2} \cr
 T_{1,2} := \frac{t}{2} \cr
 \end{array}
 \right. .
\eeq
Let us now verify that one gets a Hamiltonian system driving the time evolution of the spectral Darboux coordinates. Identifying the value of the zero of $W(x)$ in both matrices as well as the value of $P(x)$ at the corresponding point, one gets
\beq
\left\{
\begin{array}{l}
q = -\frac{1}{q_{\text{IMS}}} \cr
p = \frac{t}{2} - \frac{\theta_\infty}{2 q} + \left(p_{\text{IMS}}- \frac{t}{2} \right) \frac{1}{q^2} \cr
\end{array}
\right. 
\eeq
 where $(q_{\text{IMS}},p_{\text{IMS}})$ refers to the variables denoted by $(q,p)$ in \cite{MarchalIwaki}. Under this identification, they obtained an isomonodromic system with Hamiltonian
 \beq
 H^{\text{IMS}}:=\frac{1}{t} \left[ 2 q_{\text{IMS}}^2 p_{\text{IMS}}^2 + 2 p_{\text{IMS}} \left(- t q_{\text{IMS}}^2 + \theta_\infty q_{\text{IMS}} +t \right)  - \left(\theta_0+\theta_\infty\right) tq_{\text{IMS}} - t^2 - \frac{1}{4} (\theta_0^2-\theta_\infty^2) \right].
 \eeq
 After the identification, the latter reads in our notations
  \beq
 H^{\text{IMS}}= \frac{2}{t} H_{-2} - \frac{t}{2} - \frac{\theta_0^2+\theta_\infty^2}{4t}.
  \eeq
 From \cite{MarchalIwaki}, one has $\hbar \frac{\partial q_{\text{IMS}}}{\partial t} = \frac{\partial  H^{\text{IMS}}}{\partial p_{\text{IMS}}}$ and $\hbar \frac{\partial p_{\text{IMS}}}{\partial t} = - \frac{\partial  H^{\text{IMS}}}{\partial q_{\text{IMS}}}$. Since
 \beq
 dp_{\text{IMS}} \wedge dq_{\text{IMS}} = - dp \wedge dq,
 \eeq
 this just amounts to a change of Darboux coordinates and one has  $\hbar \frac{\partial q}{\partial t} = - \frac{\partial  H_{-2}}{\partial p}$ and $\hbar \frac{\partial p}{\partial t} = \frac{\partial  H_{-2}}{\partial q}$. This implies that, up to a simple shift, $\om^{-\frac{1}{2}} Z$ is a tau-function.
 
 Finally, $q_{\text{IMS}} = - q^{-1}$ is solution to Painlev\'e $3$ equation
 \beq
 \hbar^2 \frac{\partial^2 q_{\text{IMS}}}{\partial t^2} = \frac{\hbar^2}{q_{\text{IMS}}} \left(\frac{\partial q_{\text{IMS}}}{\partial t}\right)^2 - \frac{\hbar^2}{t} \frac{\partial q_{\text{IMS}}}{\partial t} + \frac{4}{t} \left( - 2 T_{1,1} q_{\text{IMS}}^2 - T_{\infty,1}  \right) + 4 q_{\text{IMS}}^3 - \frac{4}{q_{\text{IMS}}}.
 \eeq

\subsection{Painlev\'e 4}

Let us now consider the case $n=1$, $n_\infty = 0$, $r_1 = 1$ and $r_\infty=3$. The Wronskian reads
\beq
W(x) =w\frac{(x-q)}{(x-X_1)}=w+\frac{w(X_1-q)}{x-X_1}
\eeq
with
\bea w&=& - 2 T_{\infty,3} \exp \left(A_{\infty,0}^+ + A_{\infty,0}^-\right)\cr
w(X_1-q)&=&2 T_{1,1} \exp \left(A_{1,0}^+ + A_{1,0}^-\right)
\eea
We have also
\beq
q = X_1 - \frac{T_{\infty,2}}{T_\infty,3} - \left( A_{\infty,1}^++ A_{\infty,1}^- \right).
\eeq

Theorem \ref{QuantumCurveTheorem} gives
\bea
\left[\hbar^2 \frac{\partial^2}{\partial x^2} - \hbar^2 \left[\frac{1}{x-q} - \frac{1}{x} \right] \frac{\partial }{\partial x} - \frac{\hbar p}{x-q} -
T_{\infty,3}^2 x^2 - 	2 T_{\infty,2} T_{\infty,3} x - T_{\infty,2}^2 - 2 T_{\infty,1} T_{\infty3} \right.\cr
\qquad  \left. - \frac{H_{-1} }{x} - \frac{T_{1,1}^2}{x^2} - \hbar \frac{T_{\infty,3} q}{x} \right] \Psi_\pm(x) = 0 
\eea
where
\beq
H_{-1} = \hbar^2 \frac{\partial \ln \hat{Z}}{\partial t_{1,0}}  + 2 T_{\infty,1} T_{\infty,2} - \hbar T_{\infty,3} q
\eeq	
with 
\beq
\hat{Z}:=  \exp \left( \frac{A_{1,0}^+ + A_{1,0}^-}{2} \right) Z .
\eeq

One can compute
\beq
P(x) = \frac{T_{\infty,3} x^2 +T_{\infty,2} x +\left(q-X_1\right) p - T_{\infty,3} q^2 -  T_{\infty,2} q}{(x-X_1)} .
\eeq
The $\hbar$-deformed spectral curve reads
\beq
\frac{\phi_\hbar}{(dx)^2} = T_{\infty,3}^2 x^2 + 	2 T_{\infty,2} T_{\infty,3} x + T_{\infty,2}^2 + 2 \left(T_{\infty,1} + \frac{\hbar}{2} \right) T_{\infty,3} + \frac{H_{-1}}{x-X_1} + \frac{T_{1,1}^2}{(x-X_1)^2} 
\eeq
and the constraint on the spectral Darboux coordinates reads
\beq
H_{-1} = (q-X_1) \left[ p^2 -T_{\infty,3}^2 q^2 -	2 T_{\infty,2} T_{\infty,3} q - T_{\infty,2}^2 - 2 \left(T_{\infty,1} + \frac{\hbar}{2}\right) T_{\infty3} - \frac{T_{1,1}^2}{(q-X_1)^2} \right] .
\eeq

To recover the Hamiltonian structure more simply, let us select a time $t$ defined by the corresponding vector field
\beq
\frac{\partial}{\partial t}:= \frac{ \partial }{\partial x} + \frac{\partial }{\partial t_{1,0}}
\eeq
and set $X_1 = 0$ and $T_{\infty,2} = t$ in order to have a solution to the isomonodromicity condition.
The compatibility of the system evaluated at $x=q$ recovers the Hamiltonian representation of the evolution equations 
\beq
\left\{
\begin{array}{l}
\hbar \frac{\partial q}{\partial t} = \frac{\partial H_{-1} }{\partial p} \cr
\hbar \frac{\partial p}{\partial t} = - \frac{\partial H_{-1}}{\partial q} \cr
\end{array}
\right.
\eeq
leading to the fact that $\hat{Z}$ is a corresponding isomonodromic tau function.

This leads to the differential equation
\beq
q \hbar^2 \frac{\partial^2 q}{\partial t^2} = \frac{1}{2} \left(\hbar \frac{\partial q}{\partial t} \right)^2 - 2 T_{1,1} + 2 \left( t^2 + 2 \left(T_{\infty,1} + \frac{\hbar}{2} \right) T_{\infty,3} \right) q^2 + 8 t T_{\infty,3} q^3 + 6 T_{\infty,3}^3 q^4.
\eeq
One recovers a known representation of Painlev\'e $4$ by setting $T_{\infty,3} = 1$, $T_{\infty,2} = t$, $T_{\infty,1} + \frac{\hbar}{2} = \theta_\infty$ and $T_{1,1} = \theta_0$.

This also gives
\beq
p = \frac{\hbar}{2 q} \frac{\partial q}{\partial t_{1,0}}.
\eeq

\subsection{Painlev\'e 5}

Let us now consider the case $n=2$, $n_\infty = 0$, $r_1 = r_2 = 1$ and $r_\infty=2$. For simplicity, we consider $X_1 = 0$ and $X_2 = 1$. The Wronskian reads
\beq
W(x) = w \frac{(x-q)}{x(x-1)}=\frac{wq}{x}+\frac{w(q-1)}{x-1}
\eeq
with
\bea
 w&=&- 2 T_{\infty,2} \exp \left(A_{\infty,0}^+ + A_{\infty,0}^-\right) \cr
qw&=&2 T_{1,1} \exp \left(A_{1,0}^+ + A_{1,0}^-\right) \cr
(q-1)w&=&2 T_{2,1} \exp \left(A_{2,0}^+ + A_{2,0}^-\right) 
\eea

From Theorem \ref{QuantumCurveTheorem}, one has
\bea
Q(x) &=& \frac{p}{x-q} + \frac{\hbar}{x} \frac{T_{1,1}}{2} \frac{\partial (A_{1,0}^++A_{1,0}^-)}{\partial T_{1,2}} + \frac{\hbar}{x-1} \frac{T_{2,1}}{2} \frac{\partial (A_{2,0}^++A_{2,0}^-)}{\partial T_{2,2}} \cr
&=& \frac{p}{x-q} + \frac{\hbar}{x} \frac{T_{1,1}}{2} \frac{\partial \log [qw]}{\partial T_{1,2}} + \frac{\hbar}{x-1} \frac{T_{2,1}}{2} \frac{\partial \log [(q-1)w]}{\partial T_{2,2}} 
.
\eea
From \eq{eq-Q-times} and Appendix \ref{app-asymptot}, one knows that  
\beq
Q(x)  =  \frac{\hbar}{x} {T_{1,1}} \frac{\partial \log [w]}{\partial T_{1,2}} + \frac{\hbar}{x} {T_{2,1}} \frac{\partial \log [w]}{\partial T_{2,2}}+  O(x^{-2})
\eeq
 as $x \to \infty$.

One thus have
\beq
p = - \hbar \frac{T_{1,1}}{2} \frac{\partial \log \frac{q}{w}}{\partial T_{1,2}} -\hbar \frac{T_{2,1}}{2} \frac{\partial \log \frac{(q-1)}{w}}{\partial T_{2,2}}  
\eeq
and the quantum curve reads
\beq
\left\{\hbar^2 \frac{\partial^2}{\partial x^2} - \hbar^2 \left[\frac{1}{x-q} - \frac{1}{x} - \frac{1}{x-1} \right] \frac{\partial }{\partial x} - \frac{\hbar p}{x-q}  - T_{\infty,2}^2 - \frac{C_0}{x} - \frac{C_1}{x-1}  - \frac{ T_{1,1}^2}{x^2} - \frac{T_{2,1}^2}{(x-1)^2} \right\} \Psi_\pm(x) = 0 
\eeq
where the coefficients of $x^{-1}$ and $x^{-2}$ in the expansion around $x =\infty$ give
\beq
C_0+ C_1:= 2 T_{\infty,1} T_{\infty,2} - \hbar p + {\hbar^2} {T_{1,1}} \frac{\partial \log [w]}{\partial T_{1,2}} + {\hbar^2}{T_{2,1}} \frac{\partial \log [w]}{\partial T_{2,2}}
\eeq
and
\beq
C_1 = T_{\infty,1}^2 - T_{1,1}^2 - T_{2,1}^2 +{\hbar T_{2,1}} \frac{\partial \log [(q-1)^{-\frac{1}{2}} w^{-\frac{1}{2}}Z ]}{\partial T_{2,2}}.
\eeq

Let us write 
\beq
P(x)  = \frac{ax^2+bx+c}{x(x-1)}
\eeq
where $P(q) = p$ imposes that
\beq
c = q (q-1) p - a q^2 - bq.
\eeq
In terms of these coefficients, the $\hbar$-deformed spectral curve reads
\beq
\frac{\phi_\hbar}{(dx)^2}  =  T_{\infty,2}^2 + \frac{C_0}{x} + \frac{C_1}{x-1}  + \frac{ T_{1,1}^2}{x^2} + \frac{T_{2,1}^2}{(x-1)^2} - \hbar \frac{(p-a) q}{x} + \hbar \frac{p(1+q)+a(1-q)}{x-1}.
\eeq
The expansion around $x=\infty$ gives
\beq
a = T_{\infty,2}  \qquad \hbox{and} \qquad b = -T_{\infty,2} + T_{\infty,1} + \frac{\hbar}{2} + \frac{\hbar^2}{2 T_{\infty,2}} \left[T_{1,1} \frac{\partial\log w}{\partial T_{1,2}} + T_{2,1} \frac{\partial\log w}{\partial T_{2,2}} \right]
\eeq
so that 
\beq
\frac{\phi_\hbar}{(dx)^2}  =  T_{\infty,2}^2 + \frac{H_0}{x} + \frac{H_1}{x-1}  + \frac{ T_{1,1}^2}{x^2} + \frac{T_{2,1}^2}{(x-1)^2} 
\eeq
with 
\beq 
\left\{
\begin{array}{l}
H_0+H_1 = 2 T_{\infty,2} \left(T_{\infty,1}+\frac{\hbar}{2} \right) + \hbar^2 \left[T_{1,1} \frac{\partial\log w}{\partial T_{1,2}} + T_{2,1} \frac{\partial\log w}{\partial T_{2,2}} \right]
\cr
H_1 = T_{\infty,1}^2 - T_{1,1}^2 - T_{2,1}^2 + {\hbar T_{2,1}} \frac{\partial \log [(q-1)^{-\frac{1}{2}} w^{-\frac{1}{2}} Z] }{\partial T_{2,2}}+ \hbar p(1+q)+T_{\infty,2} (1-q) \cr
\end{array}
\right. .
\eeq
On the other hand, one can replace the second equation by the fact that $(p,q)$ belongs to the spectral curve:
\beq
p^2 = T_{\infty,2}^2 + \frac{H_0}{q} + \frac{H_1}{q-1}  + \frac{ T_{1,1}^2}{q^2} + \frac{T_{2,1}^2}{(q-1)^2} .
\eeq

Let us now recover the associated isomonodromic system. For this purpose, we can identify our Lax matrix with the one of \cite{RMatrixHarnad} by setting
$T_{\infty,2} := t$,
$T_{1,1}:= -\frac{\kappa_1}{2}$, where $\kappa_1$ is the monodromy at $0$, $T_{2,1}:= -\frac{\kappa_2}{2}$, where $\kappa_2$ is the monodromy at $1$ and $H_0+H_1 = a$ is related to the monodromy at $\infty$. In this identification, $\hbar^2 \left[T_{1,1} \frac{\partial\log w}{\partial T_{1,2}} + T_{2,1} \frac{\partial\log w}{\partial T_{2,2}} \right]$ has to vanish because we are working on a reduced phase space (see \cite{RMatrixHarnad}) for details so that $H_0+H_1 = 2 T_{\infty,2} \left(T_{\infty,1}+\frac{\hbar}{2}\right) = a$.

After this identification, \cite{RMatrixHarnad} proves that the isomonodromy condition is fulfilled and one has an Hamiltonian system with Hamiltonian $H_0$ so that $u = \frac{1-q}{1+q}$ is solution to Painlev\'e $5$ equation
\beq
\hbar^2 \frac{\partial^2 u}{\partial t^2} = \hbar^2 \left(\frac{1}{2u} + \frac{1}{u-1}\right) \left(\frac{\partial u}{\partial t}\right)^2 - \frac{\hbar^2}{t} \frac{\partial u}{\partial t} + \frac{(\alpha u^2+\beta)(u-1)^2}{t^2 u} + \frac{\gamma u}{t} + \frac{\delta u (u+1)}{u-1}
\eeq
where $\alpha = 2 T_{2,1}^2$, $\beta = -2 T_{1,1}^2$, $\gamma = 2 T_{\infty,2} T_{\infty,1}$ and $\delta = -2$.

\subsection{Painlev\'e 6}

Let us consider the case $n=3$, $n_\infty =0$, $r_\infty =r_1 = r_2 = r_3 = 1$, $X_1=0$ and $X_2 = 1$. In this Fuchsian system, our isomonodromic time will be $X_3 = t = t_{3,0}$. The Wronskian reads
\beq
W(x) = w \frac{x-q}{x(x-1)(x-t)}=w\left(-\frac{q}{tx}+\frac{q-1}{(t-1)(x-1)}-\frac{q-t}{t(t-1)(x-t)}\right)
\eeq
where using Appendix \ref{app-asymptot}:
\bea
w &=&  - 2 T_{\infty,1} \exp \left(A_{\infty,0}^+ + A_{\infty,0}^-\right)\cr
-\frac{q}{t}w&=&2T_{1,1} \exp \left(A_{1,0}^+ + A_{1,0}^-\right)\cr
\frac{q-1}{(t-1)}w&=&2T_{2,1} \exp \left(A_{2,0}^+ + A_{2,0}^-\right)\cr
-\frac{q-t}{t(t-1)}w&=&2T_{3,1} \exp \left(A_{3,0}^+ + A_{3,0}^-\right)
\eea
and
\beq
q = 1+t - \frac{A_{\infty,1}^+-A_{\infty,1}^-}{2 T_{\infty,1}} - A_{\infty,1}^++A_{\infty,1}^-.
\eeq

From Theorem \ref{QuantumCurveTheorem}, one has
\beq
Q(x) = \frac{p}{x-q} + \frac{\hbar}{2(x-t)} \frac{\partial}{\partial t} \left[\log \left(\frac{(q-t)w}{t(t-1)}\right)\right].
\eeq
On the other hand, Appendix \ref{app-asymptot} gives that $Q(x) = \frac{\hbar}{x} \frac{\partial \log w}{\partial t} + O(x^{-2})$, leading to 
\beq
p = -\frac{\hbar}{2(x-t)} \frac{\partial}{\partial t} \left[\log \left(\frac{(q-t)}{t(t-1)w}\right)\right].
\eeq
On the other hand, using the notations of \eq{change-times-nu}, the classical spectral curve reads
\beq
\frac{\phi_0}{(dx)^2} = \frac{T_{1,1}^2}{x^2} + \frac{T_{2,1}^2}{(x-1)^2}+ \frac{T_{3,1}^2}{(x-t)^2} + \frac{\hbar^2}{x} \frac{\partial \om_{0,0}}{\partial t_{1,0}} + \frac{\hbar^2}{x-1} \frac{\partial \om_{0,0}}{\partial t_{2,0}} + \frac{\hbar^2}{x-t} \frac{\partial \om_{0,0}}{\partial t}.
\eeq
The asymptotic expansion at infinity $\frac{\phi_0}{(dx)^2} = T_{\infty,1}^2 x^{-2} + O(x^{-3})$ leads to the conditions
\beq
\hbar^2 \frac{\partial \om_{0,0}}{\partial t_{1,0}} + \hbar^2 \frac{\partial \om_{0,0}}{\partial t_{2,0}} + \hbar^2 \frac{\partial \om_{0,0}}{\partial t} = 0
\eeq
and
\beq
T_{1,1}^2 + T_{2,1}^2 + T_{3,1}^2 + \hbar^2 \frac{\partial \om_{0,0}}{\partial t_{2,0}} +t \hbar^2 \frac{\partial \om_{0,0}}{\partial t}  = T_{\infty,1}^2.
\eeq
More generally, as vector fields in $\mathcal{Q}(\mathbb{P}^1,-\infty-0-1-t,0,\mathbf{T})$, the vector fields $\hbar^2 \frac{\partial }{\partial t_{1,0}} + \hbar^2 \frac{\partial}{\partial t_{2,0}} + \hbar^2 \frac{\partial }{\partial t}$ and $\hbar^2 \frac{\partial}{\partial t_{2,0}} + \hbar^2 t \frac{\partial }{\partial t}$ are vanishing.
The quantum curve reads
\bea
\left[\hbar^2 \frac{\partial^2}{\partial x^2} - \hbar^2 \left[\frac{1}{x-q} - \frac{1}{x} - \frac{1}{x-1} -\frac{1}{x-t}\right] \frac{\partial }{\partial x} - \frac{\hbar p}{x-q}  -\frac{C_0}{x} - \frac{C_1}{x-1}- \frac{C_t}{x-t} \right. \cr
\left. - \frac{T_{1,1}^2}{x^2} - \frac{T_{2,1}^2}{(x-1)^2}- \frac{T_{3,1}^2}{(x-t)^2}  \right] \Psi_\pm =0
\eea
where
\beq
\left\{
\begin{array}{l}
C_0 = \hbar^2 \frac{\partial \log Z}{\partial t_{1,0}} \cr
C_1 = \hbar^2 \frac{\partial \log Z}{\partial t_{2,0}} \cr
C_t = \hbar^2 \frac{\partial \log \left[wZ\right]}{\partial t} - \hbar p\cr
\end{array}
\right. .
\eeq

Let us write
\beq
P(x) = \frac{ax^2+bx+c}{x(x-1)(x-t)}.
\eeq
In terms of these coefficients, the $\hbar$-deformed spectral curve reads
\beq
\frac{\phi_\hbar}{(dx)^2} = \frac{H_0}{x} + \frac{H_1}{x-1} + \frac{H_t}{x-t} 
+ \frac{T_{1,1}^2}{x^2} + \frac{T_{2,1}^2}{(x-1)^2} + \frac{T_{3,1}^2}{(x-t)^2} 
\eeq
where
\beq
\left\{
\begin{array}{l}
H_0 =\hbar^2 \frac{\partial \log Z}{\partial t_{1,0}}+  \hbar \frac{pt-q(a+p(1+t))}{t} \cr
H_1 = \hbar^2 \frac{\partial \log Z}{\partial t_{2,0}} +\hbar \frac{a(1-q)-qp(1+t)}{1-t} \cr
H_t =  \hbar^2 \frac{\partial \log \left[wZ\right]}{\partial t} - \hbar p +\hbar  \frac{a (t-q)- qp(1+t)}{t(t-1)} \cr
\end{array}
\right. 
\eeq
satisfy the conditions 
\beq
\left\{
\begin{array}{l}
H_0+H_1+H_t =0\cr
H_1+tH_t +T_{1,1}^2 + T_{2,1}^2 + T_{3,1}^2  = \hbar a  \cr
\end{array}
\right. .
\eeq

The expansion around $x=\infty$ imposes that $a$ is solution to 
\beq
a^2 = T_{\infty,1}^2 + \hbar a .
\eeq

This means that we have a Fuchsian system with monodromies at $0$, $1$, $t$ and $\infty$ given by $\theta_0 = 2 T_{1,1}$, $\theta_1 = 2 T_{2,1}$ $\theta_t = 2 T_{3,1}$ and $\theta_\infty  = 2a$ and $q$ satisfies the associated Painlev\'e $6$ equation
\bea
\hbar^2 \frac{\partial^2 q}{ \partial t^2} &=& \frac{\hbar^2}{2} \left[\frac{1}{q} + \frac{1}{q-1}+\frac{1}{q-t} \right] \left(\frac{\partial q}{\partial t}\right)^2
- \hbar^2 \left[ \frac{1}{t} + \frac{1}{t-1}+\frac{1}{q-t} \right] \frac{\partial q}{\partial t} \cr
&& \quad + \frac{q(q-1)(q-t)}{t^2(t-1)^2} \left[2 T_{\infty,1}^2+\frac{\hbar^2}{2} - 2 T_{1,1}^2 \frac{t}{q^2} + 2 T_{2,1}^2 \frac{t-1}{(q-1)^2} - \left(2T_{3,1}-\frac{\hbar^2}{2} \right) \frac{t(t-1)}{(q-t)^2} \right] .\cr
&& 
\eea

\subsection{Second equation of the Painlev\'e 2 hierarchy}

Let us consider the case $n=0$, $r_\infty = 5$ and $n_\infty = 0$. The Wronskian takes the form
\beq
W(x) = w (x-q_1) (x-q_2)
\eeq
where
\bea
w &=& -2T_{\infty,5}\exp(A_{\infty,0}^++A_{\infty,0}^-)\cr
q_1 +q_2 &=&-(A_{\infty,1}^++A_{\infty,1}^-)-\frac{T_{\infty,4}}{T_{\infty,5}}\cr
q_1 q_2 &=&(A_{\infty,2}^++A_{\infty,2}^-)+\frac{1}{2}(A_{\infty,1}^++A_{\infty,1}^-)^2+\frac{T_{\infty,4}}{T_{\infty,5}}(A_{\infty,1}^++A_{\infty,1}^-)+\frac{T_{\infty,3}}{T_{\infty,5}}
\eea

The quantum curve reads
\beaa
&&\Bigg[\hbar^2 \frac{\partial^2}{\partial x^2} - \hbar^2 \left[\sum_{i=1}^2 \frac{1}{x-q_i} \right] \frac{\partial }{\partial x} - \sum_{i=1}^2 \frac{\hbar p_i}{x-q_i} 
- T_{\infty,5}^2 x^6 - 2 T_{\infty,5} T_{\infty,4} x^5  \cr
&&  - \left( 2 T_{\infty,5} T_{\infty,3} + T_{\infty,4}^2 \right) x^4 - \left( 2 T_{\infty,5} T_{\infty,2} + 2 T_{\infty,4} T_{\infty,3} \right) x^3 
- \left( 2 T_{\infty,5} T_{\infty,1} + 2 T_{\infty,4} T_{\infty,2} + T_{\infty,3}^2 \right) x^2 \cr
&& \qquad \qquad \qquad - C_1 x - C_0 \Bigg] \Psi_\pm =0  \cr
\eeaa
where
\beq
\left\{
\begin{array}{l}
C_1 =
\hbar^2T_{\infty,5}\frac{\partial}{\partial T_{\infty,2}}\log \left(w^{\frac{1}{2}} Z\right) +2T_{\infty,4}T_{\infty,1}+2T_{\infty,3}T_{\infty,2}  \cr
C_0 = 
\hbar^2T_{\infty,4}\frac{\partial}{\partial T_{\infty,2}}\log \left(w^{\frac{1}{2}}Z\right)
+2\hbar^2T_{\infty,5}\frac{\partial}{\partial T_{\infty,3}}\log \left(w^{\frac{1}{2}} Z\right) 
+\frac{\hbar^2}{2}T_{\infty,5}\frac{\partial \log(q_1+q_2)}{\partial T_{\infty,2}}+ T_{\infty,2}^2+2T_{\infty,3}T_{\infty,1} \cr
\end{array}
\right. .
\eeq

The diagonal term of the Lax matrix reads
\beq
P(x) = T_{\infty,5} x^3 + T_{\infty,4} x^2 + \alpha x + \beta
\eeq
where
\beaa
\alpha&=&\frac{p_1-p_2}{q_1-q_2}-T_{\infty,5}(q_2^2+q_1^2+q_1q_2)-T_{\infty,4}(q_1+q_2),\cr
\beta&=&\frac{p_2q_1-p_1q_2}{q_1-q_2}+T_{\infty,5}q_1q_2(q_1+q_2)+T_{\infty,4}q_1q_2 .\cr
\eeaa

The $\hbar$-deformed spectral curve reads
\bea
\frac{\phi_\hbar}{(dx)^2} &=& T_{\infty,5}^2 x^6 + 2 T_{\infty,5} T_{\infty,4} x^5 + \left( 2 T_{\infty,5} T_{\infty,3} + T_{\infty,4}^2 \right) x^4 + \left( 2 T_{\infty,5} T_{\infty,2} + 2 T_{\infty,4} T_{\infty,3} \right) x^3 \cr
&& + \left( 2 T_{\infty,5} T_{\infty,1} + 2 T_{\infty,4} T_{\infty,2} + T_{\infty,3}^2 \right) x^2 + H_1 x + H_0 
\eea
where
\beq
\left\{
\begin{array}{l}
H_1 = C_1 - \hbar T_{\infty,5} (q_1+q_2) \cr
H_0 = C_0 - \hbar T_{\infty,5} (q_1^2+q_2^2) - \hbar T_{\infty,4} (q_1+q_2)\cr
\end{array}
\right. .
\eeq

On the other hand, the fact that the spectral Darboux coordinates belong to the $\hbar$-deformed spectral curve implies that $H_0$ and $H_1$ are subject to the constraint 
\beq \label{eq-P22-Ham}
\begin{array}{rcl}
H_1 q_i + H_0 &=& p_i^2 - T_{\infty,5}^2 q_i^6 - 2 T_{\infty,5} T_{\infty,4} q_i^5 - \left( 2 T_{\infty,5} T_{\infty,3} + T_{\infty,4}^2 \right) q_i^4 - \left( 2 T_{\infty,5} T_{\infty,2} + 2 T_{\infty,4} T_{\infty,3} \right) q_i^3 \cr
&& - \left( 2 T_{\infty,5} T_{\infty,1} + 2 T_{\infty,4} T_{\infty,2} + T_{\infty,3}^2 \right) q_i^2 \cr
\end{array}
\eeq
for $i\in\{1,2\}$.  This means that
\beaa
H_1&=&\frac{p_1^2-p_2^2}{q_1-q_2}-T_{\infty,5}^2(q_1+q_2)(q_1^4+q_1^2q_2^2+q_2^4)-2T_{\infty,5}T_{\infty,4}(q_1^4+q_1^3q_2+q_1^2q_2^2+q_1q_2^3+q_2^4)\cr
&&-(2T_{\infty,5}T_{\infty,3}+T_{\infty,4}^2)(q_1+q_2)(q_1^2+q_2^2)-(2T_{\infty,5}T_{\infty,2}+2T_{\infty,4}T_{\infty,3})(q_1^2+q_1q_2+q_2^2)\cr
&&-(2T_{\infty,5}T_{\infty,1}+2T_{\infty,4}T_{\infty,2}+T_{\infty,3}^2)(q_1+q_2),\cr
H_0&=&\frac{q_1p_2^2-q_2p_1^2}{q_1-q_2}+q_1q_2\Big[T_{\infty,5}^2(q_1^4+q_1^3q_2+q_1^2q_2^2+q_1q_2^3+q_2^4)+2T_{\infty,5}T_{\infty,4}(q_1^3+q_1^2q_2+q_1q_2^2+q_2^3)\cr
&&+(T_{\infty,4}^2+2T_{\infty,5}T_{\infty,3})(q_1^2+q_1q_2+q_2^2)+(2T_{\infty,4}T_{\infty,3}+2T_{\infty,5}T_{\infty,2})(q_1+q_2)+T_{\infty,3}^2\cr
&&+2T_{\infty,4}T_{\infty,2}+2T_{\infty,5}T_{\infty,1}\Big].
\eeaa

As explained for example in Section 6.3.1 of \cite{MO}, it is possible to introduce an explicit dependence in the times  $t_{\infty,1}$ and $t_{\infty,2}$ in such a way that they give rise to isomonodromic deformations by solving \eq{isomono-condition}.

This implies that, after having made the Lax matrix $L(x)$ explicitly dependent in these times, one obtains Hamilton's equations
\beq
\forall \,i\in\{1,2\} \, , \; \left\{
\begin{array}{l}
2 \hbar \frac{\partial q_i}{\partial t_{\infty,1}} = - \frac{\partial {H}_0}{\partial p_i} \cr
2 \hbar \frac{\partial q_i}{\partial t_{\infty,2}} = - \frac{\partial {H}_1}{\partial q_i}  \cr
2 \hbar \frac{\partial p_i}{\partial t_{\infty,1}} =  \frac{\partial {H}_0}{\partial q_i} \cr
2 \hbar \frac{\partial p_i}{\partial t_{\infty,2}} =  \frac{\partial {H}_1}{\partial q_i}  \cr
\end{array}
\right. .
\eeq

The second equation gives 
\beq
\left\{
\begin{array}{l}
p_1 = - (q_1-q_2) \frac{\partial q_1}{\partial t_{\infty,2}}\cr
p_2 = -(q_1-q_2) \frac{\partial q_2}{\partial t_{\infty,2}}\cr
\end{array}
\right. .
\eeq


\eop

\appendix

\numberwithin{equation}{section}

\section{PDE for the perturbative wave functions}
\label{app-proof-PDE}

This section is devoted to the proof of Theorem \ref{th-PDE}.

\subsection{Proof of Lemma \ref{lemma-PDE-1}}
\label{app-proof-PDE-1}

In this section, we prove Lemma \ref{lemma-PDE-1}. For this purpose, we generalize the procedure of \cite{Quantum} used also in \cite{Iwaki}. This is simply obtained by writing the integration of the RHS of the topological recursion formula along the boundary of a fundamental domain $\mathcal{D}$ obtained by cutting along the $\mathcal{A}$ and $\mathcal{B}$ cycles considered. 

\subsubsection{Case $2h-2+n \geq 2$.}

In this case, it reads, after integrating the variables $z_2,\dots,z_n$ along paths from $z_i$ to $\sigma(z_i)$ as (since the poles of the integrand are either ramification points or coinciding points)
\bea\label{eq-contour-TR}
\frac{1}{ 2 \pi i} \oint_{ z \in \delta \mathcal{D}} K(z_1,z) R_{h,n}(z,z_2,\dots,z_n) &=& \sum_{a \in \mathcal{R}} \Res_{z \to a} K(z_1,z) R_{h,n}(z,z_2,\dots,z_n)\cr
&&+ \sum_{i=1}^n  \Res_{z \to z_i,\sigma(z_i)} K(z_1,z) R_{h,n}(z,\dots,z_n) 
\eea
where, for $2h-2+n \geq 1$
\bea
 R_{h,n}(z_1,\dots,z_n) &:=& d_{u_1} d_{u_2}\left[ F_{h-1,n+1} (u_1,u_2,z_2,\dots,z_n)  \right.\cr
&& \left.   \left. + \overset{\mathrm{stable}}{\sum_{\begin{array}{c}
h_1+h_2 = h \cr
A\sqcup B = \{z_2,\dots,z_n\} \cr
\end{array}}}  F_{h_1,|A|+1}(u_1,A) F_{h_2,|B|+1}(u_2,B)\right] \right|_{u_1 = z_1 \, , \; u_2= \sigma(z_1)} \cr
&& + \sum_{j=2}^n \frac{1}{2} \int_{\sigma(z_j)}^{z_j} \om_{0,2}(z_1, \cdot) \; d_{\sigma(z_1)}  F_{h,n-1}(\sigma(z_1),\mathbf{z}_{\{2,\dots,n\}\setminus \{j\}}) \cr
&& + \sum_{j=2}^n \frac{1}{2} \int_{\sigma(z_j)}^{z_j} \om_{0,2}(\sigma(z_1), \cdot) \; d_{z_1}  F_{h,n-1}(z_1,\mathbf{z}_{\{2,\dots,n\}\setminus \{j\}}) \cr
&&\eea
where $d_u$ refers to the exterior derivative with respect to the variable $u$ (which has nothing to do with a local coordinate),
\beq
K(z_1,z) :=  \frac{\int_{\sigma(z)}^z \om_{0,2}(z_1,\cdot)}{2 (\om_{0,1}(z)-\om_{0,1}(\sigma(z)))}
\eeq
 and  $d_u d_v F_{0,2}(u,v) := \om_{0,2}(u,v)$.
 In order to derive this expression, one has used that for $(h,n) \neq (0,2)$
 \beq
 d_{z_1} F_{h,n}(z_1,\dots,z_n) = \frac{1}{2^{n-1}} \int_{\sigma(z_2)}^{z_2} \dots \int_{\sigma(z_n)}^{z_n} \om_{h,n}(z_1,\cdot,\dots,\cdot).
 \eeq

The first term of the right hand side is the recursive definition of $d_{z_1} F_{h,n}(z_1,\dots,z_n)$. 

The other terms get contributions only from the poles of $\om_{0,2}$. First of all, thanks to \eq{eq-skew-sym} and \eq{eq-skew-sym-int}, one can observe that 
\beq
K(z,z_1) R_{h,n}(z,z_2,\dots,z_n) = K(\sigma(z),z_1) R_{h,n}(\sigma(z),z_2,\dots,z_n) 
\eeq
meaning that, for $i\in \llbracket1,n\rrbracket$,
\beq
\Res_{z \to z_i,\sigma(z_i)} K(z_1,z) R_{g,n}(z,\dots,z_n) = 2 \Res_{z \to z_i} K(z_1,z) R_{g,n}(z,\dots,z_n).
\eeq
The same properties imply that 
\bea
 R_{h,n}(z_1,\dots,z_n) &:=& - d_{u_1} d_{u_2}\Bigg[ F_{h-1,n+1} (u_1,u_2,z_2,\dots,z_n) + \cr
&& \left.   \left. + \overset{\mathrm{stable}}{\sum_{\begin{array}{c}
h_1+h_2 = h \cr
A\sqcup B = \{z_2,\dots,z_n\} \cr
\end{array}}}  F_{h_1,|A|+1}(u_1,A) F_{h_2,|B|+1}(u_2,B)\right] \right|_{u_1 = u_2 = z_1} \cr
&& - \sum_{j=2}^n  \int_{\sigma(z_j)}^{z_j} \om_{0,2}(z_1, \cdot) \; d_{z_1}  F_{h,n-1}(z_1,\mathbf{z}_{\{2,\dots,n\}\setminus \{j\}}) \cr
&&
\eea

One has a simple pole as $z \to z_1$ giving
\bea
&&\Res_{z \to z_1,\sigma(z_1)} K(z_1,z) R_{g,n}(z,\dots,z_n)
=  \frac{1}{2 \om_{0,1}(z_1)}
d_{u_1} d_{u_2}\Bigg[ F_{h-1,n+1} (u_1,u_2,z_2,\dots,z_n) + \cr
&& \left.   \left. + \overset{\mathrm{stable}}{\sum_{\begin{array}{c}
h_1+h_2 = h \cr
A\sqcup B = \{z_2,\dots,z_n\} \cr
\end{array}}}  F_{h_1,|A|+1}(u_1,A) F_{h_2,|B|+1}(u_2,B)\right] \right|_{u_1 = u_2 = z_1} \cr
&& + \sum_{j=2}^n  \frac{\int_{\sigma(z_j)}^{z_j} \om_{0,2}(z_1, \cdot)}{2 \om_{0,1}(z_1)} \; d_{z_1}  F_{h,n-1}(z_1,\mathbf{z}_{\{2,\dots,n\}\setminus \{j\}})
\eea
where $\mathbf{z}_{\{2,\dots,n\}\setminus \{j\}}=\{z_2,\dots,z_n\}\setminus\{z_j\}$. 

One can further compute
\bea
\Res_{z \to z_j,\sigma(z_j)} K(z_1,z) R_{h,n}(z,\dots,z_n) & = &
2 \Res_{z \to z_j} K(z_1,z) R_{h,n}(z,\dots,z_n) \cr
& = &
- \frac{\int_{\sigma(z_j)}^{z_j} \om_{0,2}(z_1,\cdot)}{2 \om_{0,1}(z_j)}
d_{z_j} F_{h,n-1}(z_j,\mathbf{z}_{\{2,\dots,n\}\setminus \{j\}}) 
\eea

Combining all this, one gets
\beq
\begin{array}{rcl}
&&\frac{1}{2 \pi i} \oint_{ z \in \delta \mathcal{D}} K(z_1,z) R_{h,n}(z,\dots,z_n) = d_{z_1} F_{h,n}(z_1, \dots,z_n)  \cr
&& + {\displaystyle \sum_{j=2}^n} \int_{\sigma(z_j)}^{z_j} \om_{0,2}(z_1,\cdot)  \left[\frac{ d_{z_1} F_{h,n-1}(z_1,\mathbf{z}_{\{2,\dots,n\}\setminus \{j\}}) }{2 \om_{0,1}(z_1)} 
- \frac{ d_{z_j} F_{h,n-1}(z_j,\mathbf{z}_{\{2,\dots,n\}\setminus \{j\}}) }{2 \om_{0,1}(z_j)} \right]  \cr
 && + \frac{1}{2 \om_{0,1}(z_1)} d_{u_1} d_{u_2} \Bigg[ F_{h-1,n+1}(u_1,u_2,z_2,\dots,z_n) \cr
&& + {\displaystyle \overset{\mathrm{stable}}{\sum_{\begin{array}{c}
h_1+h_2 = h \cr
A\sqcup B = \{z_2,\dots,z_n\} \cr
\end{array}}}} 
F_{h_1,|A|+1}(u_1,\mathbf{z}_A) \, F_{h_2,|B|+1}(u_2,\mathbf{z}_B)
\Bigg]_{u_1 = u_2 = z_1} .
\cr
\end{array}
\eeq

By Riemann bilinear identity, the left hand side is an holomorphic form in $z_1$, thus concluding the proof.

\subsubsection{Case $2h-2+n = 1$.}

Let us now consider the case $(h,n) = (0,3)$. 
One has
\beq
\frac{1}{2 \pi i} \oint_{ z \in \delta \mathcal{D}} K(z_1,z) R_{0,3}(z,z_2,z_3) = 
\sum_{a \in \mathcal{R}^{\text{finite}}} \Res_{z \to a} K(z_1,z) R_{0,3}(z,z_2,z_3) + \sum_{i=1}^3 \Res_{z \to z_i,\sigma(z_i)} K(z_1,z) R_{0,3}(z,z_2,z_3) 
\eeq
where
\beq
\begin{array}{rcl}
R_{0,3}(z_1,z_2,z_3) &:=& 
\frac{1}{4} \left( \int_{\sigma(z_2)}^{z_2} \om_{0,2}(z_1,\cdot) \; \int_{\sigma(z_3)}^{z_3} \om_{0,2}(\sigma(z_1),\cdot) +  \int_{\sigma(z_3)}^{z_3} \om_{0,2}(z_1,\cdot) \; \int_{\sigma(z_2)}^{z_2} \om_{0,2}(\sigma(z_1),\cdot) \right) \cr
&=& - \frac{1}{2}  \int_{\sigma(z_2)}^{z_2} \om_{0,2}(z_1,\cdot) \; \int_{\sigma(z_3)}^{z_3} \om_{0,2}(z_1,\cdot), \cr
\end{array}
\eeq
the second equality follows from \eq{eq-skew-sym-int}.

Once again, the left hand side is holomorphic while the first term on the right hand side is the recursive definition of $d_{z_1} F_{0,3}(z_1,z_2,z_3)$.

Evaluation of the residues gives
\bea\label{Case03}
&&\frac{1}{2 \pi i} \oint_{ z \in \delta \mathcal{D}} K(z_1,z) R_{0,3}(z,z_2,z_3) = d_{z_1} F_{0,3}(z_1,z_2,z_3) + \frac{\int_{\sigma(z_2)}^{z_2} \om_{0,2}(z_1,\cdot) \; \int_{\sigma(z_3)}^{z_3} \om_{0,2}(z_1,\cdot)}{4 \om_{0,1}(z_1)}\cr
&& - \frac{\int_{\sigma(z_2)}^{z_2} \om_{0,2}(z_1,\cdot) \; \int_{\sigma(z_3)}^{z_3} \om_{0,2}(z_2,\cdot)}{4 \om_{0,1}(z_2)}
- \frac{\int_{\sigma(z_3)}^{z_3} \om_{0,2}(z_1,\cdot) \; \int_{\sigma(z_2)}^{z_2} \om_{0,2}(z_3,\cdot)}{4 \om_{0,1}(z_3)} .\cr
&&
\eea

Finally, for $(h,n) = (1,1)$, one only has contributions from poles at the ramification points, $z_1$ and $\sigma(z_1)$. This gives
\beq
\frac{1}{2 \pi i} \oint_{ z \in \delta \mathcal{D}} K(z_1,z) R_{1,1}(z) = d_{z_1} F_{1,1}(z_1)
- \frac{\om_{0,2}(z_1,\sigma(z_1))}{2 \om_{0,1}(z_1)}.
\eeq

\subsection{Conclusion of the proof of Theorem \ref{th-PDE}}
\label{app-proof-PDE-2}

\subsubsection{Case $2h-2+n \geq 2$.}

Let us apply \eq{eq-res-holo} to the holomorphic differential of \eq{eq-holo-1}.

To simplify the expression obtained, let us first note that 
\beq
\sum_{p \in \mathcal{P}} \Res_{z_2 \to p} \frac{\om(z_2) \, y(z_2)}{x(z_2)-x(z_1)}
\eeq
is vanishing for $\om(z) = \frac{\Omega(z)}{\om_{0,1}(z)}$ where $\Omega(z)$ is a quadratic differential holomorphic at the $p$'s. This allows getting rid of the contributions of this residue for all the terms proportional to $\frac{1}{\om_{0,1}(z_1)}$ in \eq{eq-holo-1}. 

One thus gets
\beq
\begin{array}{l}
d_{z_1} F_{h,n}(z_1, \dots,z_n) +
{\displaystyle \sum_{j=2}^n} \int_{\sigma(z_j)}^{z_j} \om_{0,2}(z_1,\cdot)  \left[\frac{ d_{z_1} F_{h,n-1}(z_1,\mathbf{z}_{\{2,\dots,n\}\setminus \{j\}}) }{2 \om_{0,1}(z_1)} 
- \frac{ d_{z_j} F_{h,n-1}(z_j,\mathbf{z}_{\{2,\dots,n\}\setminus \{j\}}) }{2 \om_{0,1}(z_j)} \right]  \cr
  + \frac{1}{2 \om_{0,1}(z_1)} d_{u_1} d_{u_2} \Bigg[ F_{h-1,n+1}(u_1,u_2,z_2,\dots,z_n) \cr
 \qquad + {\displaystyle \overset{\mathrm{stable}}{\sum_{\begin{array}{c}
h_1+h_2 = h \cr
A\sqcup B = \{z_2,\dots,z_n\} \cr
\end{array}}}} 
F_{h_1,|A|+1}(u_1,\mathbf{z}_A) \, F_{h_2,|B|+1}(u_2,\mathbf{z}_B)
\Bigg]_{u_1 = u_2 = z_1} 
\cr
= -\frac{dx(z_1)}{2 y(z_1)} {\displaystyle \sum_{p \in \mathcal{P}} \Res_{z \to p}} \frac{  y(z)}{x(z)-x(z_1)} \left[
d_{z} F_{h,n}(z,z_2, \dots,z_n) -
{\displaystyle \sum_{j=2}^n} \int_{\sigma(z_j)}^{z_j} \om_{0,2}(z,\cdot) 
 \frac{ d_{z_j} F_{h,n-1}(z_j,\mathbf{z}_{\{2,\dots,n\}\setminus \{j\}}) }{2 \om_{0,1}(z_j)} \right] \cr
\end{array}
\eeq

Using that, for any function $f$,
\beq
\lim_{z_1 \to z_j} \int_{\sigma(z_j)}^{z_j} \om_{0,2}(z_1,\cdot) \left[f(z_1) - f(z_j) \right] = d_{z_j} f(z_j),
\eeq
the diagonal specialization $z = z_1 = z_2 = \dots = z_n$ gives
\beq
\begin{array}{l}
 \frac{1}{n} d F_{h,n}(z,\dots,z) 
 + (n-1) d_{z_1} \left[\frac{1}{2 \om_{0,1}(z_1)} d_{z_1} F_{h,n-1}(z_1,z,z,\dots,z)\right]_{z_1 =z} \cr
 + \frac{1}{2 \om_{0,1}(z)} d_{u_1} d_{u_2} \left[ F_{h-1,n+1}(u_1,u_2,z,\dots,z) \right]_{u_1 = u_2 = z} \cr
 +  \frac{1}{2 \om_{0,1}(z)} {\displaystyle \overset{\mathrm{stable}}{\sum_{\begin{array}{c}
h_1+h_2 = h \cr
n_1+n_2 = n-1 \cr
\end{array}}}} 
\frac{(n-1)!}{n_1! \, n_2 !} \frac{d F_{h_1,n_1+1}(z,\dots,z)}{n_1+1} \, \frac{d F_{h_2,n_2+1}(z,\dots,z)}{n_2+1}
 \cr
 = -\frac{dx(z)}{2 y(z)} {\displaystyle \sum_{p \in \mathcal{P}} \Res_{z' \to p}} \frac{  y(z')}{x(z')-x(z)} 
d_{z'} F_{h,n}(z',z, \dots,z) 
 + \frac{ d_{z} F_{h,n-1}(z,\dots,z) }{4 y(z)^2 } 
{\displaystyle \sum_{p \in \mathcal{P}} \Res_{z' \to p}} \frac{  y(z')}{x(z')-x(z)} 
 \int_{\sigma(z)}^{z} \om_{0,2}(z',\cdot)   . \cr
\end{array}
\eeq
which, multiplying by $\frac{2 y(z) }{(n-1)! \, dx(z)}$, can be written
\beq\label{eq-PDE-inter-1}
\begin{array}{l}
 2 y(z) \frac{d}{dx(z)} \left[ \frac{F_{h,n}(z,\dots,z) }{n!} \right] 
  + \left(\frac{d}{dx(z)}\right)^2 \left[\frac{F_{h,n-1}(z,z,z,\dots,z)}{(n-1)!} \right] \cr
  +   {\displaystyle \overset{\mathrm{stable}}{\sum_{\begin{array}{c}
h_1+h_2 = h \cr
n_1+n_2 = n-1 \cr
\end{array}}} }
\frac{d}{dx(z)} \left[\frac{F_{h_1,n_1+1}(z,\dots,z)}{(n_1+1)!} \right] \, \frac{d}{dx(z)} \left[\frac{F_{h_2,n_2+1}(z,\dots,z)}{(n_2+1)!}\right]
 \cr
 + \frac{d}{dx(z)} \left[ \frac{F_{h,n-1}(z,\dots,z)}{(n-1)!} \right] \left(- \frac{ d \log y(z)}{dx(z)}  
- \frac{1 }{2 y(z)} {\displaystyle \sum_{p \in \mathcal{P}} \Res_{z' \to p}} \frac{  y(z')}{x(z')-x(z)} 
 \int_{\sigma(z)}^{z} \om_{0,2}(z',\cdot)  
  \right)
 \cr
 + \frac{d^2}{dx(u_1) \, dx(u_2)} \left[ \frac{F_{h-1,n+1}(u_1,u_2,z,\dots,z)}{(n-1)!} - \frac{F_{h,n-1}(u_1,u_2,z,\dots,z)}{(n-3)!} \right]_{u_1 = u_2 = z} \cr
 = - {\displaystyle \sum_{p \in \mathcal{P}} \Res_{z' \to p}} \frac{  y(z')}{x(z')-x(z)} 
\frac{d_{z'} F_{h,n}(z',z, \dots,z)}{(n-1)!} .\cr
\end{array}
\eeq

To simplify this expression, let us compute
\beq\label{eq-res-holo-1}
\begin{array}{rcl}
- {\displaystyle \sum_{p \in \mathcal{P}} \Res_{z' \to p}} \frac{  y(z')}{x(z')-x(z)} 
 \int_{\sigma(z)}^{z} \om_{0,2}(z',\cdot)  &=&   {\displaystyle \Res_{z' \to z,\sigma(z)}} \frac{  y(z')}{x(z')-x(z)} 
 \int_{\sigma(z)}^{z} \om_{0,2}(z',\cdot)   \cr
&=&   2 {\displaystyle \Res_{z' \to z}} \frac{  y(z')}{x(z')-x(z)} 
 \int_{\sigma(z)}^{z} \om_{0,2}(z',\cdot)   ,\cr
\end{array}
\eeq
where the first equality follows from the absence of boundary term and the second one from the invariance of the integrand under $z' \to \sigma(z')$.

Let us now remind that 
\beq
\int_{\sigma(z)}^{z} \om_{0,2}(z',\cdot)  = 2 d_{z'} F_{0,2}(z',z) + \frac{dx(z')}{x(z')-x(z)}
\eeq
where $d_{z'} F_{0,2}(z',z)$ is holomorphic at $z' \to z$. Plugging this expression into \eq{eq-res-holo-1} gives
\beq
\begin{array}{rcl}
- {\displaystyle \sum_{p \in \mathcal{P}} \Res_{z' \to p}} \frac{  y(z')}{x(z')-x(z)} 
 \int_{\sigma(z)}^{z} \om_{0,2}(z',\cdot)  
 &=&   2 {\displaystyle \Res_{z' \to z}} \frac{  y(z')}{x(z')-x(z)} \left[2 d_{z'} F_{0,2}(z',z) + \frac{dx(z')}{x(z')-x(z)}\right] \cr
 &=& 4 \frac{  y(z)}{dx(z)} \left.d_{z'} F_{0,2}(z',z)\right|_{z' = z} + 2 \frac{dy(z)}{dx(z)} \cr
 \end{array}
 \eeq
 and thus
 \beq\label{eq-S0-logy}
- \frac{ d \log y(z)}{dx(z)}  
- \frac{1 }{2 y(z)} {\displaystyle \sum_{p \in \mathcal{P}} \Res_{z' \to p}} \frac{  y(z')}{x(z')-x(z)} 
 \int_{\sigma(z)}^{z} \om_{0,2}(z',\cdot) =  2 \left.  \frac{ dF_{0,2}(z',z) }{dx(z')} \right|_{z' = z}  . 
 \eeq
Plugging this into \eq{eq-PDE-inter-1}, one gets
\beq
\begin{array}{l}
  \left(\frac{d}{dx(z)}\right)^2 \left[\frac{F_{h,n-1}(z,z,z,\dots,z)}{(n-1)!} \right] 
  +   {\displaystyle {\sum_{\begin{array}{c}
h_1+h_2 = h \cr
n_1+n_2 = n-1 \cr
\end{array}}} }
\frac{d}{dx(z)} \left[\frac{F_{h_1,n_1+1}(z,\dots,z)}{(n_1+1)!} \right] \, \frac{d}{dx(z)} \left[\frac{F_{h_2,n_2+1}(z,\dots,z)}{(n_2+1)!}\right]
 \cr
 + \frac{d^2}{dx(u_1) \, dx(u_2)} \left[ \frac{F_{h-1,n+1}(u_1,u_2,z,\dots,z)}{(n-1)!} - \frac{F_{h,n-1}(u_1,u_2,z,\dots,z)}{(n-3)!} \right]_{u_1 = u_2 = z} \cr
 = - {\displaystyle \sum_{p \in \mathcal{P}} \Res_{z' \to p}} \frac{  y(z')}{x(z')-x(z)} 
\frac{d_{z'} F_{h,n}(z',z, \dots,z)}{(n-1)!} .\cr
\end{array}
\eeq
  
Summing over $h$ and $n$ such that $2h-2+n  =m \geq 2$, one gets
 \beq\label{eq-Sm-res}
\frac{ \partial^2 S_{m-1}^{+ \text{ pert}}(x) }{\partial x^2} + \sum_{m_1+m_2 = m-1} \frac{\partial S_{m_1}^{+ \text{ pert}}(x)}{\partial x} \; \frac{\partial S_{m_2}^{+ \text{ pert}}(x)}{\partial x} = 
- {\displaystyle \sum_{2h-2+n = m}} {\displaystyle \sum_{p \in \mathcal{P}} \Res_{z' \to p}} \frac{  y(z')}{x(z')-x(z)} 
\frac{d_{z'} F_{h,n}(z',z, \dots,z)}{(n-1)!} .
\eeq
Let us now interpret the right hand side in terms of the variational formulas.
To do so, one shall compute the residues as $z \to p$ whose expressions in terms of local coordinates depend on whether $x(p) = \infty$ or not.
\begin{itemize}

\item For $p = b_\nu^\pm$, a local coordinate is $x(z') - X_\nu$. Thus,
\beq
\Res_{z' \to b_\nu^\pm} \frac{  y(z')}{x(z')-x(z)} \frac{d_{z'} F_{h,n}(z',z, \dots,z)}{(n-1)!} 
 = - \sum_{k=0}^\infty (x(z)- X_\nu)^{-k-1} \Res_{z' \to b_\nu^\pm} y(z') \frac{d_{z'} F_{h,n}(z',z, \dots,z)}{(n-1)!} (x(z')-X_\nu)^k.
 \eeq
Since
\beq
y(z') = \pm \sum_{l=1}^{r_\nu} T_{\nu,l} (x(z')-X_\nu)^{-l} + O(1)
\eeq
and
\beq
\forall K \geq  2 \, , \; \Res_{z' \to b_\nu^\pm} d_{z'} F_{h,n}(z',z, \dots,z) (x(z')-X_\nu)^{-K+1} = \pm (K-1) \frac{\partial  F_{h,n-1}(z, \dots,z)}{\partial T_{\nu,K}},
\eeq
one has
\beq
{\displaystyle \Res_{z' \to b_\nu^\pm}} \frac{  y(z')}{x(z')-x(z)} \frac{d_{z'} F_{h,n}(z',z, \dots,z)}{(n-1)!} 
 = - {\displaystyle \sum_{K = 2}^{r_{\nu}+1}} U_{\nu,K}(x(z)) \frac{\partial  \frac{F_{h,n-1}(z,\dots,z)}{(n-1)!}}{\partial T_{\nu,K}} 
\eeq
where 
\beq
U_{\nu,K}(x) =  (K-1) \sum_{l=K-1}^{r_{\nu}} T_{\nu,l} (x-X_\nu)^{-l+K-2} 
\eeq
is a rational function of $x$.

\item For $ p = b_\infty^\pm$ when $n_\infty =0$, a local coordinate is $x(z')^{-1}$. Let us write
\beq
\Res_{z' \to b_\infty^\pm} \frac{  y(z')}{x(z')-x(z)} \frac{d_{z'} F_{h,n}(z',z, \dots,z)}{(n-1)!}  = 
\sum_{k=0}^\infty x(z)^k \Res_{z' \to b_\infty^\pm}   y(z') \frac{d_{z'} F_{h,n}(z',z, \dots,z)}{(n-1)!} x(z')^{-k-1}.
\eeq
Reminding that
\beq
y(z)  = \mp \sum_{k=1}^{r_\infty} T_{\infty,k} x(z)^{k-2} + O(x(z)^{-2})
\eeq
and
\beq
\forall\, K \geq  2 \, , \; \Res_{z' \to b_\infty^\pm} d_{z'} F_{h,n}(z',z, \dots,z) x(z')^{K-1} = \mp (K-1) \frac{\partial  F_{h,n-1}(z, \dots,z)}{\partial T_{\infty,K}},
\eeq
one has
\beq
\Res_{z' \to b_\infty^\pm} \frac{  y(z')}{x(z')-x(z)} \frac{d_{z'} F_{h,n}(z',z, \dots,z)}{(n-1)!}  = - {\displaystyle \sum_{K = 2}^{r_{\infty}-2}} U_{\infty,K}(x(z)) \frac{\partial  \frac{F_{h,n-1}(z,\dots,z)}{(n-1)!}}{\partial T_{\infty,K}} 
\eeq
where
\beq
U_{\infty,K}(x) =  (K-1) \sum_{l=K+2}^{r_{\infty}} T_{\infty,l} x^{l-K-2} .
\eeq

\item For $ p = b_\infty$ when $n_\infty =1$, a local coordinate is $x(z)^{-\frac{1}{2}}$.
Reminding that
\beq
y(z)  = - \sum_{k=1}^{r_\infty} \frac{T_{b_\infty,k}}{2} x(z)^{k-\frac{5}{2}} + O(x(z)^{-\frac{5}{2}})
\eeq
and
\beq
\forall \,K \geq  2 \, , \; \Res_{z' \to b_\infty} d_{z'} F_{h,n}(z',z, \dots,z) x(z')^{K-\frac{3}{2}} = (2K-3) \frac{\partial  F_{h,n-1}(z, \dots,z)}{\partial T_{b_\infty,K}},
\eeq
one has
\beq
\Res_{z' \to b_\infty} \frac{  y(z')}{x(z')-x(z)} \frac{d_{z'} F_{h,n}(z',z, \dots,z)}{(n-1)!}  = - {\displaystyle \sum_{K = 2}^{r_{\infty}-2}} U_{\infty,K}(x(z)) \frac{\partial  \frac{F_{h,n-1}(z,\dots,z)}{(n-1)!}}{\partial T_{b_\infty,K}} 
\eeq
where
\beq
U_{\infty,K}(x) =  \left(K-\frac{3}{2}\right) \sum_{l=K+2}^{r_{\infty}} T_{b_\infty,l} x^{l-K-2} .
\eeq

\end{itemize}

Plugging this into \eq{eq-Sm-res} proves for $m\geq 2$:
 \bea \label{eq-Sm}
0 &=& \frac{ \partial^2 S_m^{+\text{ pert}}(x) }{\partial x^2} + \sum_{m_1+m_2 = m-1} \frac{\partial S_{m_1}^{+\text{ pert}}(x)}{\partial x} \; \frac{\partial S_{m_2}^{+\text{ pert}}(x)}{\partial x} \cr
&& -  {\displaystyle \sum_{K = 2}^{r_{\infty}-2}} U_{\infty,K}(x(z)) \frac{\partial  S_{m-1}^{+\text{ pert}}(x)}{\partial T_{\infty,K}} 
- \sum_{\nu} {\displaystyle \sum_{K = 2}^{r_{b_\nu}+1}} U_{b_\nu,K}(x(z)) \frac{\partial  S_{m-1}^{+\text{ pert}}(x)}{\partial T_{b_\nu,K}} \cr 
&& - \sum_{k=0}^\infty \delta_{m+1,2k} \left[{\displaystyle \sum_{K = 2}^{r_{\infty}-2}} U_{\infty,K}(x(z)) \frac{\partial  F_{k,0}}{\partial T_{\infty,K}} 
 \sum_{\nu} {\displaystyle \sum_{K = 2}^{r_{b_\nu}+1}} U_{b_\nu,K}(x(z)) \frac{\partial  F_{k,0}}{\partial T_{b_\nu,K}} 
\right]
 . \cr
&&
\eea

\subsubsection{Case $2h-2+n = 1$}

Let us now proceed in the same way for $(h,n) = (0,3)$. Applying \eq{eq-res-holo-1} to the holomorphic differential \eq{Case03}, one gets
\bea\label{EQ30New}
&& d_{z_1} F_{0,3}(z_1,z_2,z_3) + \frac{\int_{\sigma(z_2)}^{z_2} \om_{0,2}(z_1,\cdot) \; \int_{\sigma(z_3)}^{z_3} \om_{0,2}(z_1,\cdot)}{4 \om_{0,1}(z_1)} \cr
&&  - \frac{\int_{\sigma(z_2)}^{z_2} \om_{0,2}(z_1,\cdot) \; \int_{\sigma(z_3)}^{z_3} \om_{0,2}(z_2,\cdot)}{4 \om_{0,1}(z_2)}
- \frac{\int_{\sigma(z_3)}^{z_3} \om_{0,2}(z_1,\cdot) \; \int_{\sigma(z_2)}^{z_2} \om_{0,2}(z_3,\cdot)}{4 \om_{0,1}(z_3)} \cr
&& = - \frac{dx(z_1)}{2 y(z_1)} \sum_{p \in \mathcal{P}} \Res_{z' \to p} \frac{y(z')}{x(z')-x(z_1)} \Bigg[ d_{z'} F_{0,3}(z',z_2,z_3)  \cr
&& - \frac{\int_{\sigma(z_2)}^{z_2} \om_{0,2}(z',\cdot) \; \int_{\sigma(z_3)}^{z_3} \om_{0,2}(z_2,\cdot)}{4 \om_{0,1}(z_2)}
- \frac{\int_{\sigma(z_3)}^{z_3} \om_{0,2}(z',\cdot) \; \int_{\sigma(z_2)}^{z_2} \om_{0,2}(z_3,\cdot)}{4 \om_{0,1}(z_3)} \Bigg] .\cr
&&
\eea
Let us move the integration contour to compute the last line. One has
\beq
\begin{array}{rcl}
&&{\displaystyle \sum_{p \in \mathcal{P}} \Res_{z' \to p}} \frac{y(z')}{x(z')-x(z_1)}  \frac{\int_{\sigma(z_2)}^{z_2} \om_{0,2}(z',\cdot) \; \int_{\sigma(z_3)}^{z_3} \om_{0,2}(z_2,\cdot)}{4 \om_{0,1}(z_2)}
 = - \frac{ \int_{\sigma(z_3)}^{z_3} \om_{0,2}(z_2,\cdot)}{4 \om_{0,1}(z_2)} {\displaystyle \Res_{z' \to \begin{array}{l}z_1,\sigma(z_1) \cr z_2,\sigma(z_2)\end{array}}} \frac{y(z')}{x(z')-x(z_1)} \int_{\sigma(z_2)}^{z_2} \om_{0,2}(z',\cdot)  \cr
 &&\stackrel{(\ref{eq-skew-sym-int})}{=} - \frac{ \int_{\sigma(z_3)}^{z_3} \om_{0,2}(z_2,\cdot)}{2 \om_{0,1}(z_2)} {\displaystyle \Res_{z' \to z_1,z_2}} \frac{y(z')}{x(z')-x(z_1)} \int_{\sigma(z_2)}^{z_2} \om_{0,2}(z',\cdot) \cr
&&=- \frac{ y(z_1) \int_{\sigma(z_3)}^{z_3} \om_{0,2}(z_2,\cdot) \; \int_{\sigma(z_2)}^{z_2} \om_{0,2}(z_1,\cdot)}{2 \om_{0,1}(z_2) dx(z_1)} - \frac{\int_{\sigma(z_3)}^{z_3} \om_{0,2}(z_2,\cdot)}{2 dx(z_2) (x(z_2)-x(z_1))} .\cr
 \end{array}
\eeq
Thanks to this property, equation \eqref{EQ30New} reads
\beq
\begin{array}{l}
d_{z_1} F_{0,3}(z_1,z_2,z_3) + \frac{dx(z_1)}{4 y(z_1) } \left[ \frac{\int_{\sigma(z_2)}^{z_2} \om_{0,2}(z_1,\cdot) \; \int_{\sigma(z_3)}^{z_3} \om_{0,2}(z_1,\cdot)}{ dx(z_1)^2} 
+ \frac{ \int_{\sigma(z_3)}^{z_3} \om_{0,2}(z_2,\cdot)}{ dx(z_2) (x(z_2)-x(z_1))}
+ \frac{ \int_{\sigma(z_2)}^{z_2} \om_{0,2}(z_3,\cdot)}{dx(z_3) (x(z_3)-x(z_1))} \right]\cr
= - \frac{dx(z_1)}{2 y(z_1)} {\displaystyle \sum_{p \in \mathcal{P}} \Res_{z' \to p} } \frac{y(z')}{x(z')-x(z_1)} d_{z'} F_{0,3}(z',z_2,z_3) . \cr
\end{array}
\eeq
Observing that
\beq\label{eq-F02-berg}
\int_{\sigma(z_2)}^{z_2} \om_{0,2}(z_1,\cdot) = 2 d_{z_1} F_{0,2}(z_1,z_2) + \frac{dx(z_1)}{x(z_1)-x(z_2)}
\eeq
this simplifies to
\beq
\begin{array}{l}
d_{z_1} F_{0,3}(z_1,z_2,z_3) + \frac{ \left[d_{z_1} F_{0,2}(z_1,z_2)\right] \; \left[d_{z_1} F_{0,2}(z_1,z_3)\right] }{y(z_1) dx(z_1)} \cr
+ \frac{dx(z_1)}{2 y(z_1) } \frac{1}{x(z_1)-x(z_2)} \left[ \frac{d_{z_1} F_{0,2}(z_1,z_3)}{dx(z_1)} - \frac{d_{z_2} F_{0,2}(z_2,z_3)}{dx(z_2)} \right]
+ \frac{dx(z_1)}{2 y(z_1) } \frac{1}{x(z_1)-x(z_3)} \left[ \frac{d_{z_1} F_{0,2}(z_1,z_2)}{dx(z_1)} - \frac{d_{z_3} F_{0,2}(z_3,z_2)}{dx(z_3)} \right] \cr
= - \frac{dx(z_1)}{2 y(z_1)} {\displaystyle \sum_{p \in \mathcal{P}} \Res_{z' \to p} } \frac{y(z')}{x(z')-x(z_1)} d_{z'} F_{0,3}(z',z_2,z_3) . \cr
\end{array}
\eeq
Specializing to $z_2 = z_3 = z$, one obtains
\beq
\begin{array}{l}
d_{z_1} F_{0,3}(z_1,z,z) + \frac{ \left[d_{z_1} F_{0,2}(z_1,z)\right] \; \left[d_{z_1} F_{0,2}(z_1,z)\right] }{y(z_1) dx(z_1)} 
+ \frac{dx(z_1)}{ y(z_1) } \frac{1}{x(z_1)-x(z)} \left[ \frac{d F_{0,2}(z_1,z)}{dx(z_1)} - \left.  \frac{d F_{0,2}(z',z) }{dx(z')} \right|_{z' = z} \right] \cr
= - \frac{dx(z_1)}{2 y(z_1)} {\displaystyle \sum_{p \in \mathcal{P}} \Res_{z' \to p} } \frac{y(z')}{x(z')-x(z_1)} d_{z'} F_{0,3}(z',z,z) . \cr
\end{array}
\eeq

Finally, considering the limit $z_1 \to z$ and multiplying by $\frac{y(z)}{dx(z)}$ gives
\beq
\frac{2 y(z)}{6} \frac{d F_{0,3}(z,z,z) }{dx(z)} + \frac{1}{4} \left( \frac{d F_{0,2}(z,z)}{dx(z)} \right)^2 + \left. \frac{d^2 F_{0,2}(z_1,z)}{dx(z_1)^2}\right|_{z_1 = z}
 = - \frac{1}{2}{\displaystyle \sum_{p \in \mathcal{P}} \Res_{z' \to p} } \frac{y(z')}{x(z')-x(z)} d_{z'} F_{0,3}(z',z,z) . 
 \eeq
As before, the right hand side can be written in terms of the variational formulas to read
\beq
2 y(z) \frac{d \frac{F_{0,3}(z,z,z)}{6} }{dx(z)} +\left( \frac{d \frac{F_{0,2}(z,z)}{2}}{dx(z)} \right)^2 + \left. \frac{d^2 F_{0,2}(z_1,z)}{dx(z_1)^2}\right|_{z_1 = z} 
 = {\displaystyle \sum_{K = 2}^{r_{\infty}-2}} U_{\infty,K}(x(z)) \frac{\partial  \frac{F_{0,2}(z,z)}{2}}{\partial T_{b_\infty,K}} 
+ \sum_{\nu=1}^n {\displaystyle \sum_{K = 2}^{r_{\nu}+1}} U_{\nu,K}(x(z)) \frac{\partial  \frac{F_{0,2}(z,z)}{2}}{\partial T_{b_\nu,K}} . 
 \eeq

Applying the same procedure for $(h,n) = (1,1)$ gives
\beq\label{NEW1}
2 y(z) \frac{d F_{1,1}(z)}{dx(z)}
+ \left. \frac{d^2 F_{0,2}(z_1,z_2)}{dx(z_1) dx(z_2)} \right|_{z_1 = z_2 = z}
 = {\displaystyle \sum_{K = 2}^{r_{\infty}-2}} U_{\infty,K}(x(z)) \frac{\partial  F_{1,0}}{\partial T_{b_\infty,K}} 
+ \sum_{\nu=1}^n {\displaystyle \sum_{K = 2}^{r_{\nu}+1}} U_{\nu,K}(x(z)) \frac{\partial  F_{1,0}}{\partial T_{b_\nu,K}}.
\eeq
Using the fact that
\beq\label{NEW2}
\left. \frac{d^2 F_{0,2}(z_1,z_2)}{dx(z_1) dx(z_2)} \right|_{z_1 = z_2 = z}
+ \left. \frac{d^2 F_{0,2}(z_1,z)}{dx(z_1)^2}\right|_{z_1 = z} = 
\frac{d^2 \frac{F_{0,2}(z,z)}{2}}{dx(z)^2},
\eeq
the sum of \eq{NEW1} and \eq{NEW2} reads
\beq\label{eq-S1}
\begin{array}{rcl}
2 \frac{d S_{-1}^{+\text{ pert}}(x)}{dx} \frac{d S_{1}^{+\text{ pert}}(x)}{dx} + \left(\frac{dS_0^{+\text{ pert}}(x)}{dx} \right)^2 + \frac{d^2 S_0^{+\text{ pert}}(x)}{dx^2} &=& 
{\displaystyle \sum_{K = 2}^{r_{\infty}-2}} U_{\infty,K}(x) \frac{\partial  F_{1,0}}{\partial T_{b_\infty,K}} 
+ {\displaystyle \sum_{\nu=1}^n} {\displaystyle \sum_{K = 2}^{r_{\nu}+1}} U_{\nu,K}(x) \frac{\partial  F_{1,0}}{\partial T_{b_\nu,K}} \cr
&& +{\displaystyle \sum_{K = 2}^{r_{\infty}-2}} U_{\infty,K}(x) \frac{\partial  S_0^{+\text{ pert}}(x)}{\partial T_{b_\infty,K}} 
+ {\displaystyle \sum_{\nu=1}^n} {\displaystyle \sum_{K = 2}^{r_{\nu}+1}} U_{\nu,K}(x) \frac{\partial S_0^{+\text{ pert}}(x)}{\partial T_{b_\nu,K}}. \cr
\end{array}
\eeq

\subsubsection{Cases $2h-2+n \leq 0$}

For $2h-2+n = -1$, one has, by definition of the spectral curve,
\beq\label{eq-S-1}
\left(\frac{\partial S_{-1}^{+\text{ pert}}(x)}{\partial x} \right)^2 = \frac{\phi(x)}{dx^2}.
\eeq

For  $2h-2+n = 0$, thanks to \eq{eq-S0-logy} and the expression of the residues at poles in terms of variational formulas, one can write
\beq
2 y(z) \left. \frac{d F_{0,2}(z',z)}{dx(z')}\right|_{z'=z} + \frac{dy(z)}{dx(z)} =
 {\displaystyle \sum_{K = 2}^{r_{\infty}-2}} U_{\infty,K}(x(z)) \frac{\partial  S_{-1}^{+\text{ pert}}(x(z))}{\partial T_{\infty,K}} 
+ \sum_{\nu=1}^n {\displaystyle \sum_{K = 2}^{r_{\nu}+1}} U_{\nu,K}(x(z)) \frac{\partial S_{-1}^{+\text{ pert}}(x(z))}{\partial T_{\nu,K}},
\eeq
i.e.
\beq\label{eq-S0}
2 \frac{dS_{-1}^{+\text{ pert}}(x)}{dx} \, \frac{dS_{0}^{+\text{ pert}}(x)}{dx} + \frac{d^2 S_{-1}^{+\text{ pert}}}{dx^2} = 
{\displaystyle \sum_{K = 2}^{r_{\infty}-2}} U_{\infty,K}(x) \frac{\partial  S_{-1}^{+\text{ pert}}(x)}{\partial T_{\infty,K}} 
+ \sum_{\nu=1}^n {\displaystyle \sum_{K = 2}^{r_{\nu}+1}} U_{\nu,K}(x) \frac{\partial S_{-1}^{+\text{ pert}}(x)}{\partial T_{b_\nu,K}}.
\eeq

\subsubsection{Conclusion of the proof}

Summing the contributions coming from \eq{eq-S-1}, \eq{eq-S0}, \eq{eq-S1} and \eq{eq-Sm} for $m \geq 2$ with the appropriate $\hbar$ factors, one gets the ODE satisfied by the non-perturbative wave function as stated in Theorem \ref{th-PDE}.

\section{System of ODE for the non-perturbative wave functions}

\label{app-proof-ODE}

\subsection{Proof of Lemma \ref{lemma-wronsk-log}}
\label{app-lemma-wronsk-log}

Let us compute 
\beq
\frac{\partial W(x)}{\partial x} = \hbar \left(\frac{\partial^2 \Psi_+}{\partial x^2} \Psi_- - \Psi_+ \frac{\partial^2 \Psi_-}{\partial x^2}\right) .
\eeq
Since $\Psi_+$ and $\Psi_-$ are both solutions to \eq{eq-PDE-np}, 
\beq
\hbar \frac{\partial^2 \Psi_\pm}{\partial x^2} = \hbar  \sum_{p \in \mathcal{P}} \sum_{k \in K_p}  U_{p,k}(x)  \frac{\partial \Psi_\pm}{\partial T_{p,k}} + \hbar^{-1} \mathcal{H}(x) \Psi_{\pm}.
\eeq
plugging this back into the expression above gives
\beq
\frac{\partial W(x)}{\partial x} =   \sum_{p \in \mathcal{P}} \sum_{k \in K_p}  U_{p,k}(x) W_{{T}_{p,k}}(x)
\eeq
and the lemma follows.

\subsection{Proof of Lemma \ref{lemma-no-pole-at-branch-points-1}}
\label{app-lemma-no-pole-at-branch-points-1}

We shall prove this fundamental result by following the footsteps of \cite{Iwaki}. Let us first write down the compatibility of the system \eq{eq-syst-np}. For any $(p,k) \in \mathcal{P} \times K_p$, 
writing down the equality of $\frac{\partial^2}{\partial x^2} \left[\frac{\partial \Psi}{\partial T_{p,k}}\right] = \frac{\partial }{\partial T_{p,k}} \left[ \frac{\partial^2 \Psi}{\partial x^2}\right]$ and matching the coefficients of 
 $\Psi$ et $\frac{\partial \Psi}{ \partial x}$, one gets
\beq
0 = 2 \frac{\partial Q_{p,k}}{\partial x} - \hbar \frac{\partial R}{\partial T_{p,k}} + \hbar R_{p,k} \frac{\partial R}{\partial x} + \hbar R \frac{\partial R_{p,k}}{\partial x}
+ \hbar \frac{\partial^2 R_{p,k}}{\partial x^2}
\eeq
and
\beq
0 = 2 \left(\hbar Q + \mathcal{H}\right) \frac{\partial R_{p,k}}{\partial x} + \hbar R_{p,k} \frac{\partial Q}{\partial x} - \hbar R \frac{\partial Q_{p,k}}{\partial x} +  R_{p,k} \frac{\partial \mathcal{H}}{\partial x} - \frac{\partial }{\partial T_{p,k}} \left( \hbar Q + \mathcal{H} \right) + \hbar \frac{\partial^2 Q_{p,k}}{\partial x^2}.
\eeq

Let us now write the expansion in $\hbar$ of these equalities order by order. For this purpose, let us remark that both equations can be put under the form of a trans-series as functions of $\hbar$
\beq
\sum_{m,{\mathbf{k}}} \alpha_{m,\mathbf{k}} \hbar^m \exp\left(\hbar^{-1} \sum_{j=1}^g k_j v_j\right) = 0
\eeq
meaning that the coefficients of the trans-monomials vanish. Summing over vectors $\mathbf{k}$, this implies that the coefficient of $\hbar^m$ is vanishing for any $m$. In order to compute it, let us remarks that derivation with respect to $x$ preserves this $\hbar$ grading while differentiation with respect to the times decreases the degree by $1$.

To leading order, this reads
\beq\label{eq-compatibility-1}
0 = 2 \frac{\partial Q_{p,k}^{(0)}}{\partial x} - \frac{\partial \phi}{\partial T_{p,k}} \left.\frac{\partial \hat{R}^{(0)}(x,\mathbf{T},v)}{\partial v} \right|_{v = \frac{\phi+\rho}{\hbar}}
\eeq
and
\beq\label{eq-compatibility-2}
0 = 2  \mathcal{H}^{(0)} \frac{\partial R_{p,k}^{(0)}}{\partial x} +  R_{p,k}^{(0)} \frac{\partial \mathcal{H}^{(0)}}{\partial x} - \frac{\partial \phi }{\partial T_{p,k}}  \left. \frac{\partial \hat{Q}^{(0)}(x,\mathbf{T},v)}{\partial v} \right|_{v = \frac{\phi+\rho}{\hbar}} - \frac{\partial\mathcal{H}^{(0)} }{\partial T_{p,k}} .
\eeq

Let us now assume that the functions $Q_{p,k}^{(0)}$ and $R_{p,k}^{(0)}$ have a pole at a critical value $u_i$ such that they behave as
\beq
\left\{
\begin{array}{l}
\hat{Q}_{p,k}^{(0)}(x,\mathbf{T},v) = \frac{q_{p,k}^{(0)}}{(x-u_i)^{d_{p,k}}} + O((x-u_i)^{-d_{p,k}+1})\cr
\hat{R}_{p,k}^{(0)}(x,\mathbf{T},v) = \frac{r_{p,k}^{(0)}}{(x-u_i)^{d_{p,k}'}} + O((x-u_i)^{-d_{p,k}'+1}) \cr
\end{array}
\right.
\eeq
for some positive degrees $(d_{p,k},d_{p,k}')$ and write in the same way
\beq
\left\{
\begin{array}{l}
\hat{Q}^{(0)}(x,\mathbf{T},v) = \frac{q^{(0)}}{(x-u_i)^{d}} + O((x-u_i)^{-d+1}) , \cr
\hat{R}^{(0)}(x,\mathbf{T},v) = \frac{r^{(0)}}{(x-u_i)^{d'}} + O((x-u_i)^{-d'+1}). \cr
\end{array}
\right.
\eeq

The leading order in $x \to u_i$ of \eq{eq-compatibility-1} reads
\beq
- 2 d_{p,k} \frac{q_{p,k}^{(0)}}{(x-u_i)^{d_{p,k}+1}}  = \frac{\partial \phi}{\partial T_{p,k}} \left.\frac{\partial r^{(0)}}{\partial v} \right|_{v = \frac{\phi+\rho}{\hbar}}  \frac{1 }{(x-u_i)^{d'}}
\eeq
meaning that $d_{p,k}$ is independent of $(p,k)$ and $d_{p,k} = d'-1$ for any pair $(p,k)$. From the definition of $Q$, it behaves as
\beq
\hat{Q}^{(0)}(x,\mathbf{T},v) =  \left[\sum_{p \in \mathcal{P}} \sum_{k \in K_p} U_{p,k}(u_i) q_{p,k}^{(0)} \right]\frac{1}{(x-u_i)^{d'-1}} + O((x-u_i)^{-d'+2}) .
\eeq

The leading order of \eq{eq-compatibility-2} reads
\beq
\left(- 2 d_{p,k} +1\right) \left.\left[\frac{\partial \mathcal{H}^{(0)}}{\partial x}\right]\right|_{x=u_i} \frac{r_{p,k}^{(0)}}{(x-u_i)^{d_{p,k}'+1}}
 = \frac{\partial \phi}{\partial T_{p,k}} \left. \frac{\partial q^{(0)}}{\partial v} \right|_{v = \frac{\phi+\rho}{\hbar}} \frac{1}{(x-u_i)^d}
 \eeq
which implies that $d_{p,k}' = d-1 = d'-2$ for any pair $(p,k)$ which contradicts the definition of $R$.

Thus $R^{(0)}(x)$ and $Q_{p,K}^{(0)}(x)$ are holomorphic at the critical values. This implies that $Q^{(0)}(x)$ is holomorphic as well at these points. Finally, \eq{eq-compatibility-2} together with the fact that $Q^{(0)}(x)$ is holomorphic implies that $R_{p,K}^{(0)}(x)$ is holomorphic as well at these points.

We shall now proceed by induction on $h$  for proving that the coefficients $R_{p,K}^{(h)}$ and $Q_{p,K}^{(h)}$ are holomorphic at the critical values. For this purpose, one can write the $h$'th order of the compatibility conditions as
\beq
2 \frac{\partial Q_{p,K}^{(h)}}{\partial x} - \hbar \frac{\partial \phi}{\partial T_{p,K}} \left.\frac{\partial \tilde{R}^{(h)}(x,\mathbf{T},v)}{\partial v} \right|_{v = \frac{\phi}{\hbar}} = \hbox{lower order terms}
\eeq
and
\beq
2  \mathcal{H}^{(0)} \frac{\partial R_{p,K}^{(h)}}{\partial x} +  R_{p,K}^{(h)} \frac{\partial \mathcal{H}^{(0)}}{\partial x} - \frac{\partial \phi }{\partial T_{p,K}}  \left. \frac{\partial \tilde{Q}^{(h)}(x,\mathbf{T},v)}{\partial v} \right|_{v = \frac{\phi}{\hbar}} - \frac{\partial\mathcal{H}^{(h)} }{\partial T_{p,K}} = \hbox{lower order terms}
\eeq
where the right hand sides are lower order terms in the $\hbar$ expansion which are holomorphic at the critical values by induction. Using the same argument as for the leading order, one can conclude that any $R_{p,K}^{(h)}$ and $Q_{p,K}^{(h)}$ are holomorphic at these points.

\section{Expansions of $S^\pm$ and Wronskians around $X_{\nu}$ and $\infty$}
\label{app-asymptot}

In this appendix, we recall the expansions around the poles of certain useful quantities.

\begin{itemize}

\item {\bf{Around $x = X_\nu$, for $\nu\in\llbracket 1 ,n\rrbracket$.} }

The logarithm of the wave functions reads
\beq
S_\pm(x)  = \mp \hbar^{-1} \sum_{k=2}^{r_{b_\nu}} \frac{T_{\nu,k}}{k-1} \frac{1}{(x-X_\nu)^{k-1}} \pm \hbar^{-1} T_{\nu,1} \log(x-X_\nu) + \sum_{k=0}^\infty A_{\nu,k}^\pm (x-X_\nu)^k.
\eeq 

The Wronskian in $x$ behaves as 
\beq
W(x) = \frac{2 T_{\nu,r_\nu}}{(x-X_\nu)^{r_\nu}} \exp\left(A_{\nu,0}^+ + A_{\nu,0}^-\right) + O\left((x-X_\nu)^{-r_\nu+1}\right).
\eeq
The Wronskians in the spectral times at finite pole read
\bea
&&\forall\, \nu' \in\llbracket 1,n\rrbracket \, , \; \forall\, k \geq 2 \, : \; W_{T_{\nu',k}}(x) \cr
&&= \left[- \frac{2\delta_{\nu,\nu'}}{(k-1)(x-X_\nu)^{k-1}} + \hbar \frac{ \partial \left(A_{\nu,0}^+ - A_{\nu,0}^-\right)}{\partial T_{\nu',k}} + O\left((x-X_\nu)\right) \right] \Psi_+(x)  \Psi_-(x),
\eea
i.e.
\beq
\forall\, k \geq 2 \, , \;  W_{T_{\nu,k}}(x) = -  \frac{2}{k-1} \frac{\exp\left(A_{\nu,0}^+ + A_{\nu,0}^-\right) }{(x-X_\nu)^{k-1}} + O\left((x-X_\nu)^{-k+2}\right)
\eeq
and
\beq
\forall \,\nu' \neq \nu \, , \; \forall\, k \geq 2 \, : \; W_{T_{\nu',k}}(x) = \hbar \frac{ \partial \left(A_{\nu,0}^+ - A_{\nu,0}^-\right)}{\partial T_{\nu',k}} \exp\left(A_{\nu,0}^+ + A_{\nu,0}^-\right) + O\left((x-X_\nu)\right) .
\eeq
The Wronskians in the spectral times at $\infty$ read
\beq
 \forall\,  k \geq 2 \, : \; W_{T_{\infty,k}}(x) = \hbar \frac{ \partial \left(A_{\nu,0}^+ - A_{\nu,0}^-\right)}{\partial T_{\infty,k}} \exp\left(A_{\nu,0}^+ + A_{\nu,0}^-\right) + O\left((x-X_\nu)\right) .
\eeq

\item {\bf Around $x = \infty$ if $n_\infty = 0$.}
The logarithms of the wave functions read
\beq
S_\pm(x)  = \mp \hbar^{-1} \sum_{k=2}^{r_\infty} \frac{T_{\infty,k}}{k-1} x^{k-1} \mp \hbar^{-1} T_{\infty,1} \log(x) - \frac{\log x}{2} +  \sum_{k=0}^\infty A_{\infty,k}^\pm x^{-k}.
\eeq
The Wronskian in $x$ behaves as 
\beq
W(x)= - 2 T_{\infty,r_\infty} \exp \left( A_{\infty,0}^+ + A_{\infty,0}^-\right) \, x^{r_\infty-3} + O\left(x^{r_\infty-4}\right).
\eeq
The Wronskians in the spectral times at finite poles read
\beq
\forall \,\nu\in\llbracket 1,n\rrbracket \, , \; \forall\, k \geq 2 \, : \; W_{T_{\nu,k}}(x) = O\left(x^{-1}\right).
\eeq
The Wronskians in the spectral times at $\infty$ read
\beq
 \forall \,k \geq 2 \, , \; W_{T_{\infty,k}}(x) = - \frac{2}{k-1} \exp \left( A_{\infty,0}^+ + A_{\infty,0}^-\right) \, x^{k-2}.
 \eeq

\item {\bf Around $x = \infty$ if $n_\infty = 1$.}
\beq
S_\pm(x)  = \mp \hbar^{-1} \sum_{k=2}^{r_\infty} \frac{T_{\infty,k}}{2k-3} x^{\frac{2k-3}{2}} \mp \hbar^{-1} T_{\infty,1} \log(x) - \frac{\log x}{4} + \sum_{k=1}^\infty A_{\infty,k}^\pm x^{-\frac{k}{2}}.
\eeq

The Wronskian in $x$ behaves as 
\beq
W(x) = -  T_{\infty,r_\infty} \exp \left( A_{\infty,0}^+ +  A_{\infty,0}^-\right) \, x^{r_\infty-3} + O\left(x^{r_\infty-3-\frac{1}{2}}\right).
\eeq
The Wronskians in the spectral times at finite poles read
\beq
\forall \,\nu\in\llbracket 1,n\rrbracket \, , \; \forall \, k \geq 2 \, : \; W_{T_{\nu,k}}(x) = O\left(x^{-\frac{1}{2}}\right).
\eeq
The Wronskians in the spectral times at $\infty$ read
\beq
 \forall\, k \geq 2 \, : \; W_{T_{\infty,k}}(x) = - \frac{2}{2k-3} \exp \left( A_{\infty,0}^+ + A_{\infty,0}^-\right) \, x^{k-2}.
 \eeq
\end{itemize}

Note that one may obtain the next orders in the expansions of the Wronskians from those of $S_\pm$ with the formula:
\bea W(x)&=&\hbar\left(\frac{\partial S_+(x)}{\partial x}-\frac{\partial S_-(x)}{\partial x}\right)\text{exp}\left(S_+(x)+S_-(x)\right)\cr
W_{T_{\nu,k}}(x)&=&\hbar\left(\frac{\partial S_+(x)}{\partial T_{\nu,k}}-\frac{\partial S_-(x)}{\partial T_{\nu,k}}\right)\text{exp}\left(S_+(x)+S_-(x)\right)
\eea


\begin{thebibliography}{99}


\bibitem{AHH2} M.R.~Adams, J.~Harnad, J.~Hurtubise, ``Isospectral Hamiltonian Flows in Finite and Infinite Dimensions'',  \textit{Communications in Mathematical Physics}, Vol. \textbf{134}, 555-585, 1990.

\bibitem{Darboux} M.R.~Adams, J.~Harnad, J.~Hurtubise, ``Spectral Darboux Coordinates and Liouville-Arnold Integration in Loop Algebras'', \textit{Communications in Mathematical Physics}, Vol. \textbf{155}, Issue 2, 385-413, 1993.


\bibitem{LoopLie}R.~Belliard, B.~Eynard, O.~Marchal, ``Integrable differential systems of topological type and reconstruction by the topological recursion'', \textit{Annales Henri Poincar\'{e}}, Vol. \textbf{18}, Issue 10, 3193-3248, 2017.

\bibitem{Bertola-Korotkin} M.~Bertola, D.~Korotkin and C.~Norton, ``Symplectic geometry of the moduli space of projective structures in homological coordinates'', \textit{Inventiones mathematicae}, Vol. \textbf{210}, Issue 3, 759-814, 2015.

\bibitem{Borot-review} G.~Borot, ``Lecture notes on topological recursion and geometry'',  \href{http://arxiv.org/abs/1705.09986}{\textit{arxiv:1705.09986}}, 2017.


\bibitem{BEInt} G.~Borot, B.~Eynard, ``Geometry of spectral curves and all order dispersive integrable system'', \textit{Symmetry, Integrability and Geometry: Methods and Applications}, Vol. \textbf{8}, 2012. 

\bibitem{BEKnot} G.~Borot, B.~Eynard, ``All-order asymptotics of hyperbolic knot invariants from non-perturbative topological recursion of A-polynomials'', \textit{Quantum Topology}, Vol. \textbf{6}, Issue 1, 39-138, 2015.

\bibitem{Reconstruction} V. Bouchard, B. Eynard, ``Reconstructing WKB from topological recursion'', \textit{Journal de l'Ecole Polytechnique - Math\'{e}matiques}, Vol. \textbf{4}, 845-908, 2017.


\bibitem{Bridgeland} T.~Bridgeland and I.~Smith, ``Quadratic differentials as stability conditions'', \textit{Publications math\'ematiques de l'IHES}, Vol. \textbf{121}, Issue 1, 155-278.


\bibitem{CGL} M.~Cafasso, P.~Gavrylenko, O.~Lisovyy, ``Tau functions as Widom constants'', \textit{Communications in Mathematical Physics}, Vol. \textbf{365}, {Issue 2}, 1-32, 2019.


\bibitem{DKnot} R.~Dijkgraaf, H.~Fuji and M.~Manabe, ``The Volume Conjecture, Perturbative Knot Invariants, and Recursion Relations for Topological Strings'', \textit{Nuclear Physics B}, Vol. \textbf{849}, 166, 2011.

\bibitem{Quantum} O.~Dumitrescu, M.~Mulase, ``Quantum Curves for Hitchin Fibrations and the Eynard-Orantin Theory'', \textit{Letters in Mathematical Physics}, Vol. \textbf{104}, Issue 6, 635-671, 2014. 

\bibitem{DBOSS} P.~Dunin-Barkowski, N.~Orantin, S.~Shadrin, L.~Spitz, ``Identification of the Givental formula with the spectral curve topological recursion procedure'', \textit{Communications in Mathematical Physics}, Vol. \textbf{328}, 669-700, 2014. 

\bibitem{DBGW} P.~Dunin-Barkowski, M.~Mulase, P.~Norbury, A.~Popolitov and S.~Shadrin, ``Quantum spectral curve for the Gromov-Witten theory of the complex projective line'', \textit{Journal f\"ur die reine und angewandte Mathematik}, Published Online, 2014.

\bibitem{Eyn-periods} B.~Eynard,``The Geometry of integrable systems. Tau functions and homology of Spectral curves. Perturbative definition'', \href{https://arxiv.org/abs/1706.04938}{\textit{arxiv:1706.04938}}, 2017.

\bibitem{Eyn-np} B.~Eynard and  M.~Mari\~no, ``A holomorphic and background independent partition function for matrix models and topological strings'', 
\textit{Journal of Geometry and Physics}, Vol. \textbf{61}, 1181-1202, 2011.

\bibitem{EO} B.~Eynard, N.~Orantin, ``Invariants of algebraic curves and topological expansion'', \textit{Communications in Number Theory and Physics}, Vol. \textbf{1}, 347-452, 2007.

\bibitem{Eynbook} B.~Eynard, ``Counting surfaces: CRM Aisenstadt Chair Lectures'', \textit{Progress in Mathematical Physics}, Vol. \textbf{70}, Birkh"{a}user Springer, 2016.

\bibitem{Fay} J.D.~Fay, ``Theta Functions on Riemann Surfaces'', \textit{Lecture Notes in Mathematics}, Vol. \textbf{352}, Springer-Verlag Berlin Heidelberg, 1973.

\bibitem{RMatrixHarnad} J.~Harnad, M.~Routhier,  ``R-matrix construction of electromagnetic models for the Painlev\'{e} transcendents'', \textit{Journal of Mathematical Physics}, Vol. \textbf{36}, Issue 9, 1998.

\bibitem{RMatrixHarnad2} J. Harnad, M.A. Wisse ``Moment Maps to Loop Algebras, Classical R-Matrix and Integrable Systems'', \textit{Quantum Groups Integrable Models and Statistical Systems}, 1992.

\bibitem{LoopAlgebraHarnad} J. Harnad, M. A. Wisse, ``Loop Algebra Moment Maps and Hamiltonian Models for the Painlev\'{e} Transcendants'', \textit{Fields Institute Communications}, Vol. \textbf{7}, 155-169, 1996.

\bibitem{Teschner-and-co} N.~Iorgov, O~Lisovyy, J.~Teschner, ``Isomonodromic Tau-Functions from Liouville Conformal Blocks'', \textit{Communications in Mathematical Physics}, Vol. \textbf{336}, Issue 2, 671-694, 2015.

\bibitem{IwakiExactWKB} K. Iwaki, T. Nakanishi, ``Exact WKB analysis and cluster algebras'', \textit{Journal of Physics A: Mathematical and Theoretical}, Vol. \textbf{47}, Issue 47, 2014.

\bibitem{IwakiExactWKB2} K. Iwaki, T. Nakanishi, ``Exact WKB Analysis and Cluster Algebras II: Simple Poles, Orbifold Points, and Generalized Cluster Algebras'', \textit{International Mathematical Research Notices}, Vol. \textbf{14}, 4375-4417, 2016.

\bibitem{Iwaki} K.~Iwaki, ``2-parameter $\tau$-function for the first Painlevé equation :Topological recursion and direct monodromy problem via exact WKB analysis'', \textit{Communications in Mathematical Physics}, Vol. \textbf{377}, 1047-1098, 2020.

\bibitem{P2} K.~Iwaki, O.~Marchal, ``Painlev\'e 2 equation with arbitrary monodromy parameter, topological recursion and determinantal formulas'', \textit{Annales Institut Henri Poincar\'{e}}, Vol. \textbf{18}, Issue 8, 2581-2620, 2017.

\bibitem{MarchalIwaki} K.~Iwaki, O.~Marchal, A.~Saenz, ``Painlev\'e equations, topological type property and reconstruction by the topological recursion'', \textit{Journal of Geometry and Physics}, Vol. \textbf{124}, 16-54, 2016.

\bibitem{IS} K.~Iwaki, A.~Saenz, ``Quantum Curve and the First Painlev\'e Equation'', \textit{Symmetry, Integrability and Geometry: Methods and Applications}, Vol. \textbf{12}, 2016.


\bibitem{Kont} M.~Konsevich, ``Intersection theory on the moduli space of curves and the matrix Airy function'', \textit{Communications in Mathematical Physics}, Vol. \textbf{147}, 1-23, 1992.

\bibitem{LiuHurwitz}  X.~Liu, M.~Mulase,  and A.~Sorkin, ``Quantum curves for simple Hurwitz numbers of an arbitrary base curve'', \href{https://arxiv.org/abs/1304.0015}{\textit{arxiv:1304.0015}}, 2013.

\bibitem{MO} O.~Marchal, N.~Orantin, ``Isomonodromic deformations of a rational differential system and reconstruction with the topological recursion: the $\mathfrak{sl}_2$ case'', \textit{Journal of Mathematical Physics}, Vol. \textbf{61}, 2020.

\bibitem{ReviewNorbury} P.~Norbury, ``Quantum curves and topological recursion'', \textit{Proceedings of Symposia in Pure Mathematics}, Vol. \textbf{93}, 41-65, 2016. 

\bibitem{Teschner} J.~Teschner, ``Quantization of the Hitchin moduli spaces, Liouville theory, and the geometric Langlands correspondence I'', \textit{Advances in Theoretical and Mathematical Physics}, Vol. \textbf{15}, Issue 2, 471-564, 2011.

\bibitem{Witten} E.~Witten, ``Two-dimensional gravity and intersection theory on moduli space'', \textit{Surveys in differential geometry} (Cambridge, MA), 1, Bethlehem, PA: Lehigh Univ., 243-310, 1990.

\end{thebibliography}
\end{document}